\documentclass[journal]{vgtc}                     

\onlineid{0}

\vgtccategory{Research}

\vgtcpapertype{please specify}

\usepackage{subfigure}
\newlength{\myfigbesidewidth}

\usepackage{endnotes} 
\newcounter{notelabelcount}
\renewcommand{\footnote}[1]{%
   \addtocounter{notelabelcount}{1}%
   \hyperlink{notelabel\thenotelabelcount}{%
   \endnote{%
			\addtocounter{notelabelcount}{1}%
			\hypertarget{notelabel\thenotelabelcount}%
      {\protect #1}%
   }}%
}

\newcommand*\circled[1]{\tikz[baseline=(char.base)]{
            \node[shape=circle,draw,inner sep=1pt] (char) {#1};}}
            
\newcommand{\papertitle}{Reframing \emph{Pattern}:\\ A Comprehensive Approach to a Composite Visual Variable}
\title{\papertitle}

\author{%
  \authororcid{Tingying He}{0000-0002-9670-5587},
  \authororcid{Jason Dykes}{0000-0002-8096-5763},
  \authororcid{Petra Isenberg}{0000-0002-2948-6417},
  \authororcid{Tobias Isenberg}{0000-0001-7953-8644}%
}

\authorfooter{
    \item T.\ He (\raisebox{-.5pt}{\includegraphics[height=6pt]{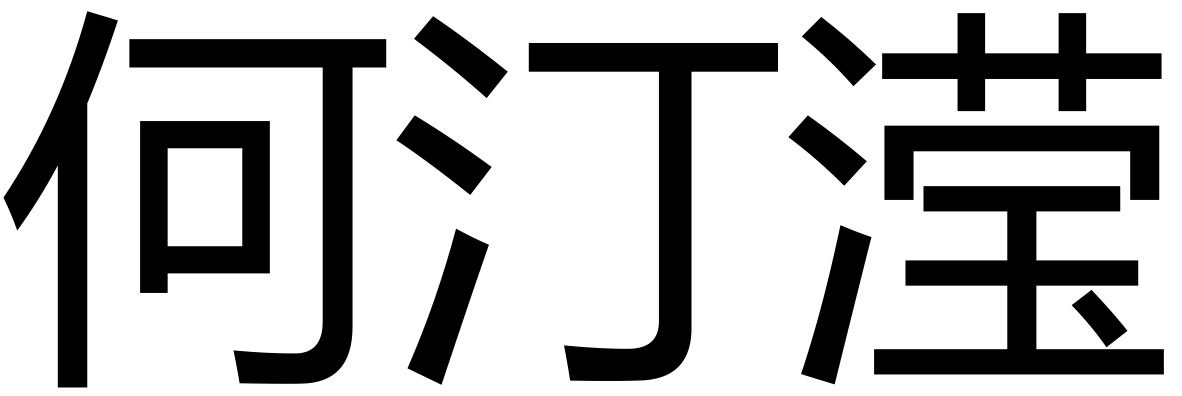}}) is with Université Paris-Saclay, CNRS, Inria, LISN, France, and the University of Utah, USA. E-mail: hetingying.hty@gmail.com.
    \item J.\ Dykes is with City, University of London, UK. E-mail: j.dykes@city.ac.uk.
    \item P.\ and T.\ Isenberg are with Université Paris-Saclay, CNRS, Inria, LISN, France. E-mail: given\_name.family\_name@inria.fr.%
}

\abstract{
We present a new comprehensive theory for explaining, exploring, and using \emph{pattern} as a visual variable in visualization.
Although patterns have long been used for data encoding and continue to be valuable today, their conceptual foundations are precarious:
the concepts and terminology used across the research literature and in practice are inconsistent, making it challenging to use patterns effectively and to conduct research to inform their use.
To address this problem, we conduct a comprehensive cross-disciplinary literature review that clarifies ambiguities around the use of ``pattern'' and ``texture’’. 
As a result, we offer a new consistent treatment of \emph{pattern} as a composite visual variable composed of structured groups of graphic primitives that can serve as marks for encoding data individually and collectively.
This new and widely applicable formulation opens a sizable design space for the visual variable \emph{pattern}, which we formalize as a new system comprising three sets of variables: the spatial arrangement of primitives, the appearance relationships among primitives, and the retinal visual variables that characterize individual primitives.
We show how our pattern system relates to existing visualization theory and highlight opportunities for visualization design. 
We further explore patterns based on complex spatial arrangements, demonstrating explanatory power and connecting our conceptualization to broader theory on maps and cartography.
An author version and additional materials are available on OSF: \osfrepo.}


\graphicspath{{figs/}{figures/}{pictures/}{images/}{./}} 

\usepackage{tabu}                      
\usepackage{booktabs}                  

\usepackage{mathptmx}                  
\usepackage{flushend}

\usepackage{cuted} 
\usepackage{ccicons} 
\usepackage{placeins}
\usepackage{dblfloatfix}
\usepackage{setspace}
\usepackage[normalem]{ulem}

\newcommand{\inlinevis}[3]{\raisebox{#1}[0pt][0pt]{\includegraphics[height=#2]{#3}}}

\newcommand{\fignumber}[1]{\raisebox{.5pt}{\small{\circled{#1}}}}

\usepackage{xspace}
\newcommand{\eg}{e.\,g.}
\newcommand{\ie}{i.\,e.}
\newcommand{\pp}{primitive\xspace} 
\newcommand{\pps}{primitives\xspace} 

\newcommand{\osfrepo}{\href{https://osf.io/9dwgq/}{\href{https://osf.io/z7ae2/}{\texttt{osf\discretionary{}{.}{.}io\discretionary{/}{}{/}z7ae2}}}}

\newcommand{\changed}[1]{\textcolor{Crimson}{#1}} 
\newcommand{\todo}[1]{\textcolor{Magenta}{#1}}
\newcommand{\hty}[1]{\textcolor{SeaGreen}{#1}} 

\renewcommand{\todo}[1]{#1}
\renewcommand{\changed}[1]{#1}
\renewcommand{\hty}[1]{#1} 
\begin{document}



\firstsection{Introduction}

\maketitle

\label{sec:introduction}


Visualization design at its very core relies on the mapping of data values to visual variables. Visual variables, also referred to as visual channels, are attributes of graphical elements---referred to as ``marks'' or ``symbols''---whose appearance can be manipulated to encode data \cite{Munzner:2015:VisualizationAnalysisAndDesign}. 
The visual variables that we draw upon are much studied and include
\emph{position}, \emph{hue}, and \emph{size} \cite{Bertin:1998:SG,bertin:1983:semiology}. Their effectiveness ranking has been the subject of much research and discussion in our field \cite{cleveland:1985:graphical,mackinlay:1986:automating,mccoleman:2021:rethinking,MacEachren:2012:Uncertainty,DiBiase:1992:Animation} \changed{with scope for encoding nominal, ordered, and numeric data \cite{roth:2017:visual}}. Among the available visual variables is one that researchers have called \emph{pattern} \cite{MacEachren:2004:HowMapsWork},
typically featuring repetitive dots or lines \inlinevis{-1pt}{1em}{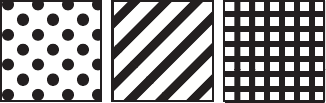}. Visualization designers often use these patterns when color is either limited or already encodes other data dimensions (\eg, \cite{bertin:1983:semiology, Bertin:1998:SG, brinton:1919:graphic, Brinton:1939:GP, tufte:1985:visual, Retchless:2015:Uncertainty, Ware:2009:Quantitative, Chan:2019:ViBr,Shenas:2005:CCT,Kumpf:2018:VCC,deBrouwer:2017:FTL,He:2023:DEB}). 

When patterns \inlinevis{-1pt}{1em}{inline-geometric-patterns} are described as visual variables, researchers also refer to them as \emph{texture}. 
This interchangeability of the terms ``pattern'' and ``texture'' may arise from the blended use of both terms in everyday language and the use of ``texture'' as a translation of Bertin's visual variable ``grain'' in the English translation of his seminal and highly influential work \cite{Bertin:1998:SG, bertin:1983:semiology}.
To add to the confusion, however, the term \emph{texture} has a diverse set of meanings in visualization research that goes beyond an understanding of texture as pattern \mbox{\inlinevis{-1pt}{1em}{inline-geometric-patterns}\hspace{1pt}.} Researchers working on 3D representations, for example, often use \emph{texture} for surface or volume characteristics of 3D objects, represented as realistic images \inlinevis{-1pt}{1em}{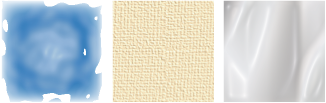} \cite{Johnson:2020:ArtifactBasedRendering,lu:2005:volume}.
These textures typically have different visual characteristics and encoding goals from \changed{the more structured repeating patterns} that are used as a visual variable in abstract data representations
. 
Even in the specific context of visual variables used for abstract data representations, researchers may interpret the term \emph{texture} as a variation of a specific dimension of a \emph{pattern}, such as ``granularity'' (Bertin's ``grain'' in French) \mbox{\inlinevis{-1pt}{1em}{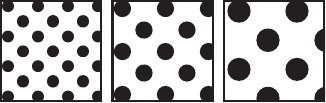}\hspace{1pt},} the spacing between the repeated elements \mbox{\inlinevis{-1pt}{1em}{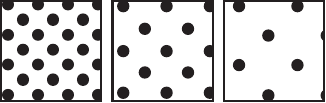}\hspace{1pt},} or the elements' shape \mbox{\inlinevis{-1pt}{1em}{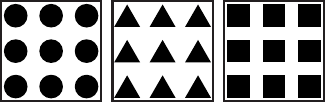}\hspace{1pt}.} This \changed{méli-mélo de terminologie},
we argue, prevents the research community from investigating \emph{pattern} as a visual variable or using its encoding effectively. Research on patterns and their practice of use \changed{is} difficult to compare and situate in the absence of a consistent terminology. 
\changed{We consider this situation to be a theoretical irritant with practical im\-pli\-ca\-tions---a persistent blind
spot that we aim to address with our work.}

Our re-reading of the literature \cite{Carpendale:2003:ConsideringVisualVariables,MacEachren:2004:HowMapsWork,Munzner:2015:VisualizationAnalysisAndDesign,Ware:2019:InformationVisualization} leads us to a consistent terminology, in which we use the term ``pattern'' to describe a composite visual variable
\inlinevis{-1pt}{1em}{inline-geometric-patterns} that consists of graphical primitives that can also serve as marks for data encoding.  \textbf{Our first contribution} is an in-depth discussion and clarification of the terms \emph{texture} and \emph{pattern} in light of existing interpretations around both terms\changed{, to address the inconsistent terminology problem.} 
As \textbf{our second contribution}, \changed{we introduce a new pattern system that extends the conceptualization of \emph{pattern} and its potential variations.} 
\changed{This system builds upon our terminology to describe a design space in ways that can be used for
encoding, exploration, and experiments.} 
We identify three sets of attributes of a pattern: the spatial arrangement of primitives, the appearance relationships among primitives, and the individual appearance characteristics of primitives. 
\hty{Furthermore, we examine more complex spatial arrangements of primitives, using the concept of pattern as a theoretical lens to compare, explain, and connect different types of visualizations and to uncover new design opportunities.}

\section{Texture and pattern}
\label{sec:texture-and-pattern}

Researchers often use the terms \emph{pattern} (\eg, \cite{krygier:2016:MakingMaps, MacEachren:2004:HowMapsWork,tyner:2017:PrinciplesOfMapDesign}) or \emph{texture} (\eg, \cite{He:2024:DCB,Ware:2009:Quantitative,ware:2022:countable}) to describe a visual variable characterized by repeated elements \inlinevis{-1pt}{1em}{inline-geometric-patterns}. \hty{Although} both terms can make sense and are understandable, Carpendale \cite{Carpendale:2003:ConsideringVisualVariables}, in her discussion of visual variables, suggests to use the term \emph{texture} for ``apparent surface quality of the material like wood or marble'' and to use \emph{pattern} for ``repetitive use of shape variations.'' 
We consider Carpendale's recommendation reasonable and useful\footnote{We agree that \emph{texture} may primarily be used for materials, but argue that the \emph{pattern} concept extends beyond the repetitive use of shape variations (\autoref{sec:texture-interpretation-pattern}).} 
due to two main issues associated with the term \emph{texture}: (1)  compared to \emph{pattern}, the term \emph{texture} has a broader meaning in visualization and related fields, can refer to different concepts (as we show in \autoref{fig:textures-in-vis} and \autoref{fig:patterns-in-vis}), making it less precise; and (2), even when \emph{texture} specifically refers to a visual variable, it is subject to different interpretations, \changed{even post-Carpendale \cite{Carpendale:2003:ConsideringVisualVariables},} as can be observed by comparing various publications that use the term \cite{kraak:2020:CartographyVisualizationOfGeospatialData, krygier:2016:MakingMaps,slocum:2022:ThematicCartography,tyner:2017:PrinciplesOfMapDesign}.
%
Below we discuss the first of these issues and clarify the use of \emph{texture} and \emph{pattern} in the visualization literature, and explain why \emph{pattern} is a more suitable term for this type of visual variable. We address the second issue in \autoref{sec:pattern-as-variable}.

\subsection{Texture: Surface characteristics}
The term \emph{texture} is often used to describe an object's ``visual or tactile surface characteristics and appearance'' \cite{mw:texture}. In computer graphics, especially in work that relates to rendering, \emph{texture} is a widely used concept. In this context, it essentially refers to a data structure that stores characteristic information (visual or other). It is typically represented as a multidimensional array. Through texture sampling, we obtain the necessary data from the texture and map it onto the corresponding location of the object. The visual texture that we ultimately observe on the object is the result of the rendering process \cite{Blinn:1976:TRC,Catmull:1974:SAC}. From an appearance standpoint, textures are often closely related to real-world materials, have a sense of depth and realism, and often look continuous.

\newlength{\imagethumbnailwidth}
\setlength{\imagethumbnailwidth}{0.295\columnwidth}
\begin{figure}[]
\centering
\setlength{\subfigcapskip}{-3ex}
\textcolor{white}{\subfigure[\hspace{.275\columnwidth}]{\label{fig:textures-in-vis:a}\includegraphics[height=\imagethumbnailwidth]{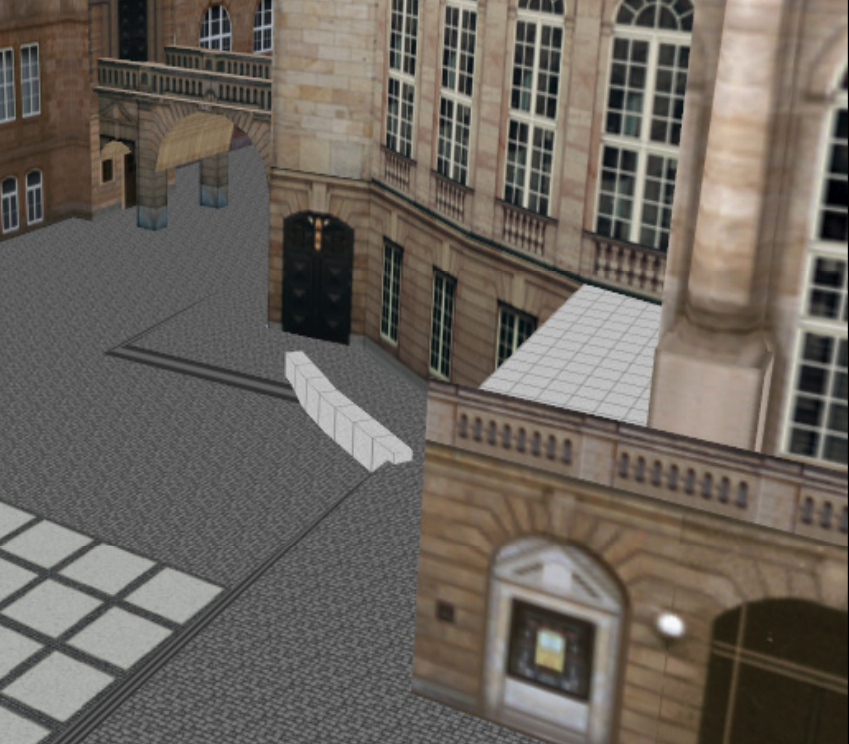}}}\hfill%
\textcolor{white}{\subfigure[\hspace{.255\columnwidth}]{\label{fig:textures-in-vis:b}\includegraphics[height=\imagethumbnailwidth]{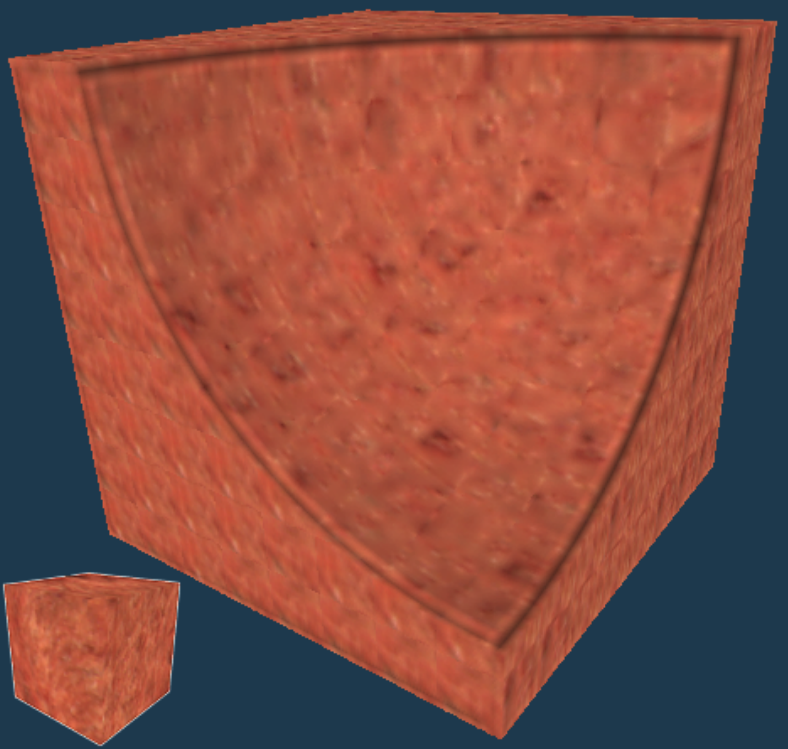}}}\hfill%
\textcolor{white}{\subfigure[\hspace{.243\columnwidth}]{\label{fig:textures-in-vis:c}\includegraphics[height=\imagethumbnailwidth]{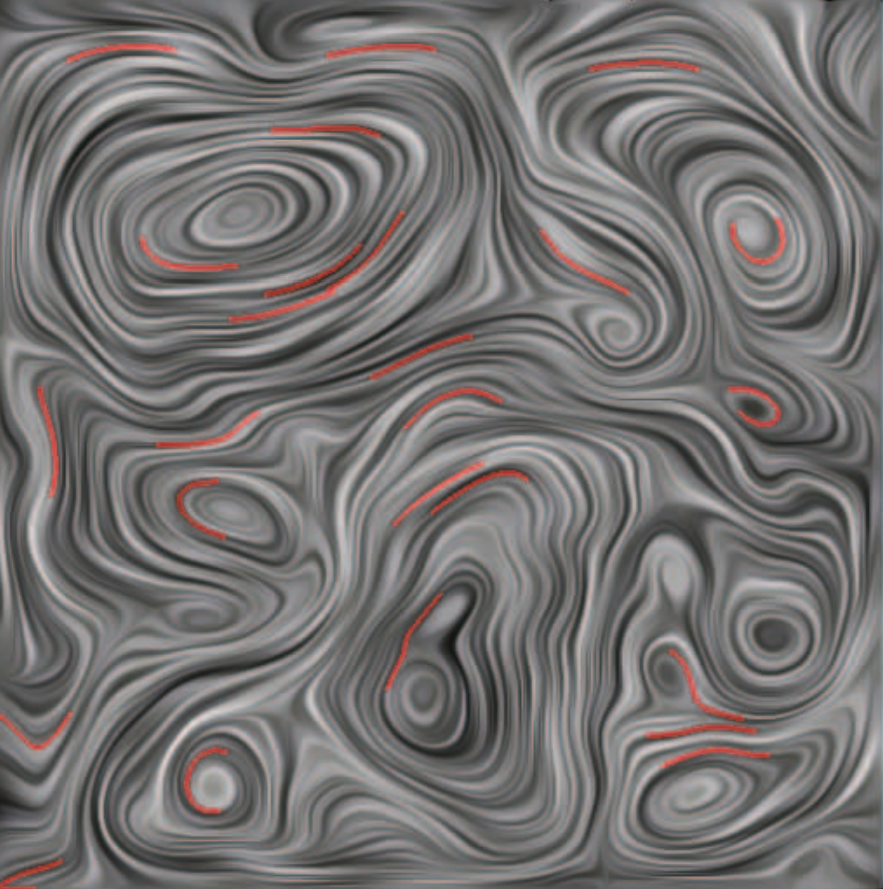}}}
\caption{Textures in (a) surface rendering \cite{Buchholz:2005:rendering}, (b) volume rendering \cite{lu:2005:volume}, and (c) flow visualization \cite{Matvienko:2015:flow}; all images \textcopyright\ IEEE; used with permission.}
\label{fig:textures-in-vis}
\end{figure}

Leveraging techniques from computer graphics, researchers in 3D visualization use \emph{texture}, for example, to depict materials of a model's surface (\eg, \autoref{fig:textures-in-vis:a}) or to define a volume's visual characteristics (\eg, \autoref{fig:textures-in-vis:b}). In flow visualization, researchers also use techniques that rely on textures, such as Line Integral Convolution (LIC) \cite{Cabral:1993:LIC} or spot noise \cite{vanWijk:1991:spot}, to represent the directionality, magnitude, and other attributes of vectors or tensors (\eg, \autoref{fig:textures-in-vis:c}).  

Texture is also used to refer to surface characteristics in other vi\-su\-a\-li\-za\-tion-re\-la\-ted fields beyond computer graphics. 
\hty{It is recognized as one of the seven elements of art}, denoting the characteristics of an object's material \cite{Gatto:1978:ExploringVisualDesign}.
In the visual arts, visual textures are called \emph{implied textures} (in contrast to \emph{actual textures}, which are tactile), \eg, to create a simulated appearance of physical materials\cite{Gatto:1978:ExploringVisualDesign}. 
In computer vision, researchers study \emph{texture analysis} techniques (\eg, \emph{texture segmentation} and \emph{classification}) to enable computers to recognize objects and understand scenes \cite{tuceryan:1993:texture}. 
In the vision sciences, researchers study \emph{texture perception} to understand how humans perceive surface qualities \cite{Rosenholtz:2015:texturePerception}.

Carpendale \cite{Carpendale:2003:ConsideringVisualVariables} brings multiple perspectives together in her discussion of visual variables for data visualization. She specifically discusses the possibility of using surface materials (\ie, the computer graphics interpretation of \emph{texture}) as a separate visual variable \cite[Table~9]{Carpendale:2003:ConsideringVisualVariables}\hty{, and illustrates it with examples using photographic images 
\cite[Table~11]{Carpendale:2003:ConsideringVisualVariables}. 
In this case, a (texture) image is applied to a chart element, and what we read is both the chart element's value and the information in the image.} 
Because both the visual appearance and the information encoding in these examples differ from the visual variable examined in our work, we recommend
\changed{using the term texture for this visual variable in line with Carpendale’s suggestion---yet a more specific term such as ``surface material texture''} would make for a clearer distinction.

\setlength{\imagethumbnailwidth}{0.300\columnwidth}
\begin{figure}[]
\centering
\setlength{\subfigcapskip}{-3ex}
\textcolor{white}{\subfigure[\hspace{.245\columnwidth}]{\includegraphics[height=\imagethumbnailwidth,trim={0 400 400 0},clip]{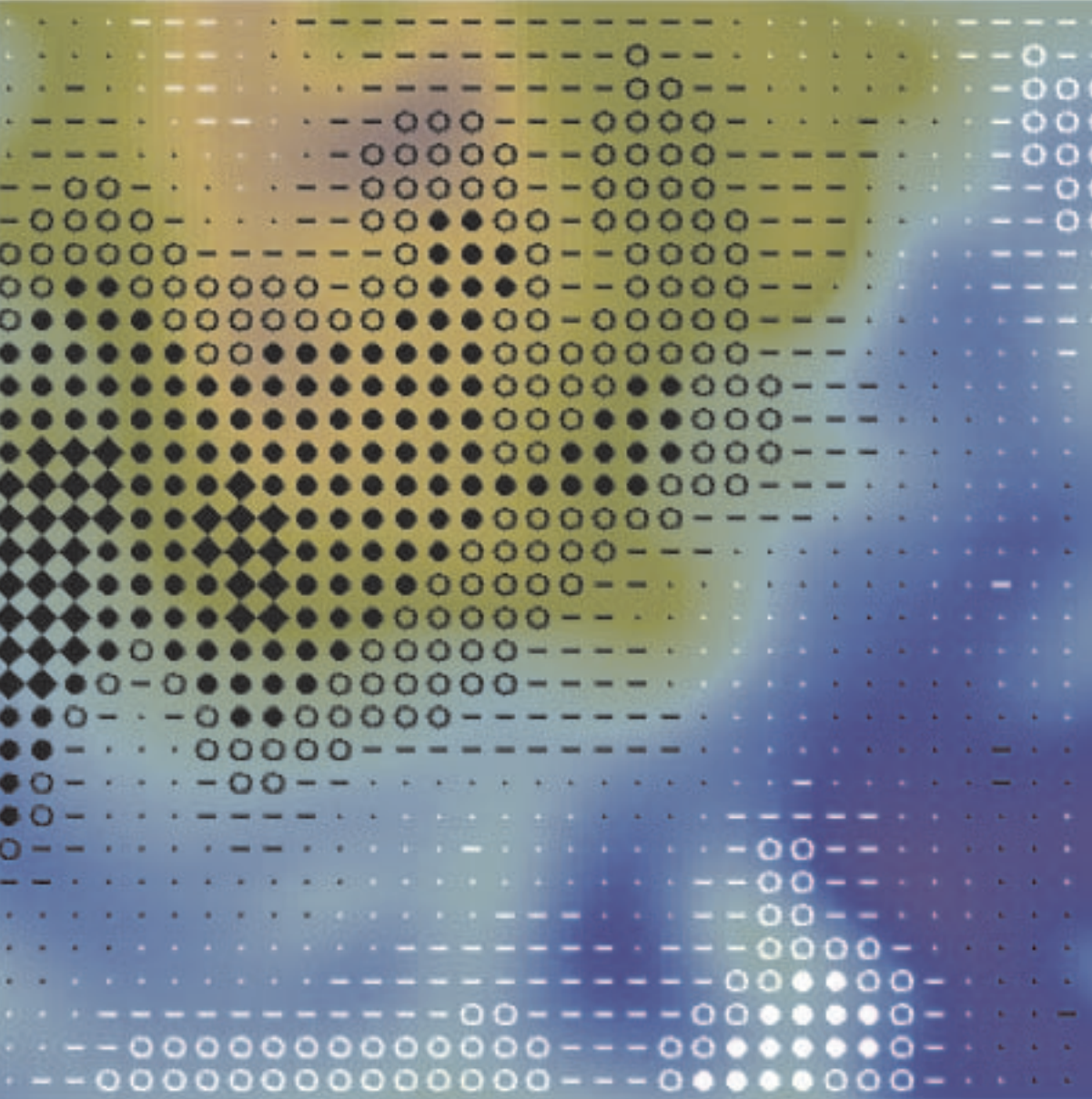}}}\hfill%
\subfigure[\hspace{.265\columnwidth}]{\includegraphics[height=\imagethumbnailwidth, trim={0 240 680 400},clip]{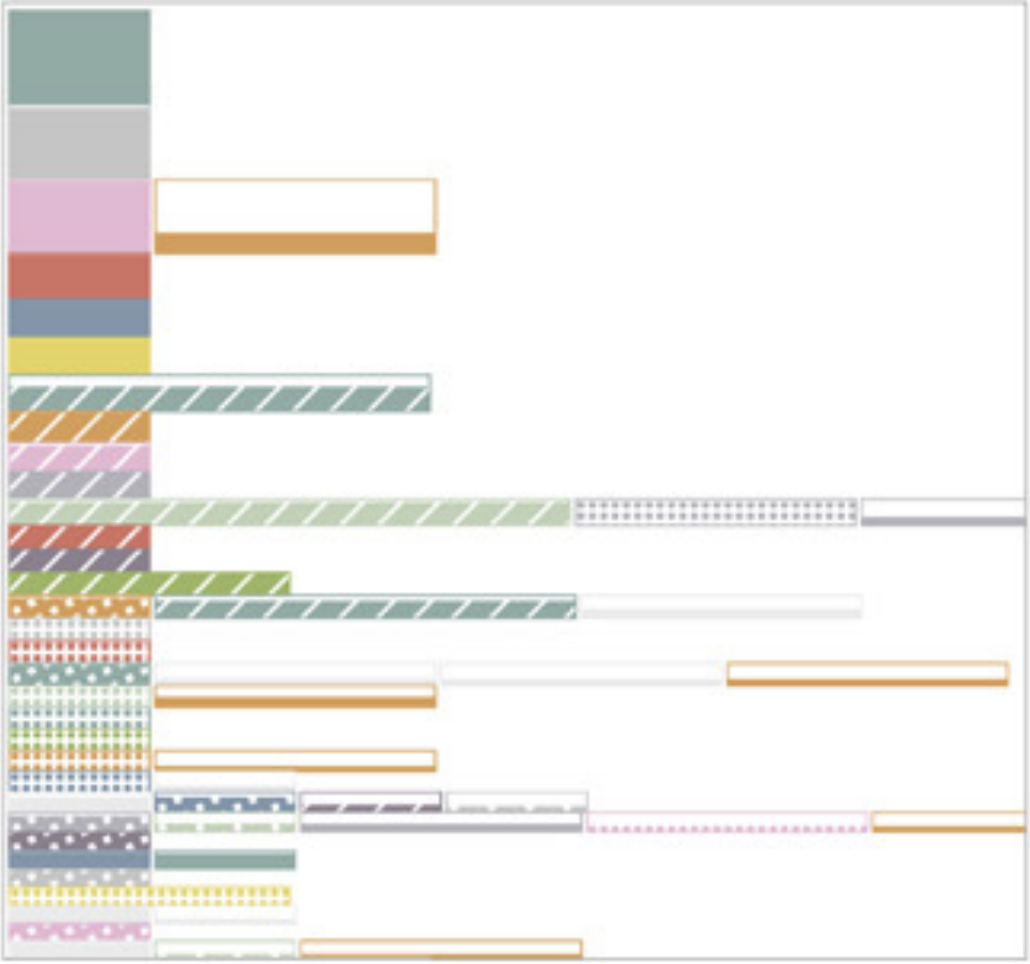}}\hfill%
\subfigure[\hspace{\columnwidth}]{\label{fig:patterns-in-vis:c}\includegraphics[height=\imagethumbnailwidth]{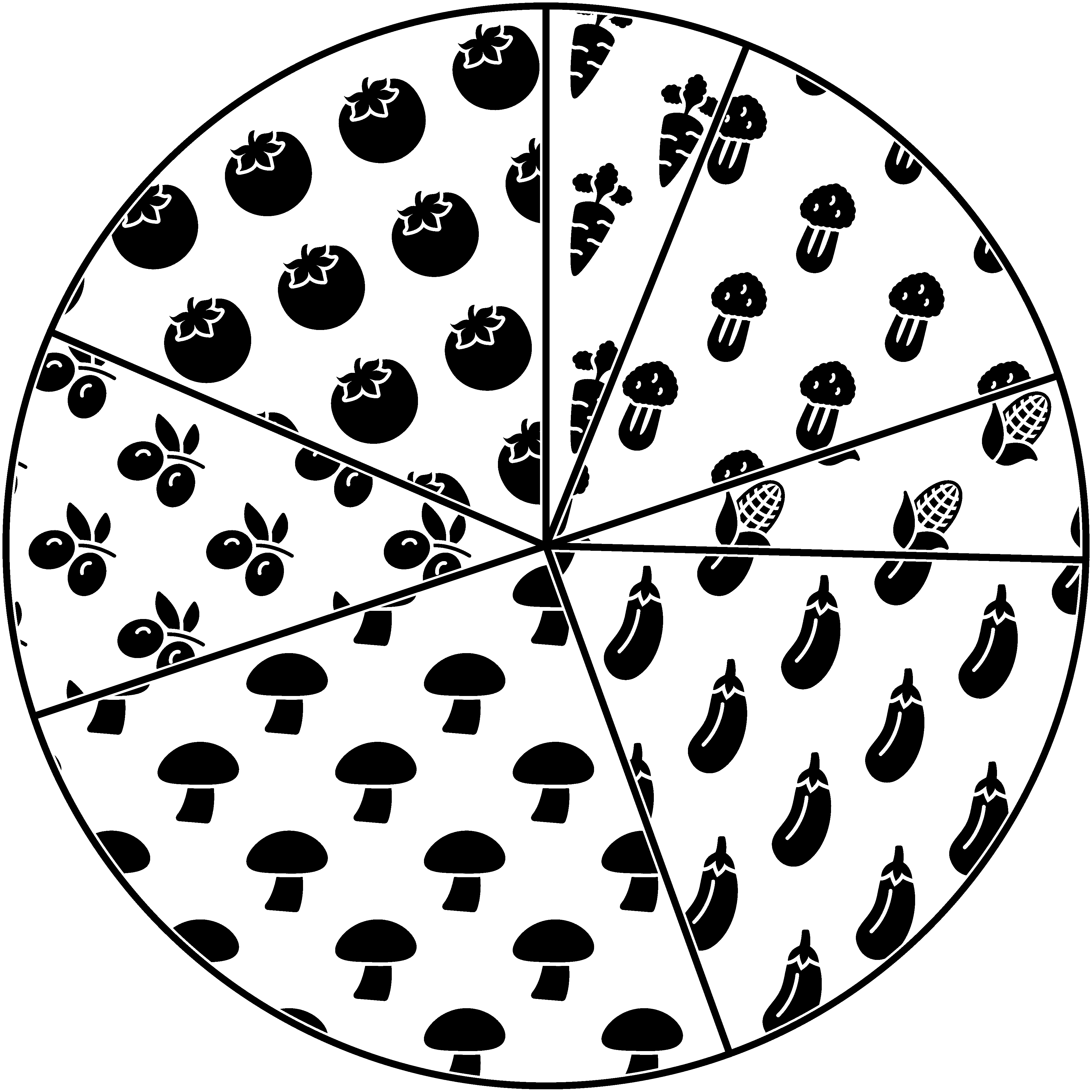}}
\caption{Patterns used in the visualization community that are called ``texture,'' from (a) \cite{Ware:2009:Quantitative}, (b)\cite{Chan:2019:ViBr} , and (c) \cite{He:2024:DCB}; (a) and (b) \textcopyright\ IEEE, (c) \href{https://creativecommons.org/licenses/by/4.0/}{\ccLogo\,\ccAttribution\ \mbox{CC BY 4.0}}; all images used with permission.}
\label{fig:patterns-in-vis}
\end{figure}

\subsection{Pattern: Repetition and structure}
\label{sec:texture-and-pattern:pattern}

\hty{A body of research and practice that has its roots in cartography and statistical graphics interprets \emph{texture} in a different manner. In \autoref{fig:patterns-in-vis} we show examples of this type of ``texture.'' 
Researchers map data dimensions to the graphical features of these textures. 
These textures typically feature clearer and more distinguishable repetitive elements than those textures used in computer graphics rendering.
\hty{even though} they may carry meaning through mimetic properties, they are generally unrelated to surface materials.
}

\hty{We describe this use of ``texture'' as \emph{pattern}---a concept that \hty{emphasizes aspects different from} \emph{texture}.} The term ``pattern'' originated from the same root as ``patron,'' derived from the Latin \emph{patronus}, meaning ``protector'' or ``defender'' \cite{Cresswell:2010:OxfordWordOrigins}. It evolved to signify ``an example to be copied'' \cite{Cresswell:2010:OxfordWordOrigins}, emphasizing the repetition of elements---rather than the tactile or perceived feeling of a material (\ie, a texture). Note that the term \emph{pattern} is not limited to visual elements, it is 
a structural concept that can be applied to the abstract\footnote{\changed{It is used as the basis for theoretical constructs in architecture \cite{Alexander:1977:APatternLanguage}, mathematics \cite{Grenander:2007:PatternTheory}, and beyond.}} as well as the physical world.\footnote{In our daily life, \eg, \emph{pattern} can refer to many physical items and abstract concepts that include repetition, such as a social/behavioral patterns, sound patterns, language structures, or chronological orders.}

We focus on the visual aspects of \emph{pattern}. The emphasis on repetition and structure makes the concept of \emph{pattern} particularly suitable for describing visual variables in the form of \changed{structured marks that repeat to fill space}, capturing both their composite encoding and abstract appearance. As Wilkinson \cite{wilkinson:2013:GrammarOfGraphics} in his discussion of visual variables mentioned, ``these [visual variables] are not
ones customarily used in computer graphics to create realistic scenes. They are not even sufficient for a semblance of realism.''
Nevertheless, patterns can also characterize a surface, suggesting that we can view patterns as a specific type of ``texture,'' one that describes a surface with distinct primitives and structured arrangement. When the repeated elements in a texture are clearly identifiable, the texture takes on the characteristics of a \emph{pattern}.
This overlap between the two concepts may explain why some researchers use the terms \emph{pattern} and \emph{texture} interchangeably.

\subsection{Summary}
The term ``texture'' is used differently in different visualization contexts, with meanings derived and used in computer graphics and in abstract data representation having some similarities, but important differences.
%
Both can characterize a surface and add visual complexity. \emph{Texture} often describes the appearance of a surface and its material properties, while \emph{pattern} emphasizes the repetition and structure of elements that involves semiotics and is frequently used in abstract data representations.
We acknowledge the overlap of the terms \emph{texture} and \emph{pattern} and are sensitive to other researchers' use of both terms. In our case of using it as a visual variable, however, we suggest that \emph{pattern} is a more precise term than \emph{texture}, as \emph{patterns} rely on repetition and are not usually meant to suggest surface material. 

\section{Pattern as a visual variable: Three interpretations under the term of ``texture''}
\label{sec:pattern-as-variable}

While \emph{pattern} is a more precise term \changed{if we think of it as a visual variable}, we frequently see ``texture'' in lists of visual variables. This \changed{use of} ``texture'' to describe a visual variable can be traced back to Bertin's seminal book, \textit{Semiology of Graphics} \cite{bertin:1983:semiology, Bertin:1998:SG}, in which he introduced the first set of visual variables---``texture'' among them. 
Subsequent literature has continued to use this term (\eg, \cite{Carpendale:2003:ConsideringVisualVariables, mackinlay:1986:automating}), however, with different interpretations. It has been referred as the variation of granularity in a \emph{pattern} (\eg, \cite{bertin:1983:semiology,Bertin:1998:SG, kraak:2020:CartographyVisualizationOfGeospatialData}), the spacing of a \emph{pattern} (\eg, \cite{dent:1999:CartographyThematicMapDesign, slocum:2022:ThematicCartography}), the shape variation of a \emph{pattern}, or a \emph{pattern} in its entirety (\eg, \cite{krygier:2016:MakingMaps,tyner:2017:PrinciplesOfMapDesign}).\footnote{These examples are all from cartography textbooks  with lists of visual variables. By observing how they interpret ``texture'' in various ways we found inconsistencies in the understanding of the term ``texture'' as a visual variable.} This inconsistency has its roots in the translation of Bertin's book, which we \changed{carefully unpick as follows}.



\setlength{\myfigbesidewidth}{.6\columnwidth}
\begin{figure}
\begin{minipage}[b]{.53\columnwidth}\includegraphics[width=\textwidth]{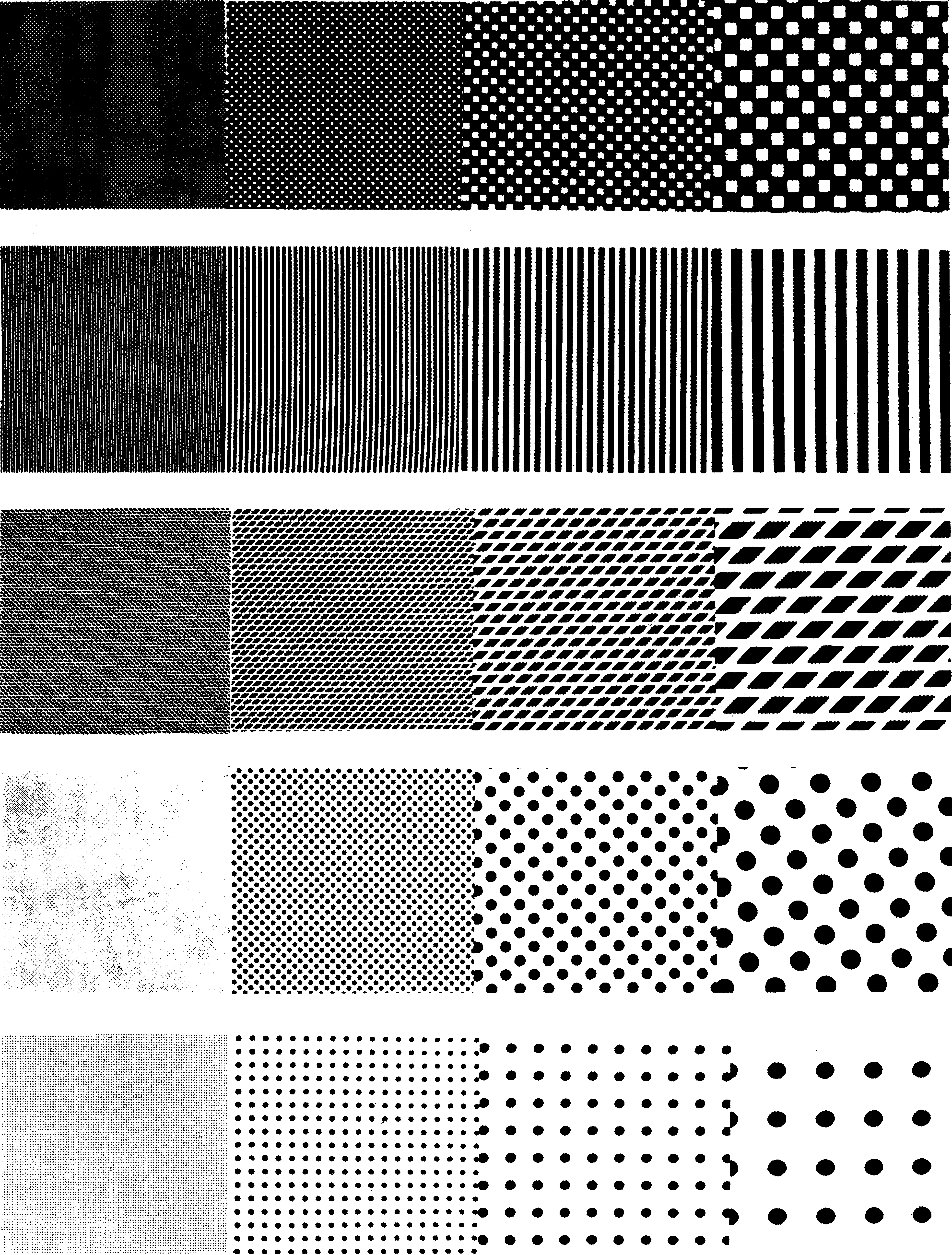}\vspace{4pt}\end{minipage}\hfill
\begin{minipage}[b]{.42\columnwidth}\caption{Bertin's diagram for granularity variation, described in French as ``Horizontalement: grain. Verticalement: valeur et forme'' [our translation: horizontal: granularity. vertical: value and shape] \cite{Bertin:1998:SG} and in English as ``texture is given horizontally; value and shape [pattern] vertically'' \cite{bertin:1983:semiology}; image \textcopyright~EHESS, used with permission.}\label{fig:BertinBook-TextureAndPattern}\end{minipage}
\end{figure}

\newlength{\fboxwidthflush}
\setlength{\fboxwidthflush}{\columnwidth}
\addtolength{\fboxwidthflush}{-2\fboxsep}
\begin{figure}[t]
    \centering
        \vspace{1ex}%
				\framebox{\includegraphics[width=\fboxwidthflush]{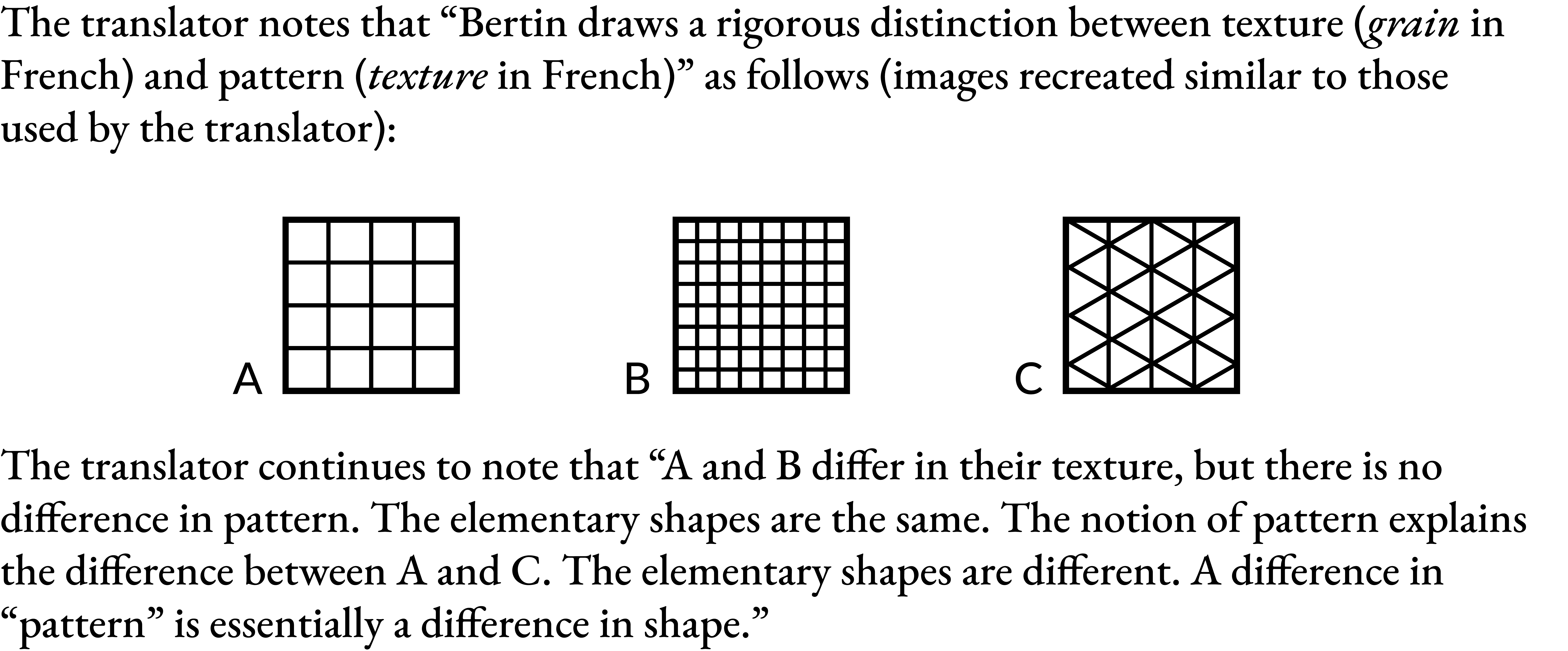}}
        \caption{Citation of the translator's note (with redrawn images similar to the original ones) in the English edition of Bertin's book \cite{Bertin:1998:SG}.}
    \label{fig:BertinBook-TranslatorNote}
\end{figure}

\subsection{Grain: The original term Bertin used}
\label{sec:texture-grain}
The French term ``grain'' is the original word that Bertin used to describe the visual variable \cite{bertin:1983:semiology}, which William J. Berg translated to ``texture'' in the English version of this book \cite{bertin:1983:semiology}. Bertin defined the visual variable as follows: ``at a given value, the [granularity]\footnote{In contrast to the official translation of the book, which uses the term ``texture,'' we intentionally changed the translation here to use ``granularity'' and also not ``grain,'' for reasons that we explain further below.} represents the number of separable marks within a unit area.'' In the ``texture'' palettes from his book that we reproduce in \autoref{fig:BertinBook-TextureAndPattern}, the variation of granularity along each \emph{horizontal} palette involves changing both the size and spacing of primitives \emph{simultaneously}, while maintaining a given ratio of black to white. As a result, the average \emph{value} of each square stays constant. This effect is similar to what can be achieved by zooming in or out of a pattern or through photographic reduction \cite{Bertin:1998:SG, bertin:1983:semiology}. 

Researchers questioned the translation of Bertin's French ``grain'' variation to the English term ``texture,'' suggesting that ``grain'' or ``granularity'' would be more precise translations. MacEachren \cite{MacEachren:2004:HowMapsWork}, \eg, suggests that the English term ``grain'' may be better to describe this variation, as it is similar to the grain in film. Similarly, Wilkinson \cite{wilkinson:2013:GrammarOfGraphics} said that Bertin ``really means granularity (as in the granularity of a photograph).''
Carpendale \cite{Carpendale:2003:ConsideringVisualVariables} also noted that this variation more closely relates to a variation of granularity and directly referred to it using the English term ``grain.''  Between ``grain'' and ``granularity,'' we recommend using ``granularity'' in English, based on the rationale that it is not the ``grain'' itself that varies but rather the size of the grain, which is more accurately described by ``granularity.'' In English, moreover, ``grain'' can refer to the longitudinal \changed{structures} of wood fibers (\ie, ``wood grain''), potentially conveying a sense of direction. The concept of direction, however, is not implied in Bertin's (French) \emph{grain}, leading us to argue that ``granularity'' is a better translation.

\subsection{Spacing: A misinterpretation in Bertin's book}
\label{sec:visual_variable:density}
Another interpretation of the word ``texture'' refers to the spacing of primitives in a \emph{pattern}. Spacing between primitives can affect \emph{density}---the smaller the spacing, the more densely packed the primitives. This variation is called ``spacing'' by Brewer \cite{brewer:2016:DesigningBetterMaps} and Slocum et al. \cite{slocum:2022:ThematicCartography}, ``density'' by Mackinlay \cite{mackinlay:1986:automating}, or ``frequency'' by Chung et al. \cite{chung:2016:ordered}.

The interpretation of ``texture'' to relate to spacing may arise from a misinterpretation in a translator's note in the English version of Bertin's book, which we reproduce in \autoref{fig:BertinBook-TranslatorNote}.
As we had just discussed, for Bertin changes in \emph{granularity} (French: ``grain,'' translated in \autoref{fig:BertinBook-TranslatorNote} to ``texture'') require that a constant average value is upheld. The trans\-la\-tor's note, however, refers to the difference between Squares A and B as a change in ``texture'' (\ie, ``grain'' in French or, for us, \emph{granulatity}). Yet, A and B do NOT share the same average value, as A has a lower black-to-white ratio than B. Therefore, A and B would not constitute a change of the (French) ``grain'' for Bertin. Instead, if we see the black lines as primitives of the pattern, we can see from the figure that A and B differ only in the \emph{spacing} between primitives, changing the black-to-white ratio while maintaining the \emph{primitive shape} (cf. \autoref{fig:BertinBook-TextureAndPattern}).

\subsection{Pattern: Not only shape variation}
\label{sec:texture-interpretation-pattern}
``Pattern,'' largely in the sense we have established in \autoref{sec:texture-and-pattern:pattern}, is a third term often used interchangeably with ``texture'' when referring to a visual variable---a confusion stemming both from the semantic overlap discussed in \autoref{sec:texture-and-pattern} and from the translator's interpretation of Bertin's work.
In the original French edition \cite{Bertin:1998:SG}, Bertin described the \emph{vertical} change between \emph{corresponding ``palette'' entries} in \autoref{fig:BertinBook-TextureAndPattern} is a variation of ``value and shape'' (French: ``valeur et forme''). 
It is unclear if this interpretation was supported by Bertin, but the English translator of the book amended this statement to ``value and shape [pattern]'' \cite{bertin:1983:semiology}. This amendment seems reasonable: we can see in \autoref{fig:BertinBook-TextureAndPattern} that the differences between palette entries on each column are not just differences in \emph{value} and \emph{shape}, but also include differences in \emph{size} of the elements, \emph{spacing} between the elements, etc.---which are all variations a \emph{pattern} can have. 
In the translator's note we just mentioned (\autoref{fig:BertinBook-TranslatorNote}), however, the translator explained that ``a difference in `pattern' is essentially a difference in shape.''%
\footnote{\changed{In this translator's note (\autoref{fig:BertinBook-TranslatorNote}), the translator also mentioned that ``pattern'' is the English equivalent for the French word ``texture.'' In the book, however, the term ``pattern'' is not actually translated from ``texture'' in French. The translator translated the following French terms from Bertin's book into English as ``pattern'': ``semis,'' ``baguettage/ligne,'' and ``trame.'' We discuss these terms in detail in \autoref{sec:more-on-pattern}.}}
Carpendale \cite{Carpendale:2003:ConsideringVisualVariables} adopts this interpretation, defining \emph{pattern} as ``repetitive use of shape variations (the use of marks upon marks),'' and equates the impact of using the visual variable \emph{pattern} in visual interpretative tasks with the visual variable \emph{shape}. 
This interpretation captures the emphasis of \emph{pattern} on repetition well and the notion of ``the use of marks upon marks'' touches an apparent inconsistency of Bertin's use of visual variables, which we discuss in \autoref{sec:bertin-inconsistency}. 
The variations of \emph{patterns}, however, should not be limited to the change of shape as we just discussed for \autoref{fig:BertinBook-TextureAndPattern}.

\subsection{Summary and our recommendation}


``Texture'' has multiple interpretations, and when used without clarification or diagrams in the context of abstract data representation, its meaning is often ambiguous. We therefore recommend reserving ``texture'' for surface materials and using ``pattern'' to describe the visual variable that features structure and repetition \inlinevis{-1pt}{1em}{inline-geometric-patterns}. Given the many subdimensions of patterns, they should be described comprehensively and consistently. To support this, we propose a consistent and expressive descriptive terminology in the remainder of the paper.

\section{Additional related work on pattern variations}
To fully understand what truly constitutes a pattern, we first review previous work on identifying \hty{subdimensions} of \emph{pattern}, then highlight an inconsistency in Bertin's use of visual variables---one that inspires our development of a comprehensive description of \emph{pattern}.


\subsection{Pattern description from two perspectives}

Researchers have investigated variations of \emph{pattern} from two perspectives, corresponding to the encoding and decoding processes. In encoding, designers use visual variables to represent differences in data; in decoding, readers perceive these variations and interpret them as data differences. The description of \emph{pattern} can thus be approached from two directions: what designers can control and what readers can perceive. Research from disciplines with a design focus has proposed \hty{subdimensions} of \emph{pattern} from a design perspective, while research from the field of vision science has explored dimensions of \emph{pattern} from a perceptual perspective. \changed{Our proposals draw upon both perspectives.}

\subsubsection{Perception perspective}
To be able to use pattern for encoding data effectively, it is vital to understand how the human visual system perceives such visual stimuli. 
Vision science researchers have tried to identify the most important perceptual dimensions that are useful for humans to judge the difference between appearance of textures (also known as texture features). 

Tamura et al. \cite{tamura:1978:textural} propose six basic texture features, namely, coarseness, contrast, directionality, line-likeness, regularity, and roughness. Amadasun and King \cite{amadasun:1989:textural} approximate five perceptual texture attributes in computational form, namely coarseness, contrast, busyness, complexity, and strength of texture. Rao and Lohse \cite{rao:1996:towards} identify a Texture Naming System with \hty{the three} most significant dimensions in natural texture perception: ``repetitive vs.\ non-repetitive; high-contrast and non-directional vs.\ low-contrast and directional; granular, coarse and low-complexity vs.\ non-granular, fine and high-complexity.'' Liu and Picard \cite{liu:1996:periodicity} identify three mutually orthogonal dimensions of texture that are important to human texture perception, namely periodicity, directionality, and randomness. Cho et al. \cite{cho:2000:reliability} extend the perceptual research and reported four texture dimensions: coarseness, contrast, lightness, and regularity. Features such as coarseness, roughness, or strength remind us of the interpretation of texture as a surface characteristic, which is understandable because vision researchers \cite{amadasun:1989:textural, cho:2000:reliability, liu:1996:periodicity, rao:1996:towards, tamura:1978:textural} have primarily focused on natural tex\-tures (\eg, the photographic textures in Brodatz' album \cite{brodatz:1966:textures}). Yet their work---de\-di\-ca\-ted to understanding how humans perceive tex\-ture---can nevertheless shed light on using pattern for data visualization. 
In particular, Ware and Knight \cite{ware:1992:orderable, ware:1995:using} iden\-ti\-fy three orderable dimensions for data displays: \emph{orientation}, \emph{size}, and \emph{contrast} (OSC). Healey and Enns \cite{healey:1998:building} build three-dimensional perceptual texture elements, called pexels, for visualizing multidimensional datasets. Pexels can be varied in three separated texture dimensions, which are height, density, and regularity, and \hty{the color} of each pexel. 
This perception perspective is highly relevant to the use of \emph{pattern} as a visual variable. 
\changed{A pattern description system developed from the design perspective can be informed by and tested in the context of the perception literature.}

\subsubsection{Design perspective} 
\hty{Researchers in design, cartography, and visualization have described the composite nature of \emph{pattern}, which has multiple dimensions that can be varied to encode data.}
%
From an architectural perspective, Caivano \cite{caivano:1994:towards, caivano:1990:visual} describes a system of patterns using the term ``texture,'' which we adopt when discussing his system. He classifies ``simple textures'' and ``complex textures,'' defining the former as ``the uniform repetition of a certain element'' \inlinevis{-1pt}{1em}{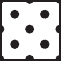} \inlinevis{-1pt}{1em}{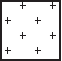} and the latter as combinations of multiple sets of these primitives  \inlinevis{-1pt}{1em}{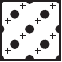} \cite{caivano:1990:visual}. 
His simple tex\-tures are essentially described by the relationship of two elements within a tiling (repeating) unit. He describes texture variation as the shape of texture elements, organization (relative positions of the two texture elements in the tiling unit), proportionality (a tiling unit's width-height ratio), and density (overall black-to-white ratio). Cavaino, however, does not intend to use his textures as a visual variable for data encoding. As a result, not all the dimensions he identified are directly manipulable, and his composition of simple textures is therefore 
unsuitable for our purpose. In addition, we identify two distinct subsets of pattern
within his simple texture classification (see \autoref{sec:more-on-caivano} for a detailed discussion). Below we thus develop our own \emph{pattern} configuration to offer an alternative that covers a wider design space, specifically aimed at encoding data.

A cartographic angle, expressed in MacEachren’s definitive \textit{How Maps Work} \cite{MacEachren:2004:HowMapsWork}, considers ``%
`pattern' as [a] higher-level visual variable, consisting of units that have shape, size, orientation, texture (in Bertin's sense of grain), and arrangement.'' In the field of visualization, Harris \cite{harris:2000:InformationGraphics}, in his book on information graphics design, suggests that ``there are many variations'' within patterns and lists factors that make up patterns: ``shape of individual elements,'' ``orientation of individual elements,'' ``texture (sometimes referred to as coarseness),'' ``size of individual elements,'' and ``spacing between individual \mbox{elements}.'' Wilkinson \cite{wilkinson:2013:GrammarOfGraphics}  wrote that ``texture includes pattern, granularity, and orientation,'' but he does not further analyze \emph{pattern}. Instead, he interprets \emph{pattern} as being ``similar to fill style in older computer graphics systems, such as GKS (Hopgood et al., 1983) or paint programs'' but does not describe the subdimensions of \emph{pattern}. In our own previous work \cite{He:2024:DCB} we identify a set of \emph{pattern} properties (as [sic] ``textures'') but also does not cover all dimensions. 

In summary, our reading of this prior work provides us with useful examples for understanding the various dimensions of patterns, but none of the individual treatments are comprehensive and the rationale that supports them is somewhat sketchy. Based on the various authors' suggestions, we thus provide a systematically defined proposal that is comprehensive in its consideration of pattern as a high-level visual va\-ri\-ab\-le---de\-ve\-lo\-ped via a structured, repeated combination of sets of visual primitives that vary in their visual characteristics to encode data.

\subsection{Inspiration from Bertin's apparent \changed{incongruity}}
\label{sec:bertin-inconsistency}
Bertin himself do not explicitly define or employ the concept of \emph{pattern}, but why then do we see many visual encodings we may intuitively call \emph{pattern} in his charts or maps?  One explanation may be that Bertin attempts to address an inherent limitations of area marks.  Area marks cannot change in \emph{size}, \emph{shape}, or \emph{orientation} \cite{Bertin:1998:SG, bertin:1983:semiology, Carpendale:2003:ConsideringVisualVariables, Munzner:2015:VisualizationAnalysisAndDesign}, without the area  changing its meaning. For example, we cannot encode an additional data attribute into the \emph{size}, \emph{shape}, or \emph{orientation} of a region (an area mark) on a map because these attributes are already used to encode geographic information. To address this limitation, 
\changed{Bertin invokes the cartographic tradition, in using a method} of adding a single additional mark with variable visual characteristics or \hty{a group of repetitive additional marks with common visual characteristics.} When he does the latter, he creates a \emph{pattern} with repetitive tiling of secondary marks (primitives)---which is why Carpendale interprets \emph{pattern} as ``the use of marks upon marks.'' Bertin explains this encoding by the fact that it is possible to change the visual characteristics of the constituents that make up an area: ``if the area is visually represented by a constellation of points or lines, these constituent points and lines can vary in size, shape, or orientation without causing the area to vary in meaning'' \cite{Bertin:1998:SG}. He also makes a similar point for the constituents of a line mark. This adjustment explains why Bertin could apply all his six retinal visual variables---including \emph{size}, \emph{shape}, and \emph{orientation}---onto line and area marks (see his overview in \autoref{fig:bertin-variables-on-three-mark-types} in \autoref{sec:additional-figures-from-bertin}), and the ``constituents'' here equate to the secondary marks (the primitives in the pattern).

\begin{figure}[]
    \centering
\setlength{\subfigcapskip}{-3ex}
        \subfigure[\hspace{\columnwidth}]{\label{fig:BertinUnconsciousRepetition:a}~~~\includegraphics[height=1.5cm]{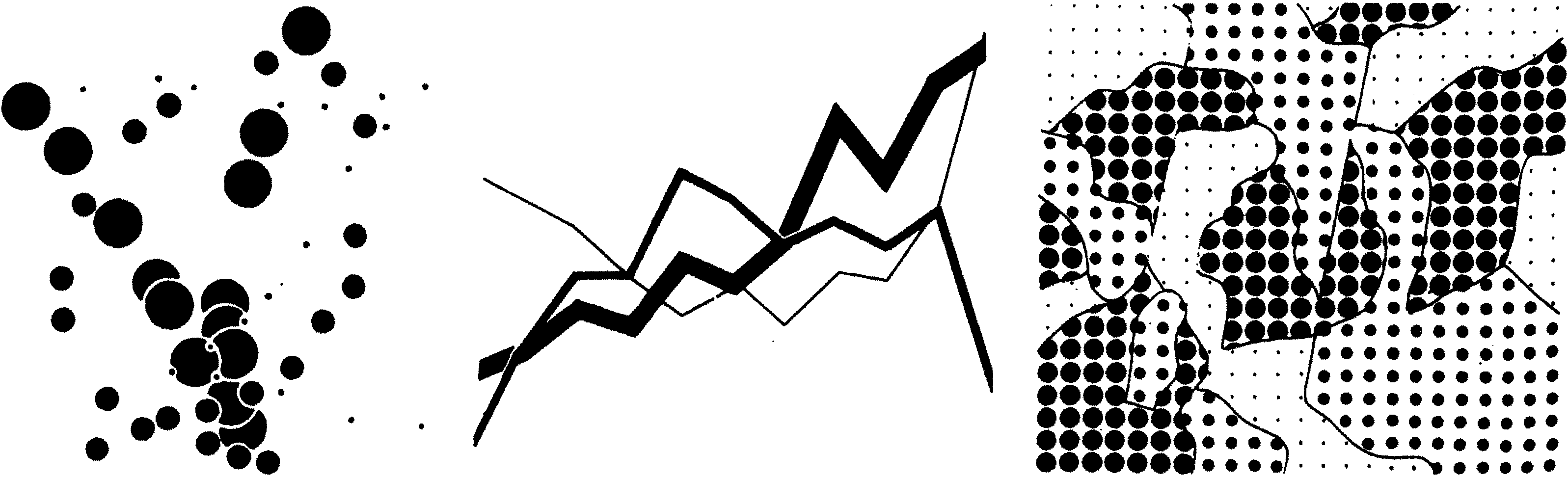}}\hfill
        \subfigure[\hspace{\columnwidth}]{\label{fig:BertinUnconsciousRepetition:b}~~~~~\includegraphics[height=1.5cm]{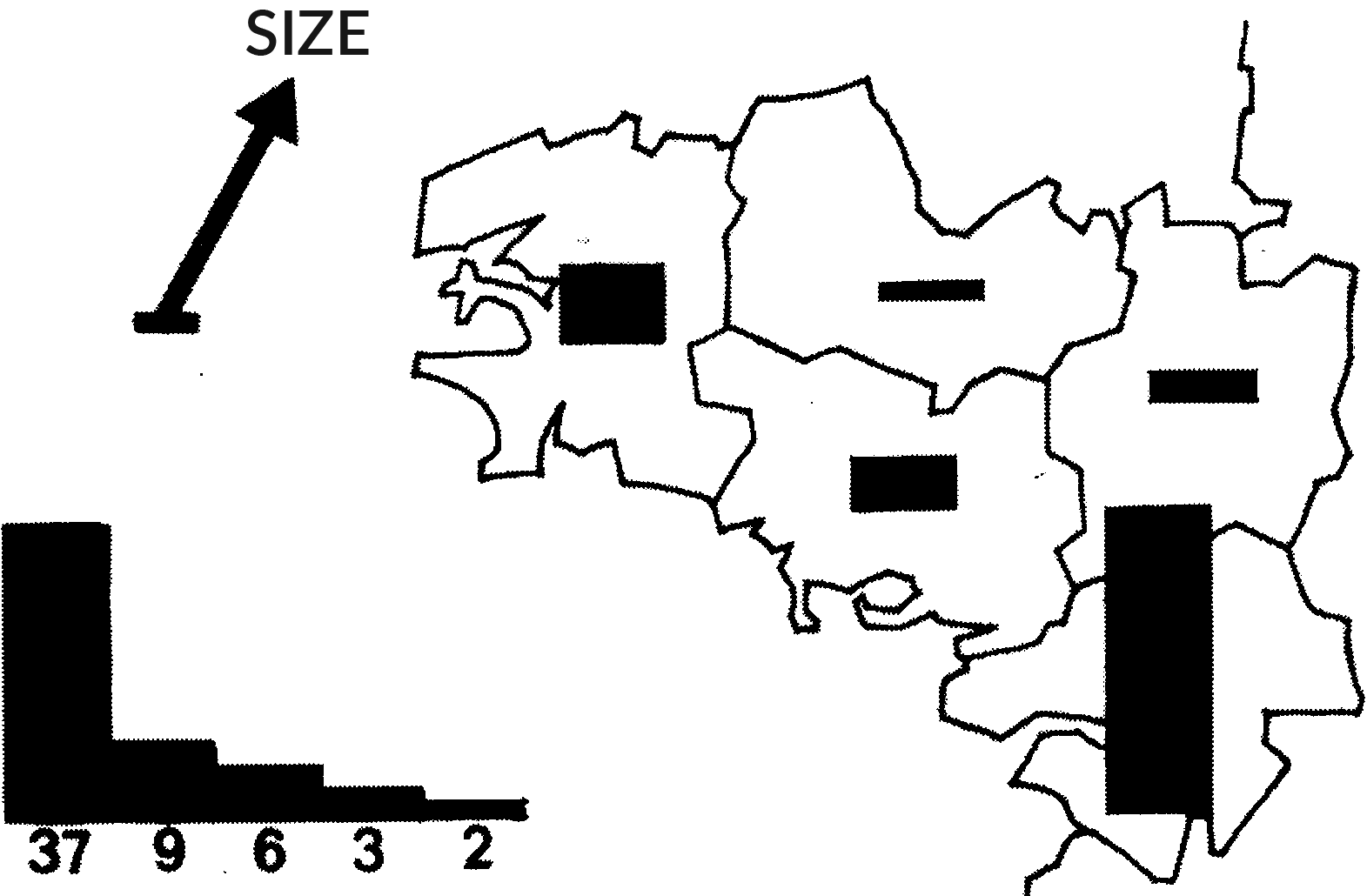}}
    \caption{Size variations from Bertin's book \cite{Bertin:1998:SG, bertin:1983:semiology}: (a) Size variations for three mark types: left and middle show Bertin’s \texttt{Approach 1}, where he directly adjusted the marks' size properties (dot size and line width); right shows Bertin’s \texttt{Approach 2.2}, where he added repetitive dots to fill the area mark and used the dots as secondary marks---he varied the dots' size properties (dot size). (b) Size variation for an area mark using Bertin’s \texttt{Approach 2.1}, where he added a single secondary mark (rectangle) and varied its size property (rectangle size). All images \textcopyright~EHESS (text translated), used with permission.}
    \label{fig:BertinUnconsciousRepetition}
\end{figure}

Let us take the visual variable \emph{size} as an example to explain how he mixes these two approaches. 
\emph{Size} is applicable to point and line marks, but not to area marks.
When applying \emph{size} to point and line marks, Bertin directly adjusts the mark's intrinsic properties, such as the dot's radius or the line's width---without introducing secondary marks (\autoref{fig:BertinUnconsciousRepetition:a}, left and middle). We call this \emph{Approach~1---Direct Encoding}: changing the graphical properties of the mark itself, which aligns with the precise definition of a visual variable. 
For area marks (such as map regions), however, \emph{size} is not applicable \changed{if the integrity of the geographic encodings is to be maintained}.
In this case, Bertin takes another approach: to introduce new mark(s)---we call this \emph{Approach~2---Secondary Marks}, which includes two options: \emph{Approach~2.1---Single Secondary Marks} adds one constituent (secondary mark), \hty{and} \emph{Approach~2.2---Repeated Secondary Marks} adds repetitive secondary marks (``constellation'') to fill the area or line mark. For example, Bertin shows the application of \emph{size} to an area mark by adding a rectangle of different size to each area mark (\autoref{fig:BertinUnconsciousRepetition:b}, which aligns with the single secondary marks approach, \changed{and this is essentially the approach favored by Cleveland and McGill in the framed-rectangle charts of their seminal graphical perception paper \cite{cleveland:1984:Graphical}}), or by repetitively filling each area mark with circles of varying sizes (\autoref{fig:BertinUnconsciousRepetition:a}, which aligns with the repeated secondary marks approach). In summary, when Bertin can use a direct encoding, he does so. When this is not possible, he automatically switches to using secondary marks, yet without clarification. In his discussion, unfortunately, he does not clearly explain why and when to select single or repeated secondary  marks.


The secondary marks that Bertin adds to the marks are typically point- or line-marks, as more visual variables can be applied to them than to area marks. With repeated secondary marks, Bertin, in fact, creates point-based and line-based patterns as we understand it, and so that is ultimately why we see many \emph{pattern} examples in Bertin's book.

Bertin's mixing of the two methods actually blurs the meaning of visual variables. As Wilkinson \cite{wilkinson:2013:GrammarOfGraphics} points out, ``Bertin uses size, shape, and orientation to characterize both the exterior form of objects (such as symbol shapes) and their interior texture pattern (such as cross-hatching).'' 
\changed{Bertin's approach does not fully explore or clearly articulate the full emerging possibilities of patterns.}
We can, in fact, apply \emph{pattern} across all visual variables and mark types, but Bertin reserved \emph{pattern}
for situations where visual variables were not applicable to certain types of marks. In addition, Bertin simply keeps the secondary marks arranged in a regular grid and ensures that each secondary mark was exactly repetitive. Similarly, Carpendale \cite{Carpendale:2003:ConsideringVisualVariables} equated a variation in patterns to the variation in shapes constituting them, as we discussed in \autoref{sec:texture-interpretation-pattern}. 
Inspired by these perspectives and having established their apparent inconsistencies, we systematically explore the opportunities for explanation and for visually encoding data offered by this new comprehensive description of `pattern.'

\section{Pattern as a visual variable: A design space}
\label{sec:pattern-design-space}

\begin{figure*}
    \centering
    \includegraphics[width=1\linewidth]{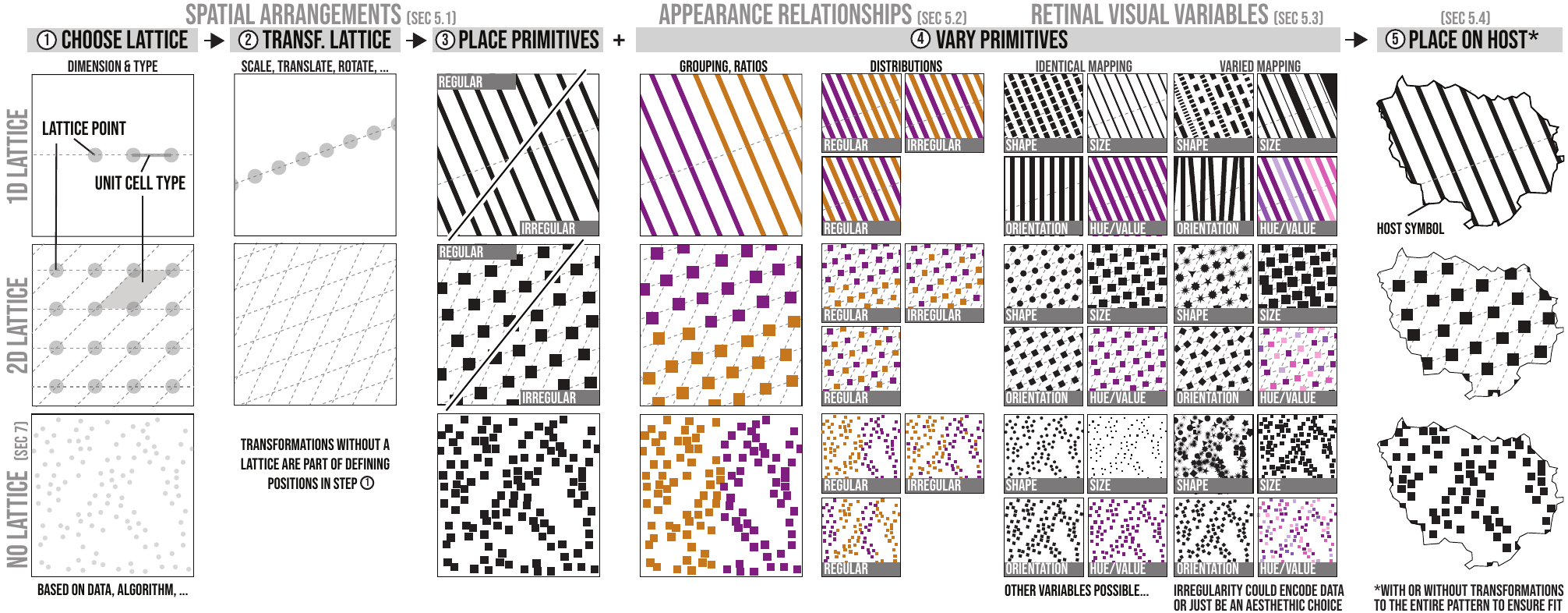}\vspace{-2pt}
    \caption{\changed{Procedure for creating a pattern by describing three sets of pattern attributes: (1) the \emph{spatial arrangement} of \pps (\autoref{sec:spatial-relationship-variables}), (2) the \emph{appearance relationships} among \pps (\autoref{sec:pattern-with-internal-variation}), and (3) the \emph{retinal visual variables} applied to each individual \pp that define its appearance (\autoref{sec:retinal}).
		We illustrate the attributes with pattern samples constructed both with a lattice (\autoref{sec:pattern-design-space}) and without one (\autoref{sec:pattern-beyond-lattice}).%
		}}
    \label{fig:pattern-procedure}
\end{figure*}


\changed{This position enables us to establish a new pattern system---based on the design perspective and with the goal of identifying and exposing the basic parameters that can be exploited to encode data. We conceptualize this system as follows:
A \emph{pattern} has a group of primitives. Primitives are graphical elements whose visual attributes can be manipulated to encode information; they can thus be considered ``marks.'' To differentiate these marks from the graphical elements (mark) to which patterns are applied, we refer to the latter as \emph{host symbols}. We thus define a \emph{pattern} as a \textbf{composite} of constituent marks (primitives) applied to a host mark (host symbol). For consistency, throughout the remainder of this paper, we use the terms \emph{primitive} and \emph{host symbol} to refer to these two respective levels of marks.}
\changed{Transitioning from a symbol as a single mark to a symbol that contains a composite of (possibly repeating) primitives that constitute a \emph{pattern} then introduces new visual attributes.
These attributes arise from the composite nature of the pattern, specifically the relationships among its primitives.}

\changed{Our approach sees pattern as a set of rules for describing the \pps, \ie, parameters that define their attributes, variation, and relationships.
By following these rules, we can generate expressive patterns that fill a host symbol. 
We identified three sets of rules that are key to this process and that also serve as attributes by which a pattern can be characterized: (1) \emph{spatial arrangement} of \pps (\autoref{sec:spatial-relationship-variables}), (2) \emph{appearance relationships} among \pps (\autoref{sec:pattern-with-internal-variation}), and (3) the \emph{retinal visual variables} applied to each individual \pp that define its appearance (\autoref{sec:retinal}).
We now discuss how these attributes can be varied to characterize a pattern. The fact that they can be deliberately and systematically exploited for encoding data enables us to speculate that they have a scope as ``visual variables,'' and we begin to explore their capacities for representing information.}

\subsection{\changed{Spatial arrangement of primitives: Lattice}}
\label{sec:spatial-relationship-variables}
\changed{Spatial arrangement attributes describe how the \pps of a pattern are spatially positioned and repeated to fill the space occupied by a host symbol.
In theory, any algorithmic method for arranging visual elements could define such spatial arrangements. We begin, however, with a pattern type that is commonly found in existing visualizations---characterized by the regular arrangement of \pps along or across the host symbol \inlinevis{-1pt}{1em}{inline-geometric-patterns}.
We can use a lattice structure to describe this regular arrangement, which consists of a set of regularly spaced points that can extend infinitely in space. Each point, known as a \textit{lattice point}, represents a predefined position for a \pp within the pattern (\autoref{fig:pattern-procedure}--\fignumber{1}, rows 1 and 2). 
We follow a method used in crystallography \cite{giacovazzo:1992:FundamentalsOfCrystallography} to define a lattice, whose central idea is to identify the unit cell of the lattice. 
The unit cell is the smallest unit of a lattice and the entire lattice can be generated by the repetitive tiling of the unit cells. We call the parameters that define the unit cell and thus the lattice structure ``lattice parameters.''
Each lattice has a unique set of parameters, such as $\theta$, $a$, and $b$ for oblique lattices\footnote{The parameters $a$ and $b$ describe the spacing between primitives within the lattice. The simplest form of a 2D lattice is a square lattice where $a=b$ and $\theta=$90\textdegree.} or $\theta=120$\textdegree\ and $a$ for hexagonal latices. These parameters are the spatial relationship attributes for characterizing a pattern.
Although the discussion of lattices relates to their use in crystallography \cite{giacovazzo:1992:FundamentalsOfCrystallography} and in tiling \cite{Kaplan:2009:ITT}, we take a more general approach here and say that all lattice definitions are allowed---including nonuniform ones. 
In this section, we focus on uniform regular lattices as examples for explaining the spatial relationship attributes of patterns. We discuss patterns composed of primitives arranged in more complex or non-lattice structures in \autoref{sec:pattern-beyond-lattice}.}

\subsubsection{\changed{Define lattice: Dimensionality and unit cell shape}}
\label{sec:composition}

\changed{\textbf{Lattice dimensionality.}}
The number of lattice dimensions affects the parameters that are required to define the lattice.
\changed{We can organize the primitives of a pattern for 2D visualizations into either 1D or 2D lattices (rows 1 and 2 in \autoref{fig:pattern-procedure})}. It is important to note that the lattice dimensionality (1D or 2D) differs from the number of dimensions that the pattern itself occupies (along a line or across an area), as well as from the number of dimensions of the host symbols onto which the pattern is applied (point, line, or area).
The lattice dimensionality is determined by the number of directions in which the lattice can extend. 

\changed{The first two rows of \autoref{fig:pattern-procedure} illustrate the two types of patterns possible in 2D representations, to which we refer as point-based and line-based patterns, respectively. 
Point-based patterns are based on 2D lattices, in which the lattice points are often equally spaced, extending in two directions (2\textsuperscript{nd} row in \autoref{fig:pattern-procedure}).
In contrast, line-based patterns rely on 1D lattices but with linear \pps that (usually) extend to infinity (1\textsuperscript{st} row in \autoref{fig:pattern-procedure}).
Both types of patterns, however, can be used on all three types of host symbols---areas, lines, and points---because all three mark types are represented by a host symbol with areal extent.\footnote{\changed{For example, a point host symbol, finally, is not a theoretical point but is represented by a small area (often a circle) with a size, which in turn has an area that can be filled with a 2D or 1D lattice.}}
If we apply the lattice to \hty{line host marks (1D elements)}, we can either treat the line as a linear host symbol with a (small) extent perpendicular to its major direction, or as a form of selection that picks a single direction from a 2D lattice or that limits the lengths of the line primitives for a 1D lattice (also see \autoref{fig:TextureComposition} in \autoref{sec:additional-figures-for-pattern-definition}).%
\footnote{Brath \cite{Brath:2021:1DTexture}, in his blog, characterized patterns in the form of \autoref{fig:TextureComposition:d} by their length, gaps, rhythm, and randomness. \changed{Using the terminology of our pattern system, length and rhythm define the \textit{appearance of primitives}; gaps define the \textit{spatial relationships} between the primitives---the \textit{lattice}; and randomness determines the \textit{positional regularity} when placing the primitives onto the lattice.}}}

\newlength{\examplewidth}
\setlength{\examplewidth}{0.15\columnwidth}

\changed{\textbf{Unit cell shape.} 
The shape of the unit cell characterizes the lattice type. In a 1D lattice, the unit cell is simply a line segment (essentially a distance) between two adjacent lattice points. In a 2D lattice, the unit cell can take many tessellating shapes, including triangles, rectangles, hexagons, and more complex geometries with nonrepetitive tilings (\eg, \cite{Penrose:1979:PCN,Smith:2024:AM}). The lattices and patterns that result in the latter case are not grid-based. In this section, we mainly use the square lattice as examples to illustrate different pattern variations. Note that we refer here to the overall geometry of the unit cell. Some shape variations can be derived through geometric transformations (\eg, shearing a square into a parallelogram), which we discuss next.}

\subsubsection{\changed{Transform lattice: Affine transformation}}
\label{sec:lattice-parameters}

\changed{The lattices can further be transformed through affine transformations, such as translation, shearing, scaling, and rotation, to achieve arrangements suitable for specific application contexts (\autoref{fig:pattern-procedure}--\fignumber{2}).
Due to the inherently infinite repetition of a lattice, a translation usually only leads to limited visual changes.\footnote{\changed{As we discuss later in \autoref{sec:transform-and-fit-pattern}, however, once primitives are placed on the lattice, we can translate the entire pattern when we apply it to a host symbol.}} Shearing, in contrast, affects the shape of the unit cell, \eg, converting a square lattice into an oblique lattice, and is thus often considered together with the unit cell shape. 
Among the various affine transformations, lattice scaling and rotation are often used for encoding data, and we discuss them in detail next.}


\changed{\textbf{Scale: Size of the unit cell.} 
Scaling changes the spacing between lattice points and, thus, the unit cell size. The spacing can be modified either uniformly across both directions or independently, the latter facilitating directional variation in the pattern. Independent adjustments in the width and length of the unit cell can significantly affect the directionality of the resulting lattice pattern. For a 1D lattice, in contrast, variations in spacing are constrained to a single dimension---along the line of the lattice. In both 1D and 2D cases, however, a variation of the spacing between primitives affects the area of the unit cell, which in turn influences the density of primitives in a pattern.\footnote{Theoretically, the range of possible spacings---or unit cell sizes---extends from zero up to the size of the entire marking. Practically, it is crucial to use a sufficient number of primitives within the visible area to ensure that the pattern and the extent of the symbol are discernible. If the primitives on the pattern are too sparse, both the symbol and the pattern may become difficult to perceive.}}

\changed{\textbf{Rotate: Orientation of the lattice.}
Rotation varies the orientation of the lattice.\footnote{\changed{Theoretically, the lattice can be rotated by any angle between 0° and 360°. Often, the center of rotation is the center of the symbol to which the pattern is applied, although it can be set to other points if needed.}} 
For a pattern arranged in a 2D lattice (\autoref{fig:pattern-procedure}, 2\textsuperscript{nd} row), it is important to distinguish between the orientation of the lattice itself (\autoref{fig:pattern-procedure}--\fignumber{2}) and the orientation of the primitives within it (\autoref{fig:pattern-procedure}--\fignumber{4}, retinal variables--orientation).\footnote{Wilkinson \cite{wilkinson:2013:GrammarOfGraphics} describes the orientation of a mark as ``rotation'' and the orientation of primitives in a pattern (he called it ``texture'') as ``orientation.'' He illustrates the concept of ``orientation'' exclusively with examples of line patterns (\autoref{fig:TextureComposition:b}) and does not address the orientation variable of the lattice, which we add here. 
Moreover, we argue that the distinction between ``rotation'' and ``orientation'' should not be based on whether they apply to marks or primitives, as Wilkinson suggests. Instead, the key difference lies in their relationship as method and result: an outcome of a given orientation of primitives is achieved through the method of rotation. We thus recommend to use both terms in their traditional meaning: rotation as the action and orientation as the status; both terms applied to either lattice, primitives, or both together.}
If both the primitives and the lattice are rotated by the same angle, this can intuitively be described as a rotation of the entire pattern (\autoref{fig:PatternVariation-2Dp2Da-orientation} in \autoref{sec:additional-figures-for-pattern-definition}).\footnote{\changed{Patterns arranged in a 1D lattice can also be manipulated both by rotation of the lattice and by rotation of the primitives. In the case of line primitives, however, the rotation of the primitives w.r.t. the base line may also lead to a perceived change in the spacing between them.}}}
\subsubsection{\textls[-5]{Place primitives: Positional regularity}}
\label{sec:pattern-variation-positional-regularity}
\changed{So far, we have established a set of predefined position points for the primitives within the pattern---the lattice points. When we place the primitives onto the lattice, however, they may deviate from these predefined positions  (\autoref{fig:pattern-procedure}--\fignumber{3}). 
To describe the extent of this deviation, we adopt the concept of \emph{positional regularity}, introduced by Morrison \cite{Morrison:1974:TFC}.\footnote{Morrison \cite{Morrison:1974:TFC} first introduced this concept of \emph{positional regularity} into visual variables. He referred to it as ``arrangement'' and added it as an additional visual variable to Bertin's list.} This variable describes the degree to which primitives can deviate from lattice points, ranging from strict adherence to the grid via structured irregularity to a fully random placement within the mark.}
For a 2D lattice, the deviation can occur in either one direction of the unit cell or both directions (\autoref{fig:PatternVariation-2Dp2Da-A-regularity} in \autoref{sec:additional-figures-for-pattern-definition}). For 1D lattice patterns, the placement along the lattice line is affected.
\emph{Positional regularity} is not an atomic variable and has subdimensions, including its range and its dispersion level. \textbf{Range} describes how far we can deviate from the predefined point, and the \textbf{dispersion level} can be understood as the standard deviation or entropy of the deviations among all primitives.




\subsection{Appearance relationship among primitives}
\label{sec:pattern-with-internal-variation}

Our intended comprehensive treatment of pattern as a visual variable requires us to consider not only the spatial relationships between primitives but also how primitive groups appear within the pattern. For noncom\-po\-site marks, describing their graphical at\-tri\-butes---such as shape, size, and co\-lor---is sufficient. These attributes are what Bertin refers to as the ``retinal variables.''
\changed{We cannot, however, directly manipulate each primitive's retinal variables in a pattern because patterns are used to fill host symbols. Depending on the size of the host symbol, the same pattern may contain different numbers of primitives. As a result, we cannot set the attributes of each primitive individually. Instead, we define rules for the primitives' attributes, and the pattern is then generated accordingly to fill the host symbol.}
We thus need to establish rules that describe the relationships between the primitives' appearances, no matter how many of them appear in a pattern. For common patterns with repeated primitives, \eg, this rule is simply ``all primitives look the same.''
%
Our consideration of patterns as composite marks that vary in defined ways to encode data opens up a large design space, revealing more and new possibilities with combinations of graphical attributes that can serve as new ways to encode data. We discuss these next, describing the internal variation in the appearances of primitives.

\subsubsection{Number of primitive groups}
\label{sec:number-of-primitive-groups}
\emph{Number of primitive groups} describes how many distinct combinations of visual variable variation, or styles
of primitive, are used within a pattern, with each group depicted using a unique encoding or encoding combination.
\changed{Traditionally, patterns with a single repeated primitive have been the most widely used,}
meaning that one or more visual variables are applied consistently across all primitives 
\changed{used in a pattern}
to encode data (\autoref{fig:pattern-procedure}--\fignumber{4}: ``identical mapping''). 
The pattern's composite nature and the enabling nature of today's digital technologies (\eg, SVG), however, allow us to be more expressive than is required by this consistency:
We can vary visual variables to associate and differentiate subsets of the primitives in a pattern (\autoref{fig:pattern-procedure}--\fignumber{4}: column ``grouping, ratios'').
Intuitively, internal variation in patterns allows us to visualize a new data facet and represent it within the mark. 
We can use the variable primitive group count to encode the number of categories associated with the facet (\eg, \cite{Jo:2019:DRM}). Its application, however, is broader: as the \emph{number of primitive groups} is an attribute just like any other visual variable, it can encode various types of data. It is ordered as it represents a numerical count, making it useful for encoding ordinal data. When using it, however, it is important to ensure that the number of primitive groups is not excessively large to maintain \changed{discriminability}.

\subsubsection{Ratio between the groups}
\label{sec:ratio-between-each-group}
If a pattern comprises multiple primitive groups (and only then), the primitive count can differ between groups, to which we refer as the \emph{ratio between groups}. \autoref{fig:pattern-procedure}--\fignumber{4}: column ``grouping, ratios'') and \autoref{fig:internal-variation-variables:b} in \autoref{sec:additional-figures-for-pattern-definition} show examples for variation in this variable.
It can encode ordered data due to its numerical nature.

\begin{figure}
    \centering%
        \setlength{\subfigcapskip}{-2.5ex}%
    \subfigure[\hspace{\columnwidth}]{\label{fig:grid-pattern-internal-variation:a}~~~~~\includegraphics[height=0.4\columnwidth]{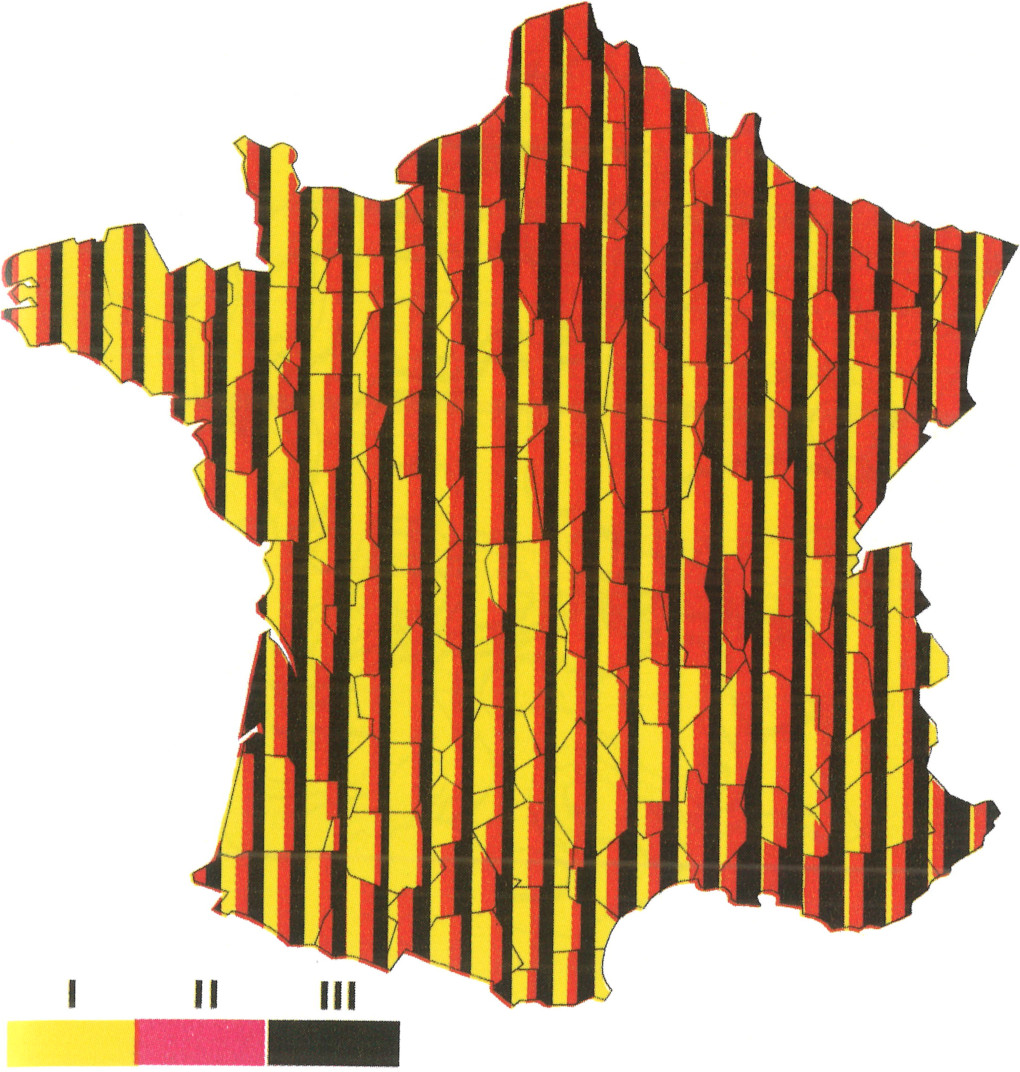}}\hfill
    \subfigure[\hspace{\columnwidth}]{\label{fig:grid-pattern-internal-variation:b}\includegraphics[height=0.4\linewidth]{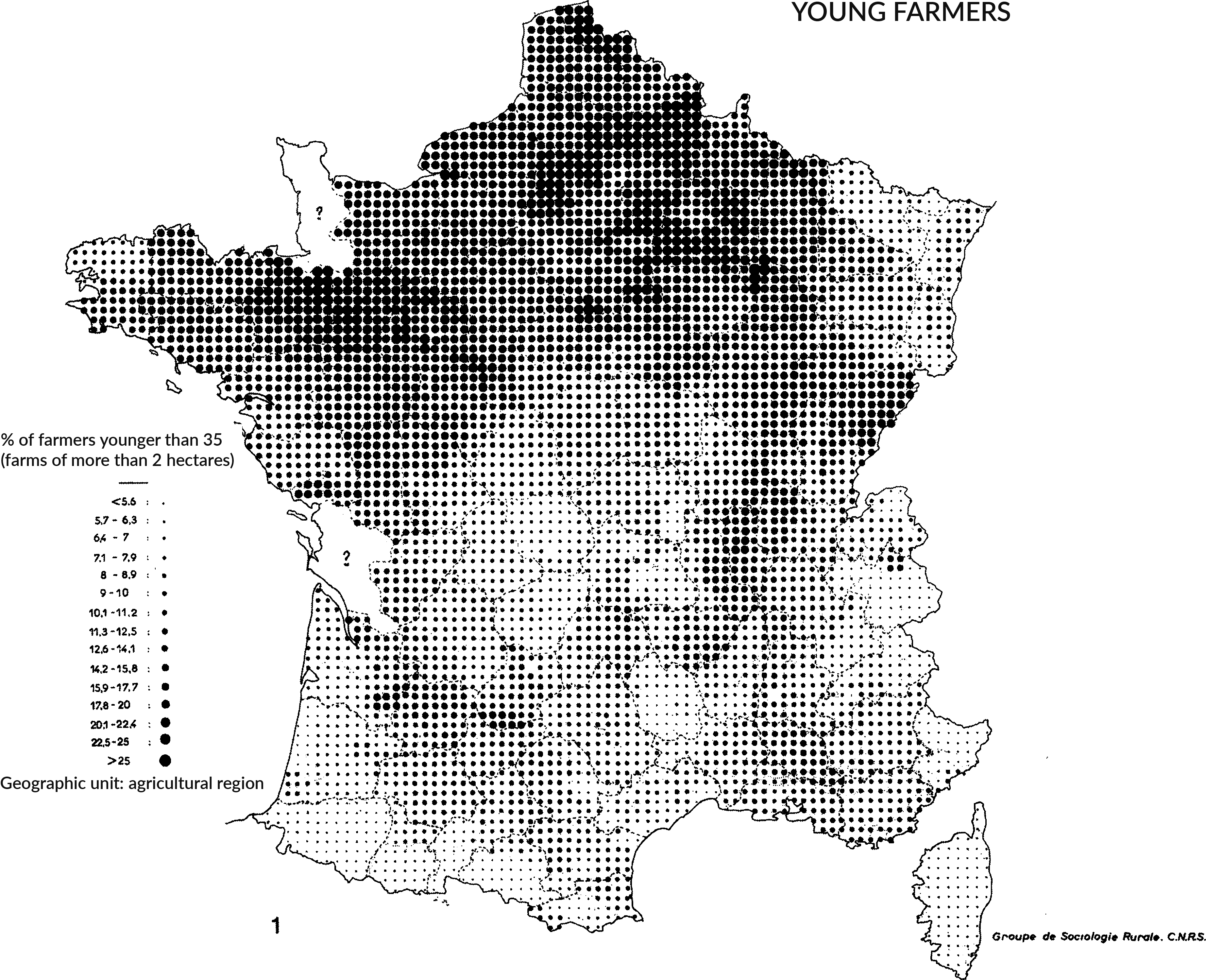}}\vspace{-2pt}
    \caption{Examples of using regular arrangement pattern with internal variation. Within the patterns, the variations (a) show a facet of data (described in \autoref{sec:ratio-between-each-group}), (b) based on geographical data (described in \autoref{sec:internal-variation-from-geographical-information}); both images \textcopyright~EHESS (text translated), used with permission.}%
    \label{fig:grid-pattern-internal-variation}
\end{figure}

\begin{figure}[!t]
    \centering
        \includegraphics[width=\columnwidth]{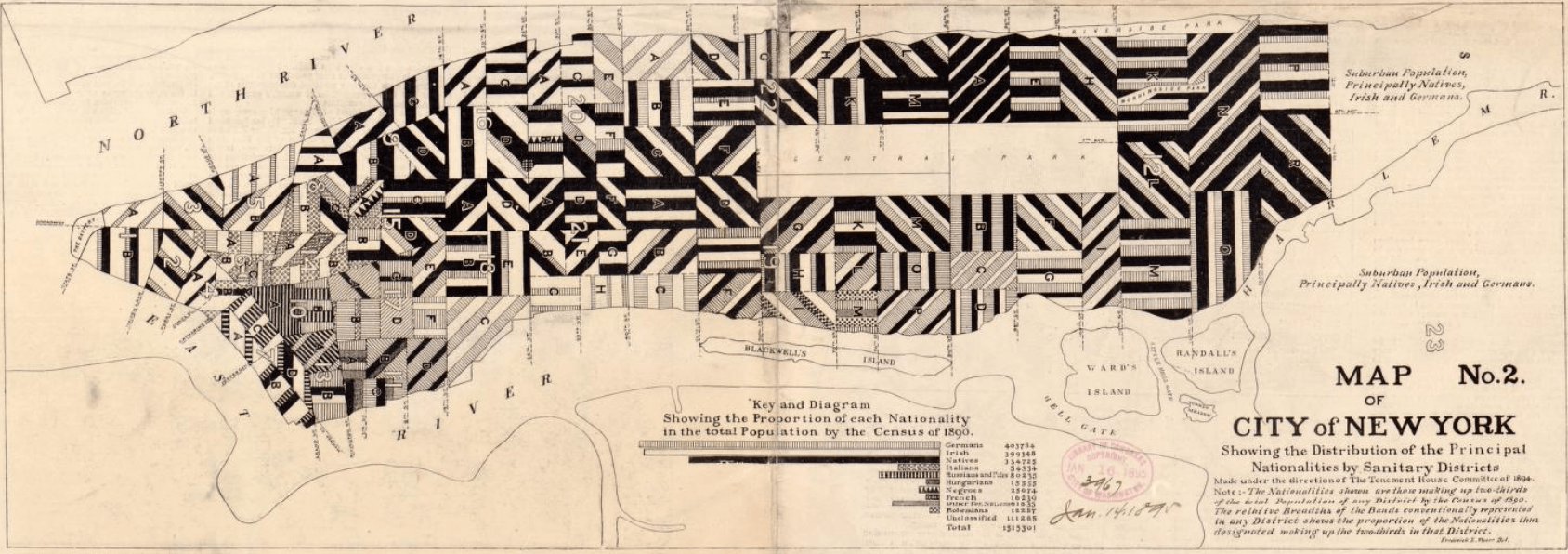}
    \caption{Map of nationality distribution in NYC \cite{NYCmap} from 1895, It depicts the distribution of different nationalities across sanitary districts in Manhattan. The designer used line pattern with variations in orientation to distinguish the city districts and used patterns on each line to distinguish different categories. \ccPublicDomain\ The image is in the public domain.}
    \label{fig:NYCmap}
\end{figure}

A straightforward way to use this facet is to form primitive groups and encode categories (\ie, keys). All purple primitives, for instance, encode data for category A, and all yellow primitives encode data for category B. If these categories have a certain distribution (\eg, 50\% of data items are of type A, and 50\% are of type B), then we could reflect this split in the primitive group ratio. \autoref{fig:grid-pattern-internal-variation:a} (reproduced from Bertin's book \cite{Bertin:1998:SG, bertin:1983:semiology}) shows a good example of using the variable to encode the distribution of categories. It shows three categories of data for France, with differently colored primitives that encode data by region. Here, the width\changed{---1D size---}of each colored primitive encodes the proportion of the respective category.
\autoref{fig:NYCmap} shows another example, employing the number of groups of primitives to encode the number of categories (IsoType, \cite{neurath:2010:hieroglyphics,haroz:2015:isotype}). The ratio between groups (indicated by line width) encodes the percentage of each category.\footnote{\changed{Unit-based visualization with subgroups within each category} can also be considered as using patterns with internal variation---where the variable \emph{shape} is used to represent the categories. \autoref{fig:great-war-isotype} in \autoref{sec:more-examples-of-patterns} shows an example of such a unit-based visualization. Here, if we view each diamond region as a pattern, it exhibits internal variation derived from a new facet, ``type of soldiers.'' Within each pattern, the number of groups of primitives encodes number of categories and the ratio between groups represents the percentage of each category.}


\subsubsection{Distribution style of different primitives}
The \emph{distribution style of each group} refers to how we \changed{arrange} each primitive group within a pattern (\autoref{fig:pattern-procedure}--\fignumber{4}, column ``distributions''). We should differentiate it from the spatial arrangement of primitives \changed{as being the allocation of primitive variations to predefined positions in the pattern}. Here, we discuss how to further specify which primitive belongs to each group after having defined all primitives' positions.

\changed{\autoref{fig:pattern-procedure}--\fignumber{4} (column ``distributions'', rows 1 and 2) shows different choices of distribution style applied to a consistent spatial arrangement of primitives, both for 1D lattice patterns (row 1) and for 2D lattice patterns (row 2). There can be regular arrangements (left part of column ``distributions'') or irregularly dispersed ones (right part of the column). Within the former, the primitives within the same group can be clustered together (the top example, respectively), as also in the IsoType image in \autoref{fig:great-war-isotype} in \autoref{sec:more-examples-of-patterns}. Alternatively, the primitives of the pattern within each group can also be dispersed (bottom examples), resulting in a uniform distribution, as also exemplified in \autoref{fig:NYCmap}.
The regularity, clustering, or dispersion of primitive groups can be used to differentiate patterns, and may also have ordinal or numeric encoding possibilities, ranging from regular to irregular, to clustered, to dispersed.%
\footnote{\changed{Haroz et al.'s work \cite{Haroz:2012:CapacityLimits} would suggest likely perceptual interactions between efforts to encode with distribution and ratio concurrently.}}}






\subsection{Retinal visual variables on each primitive}
\label{sec:retinal}
So far we have used two sets of parameters to describe the relationships between groups of primitives in a pattern: their spatial and group appearance relationships.
After establishing these rules, we still need to discuss the choice of retinal variables to characterize the appearance of primitive \changed{groups to} complete our systematic description of pattern encodings.
The patterns \inlinevis{-1pt}{1em}{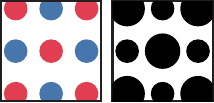}, \eg, are identical in their spatial and appearance relationships (they have the same two pattern groups, same number of primitives in each group, same distribution style, and same primitive positions) but differ in the choice of retinal variables \changed{used to characterize the groups}.
The first \inlinevis{-1pt}{1em}{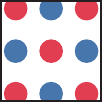} uses hue to differentiate, which could encode categories; the second \inlinevis{-1pt}{1em}{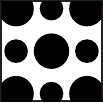} uses size, which could encode quantities. The choice of these retinal variables has a profound impact on the appearance of the pattern and we thus now explore the application of retinal variables to primitives within a pattern as well as the additional parameters and effects that arise from their use.

\subsubsection{Retinal variables for primitives}
\label{sec:retinal-variables-primitives}
Bertin \cite{Bertin:1998:SG, bertin:1983:semiology} used the term ``retinal variables'' to describe the graphical attributes that ``elevate marks above the plane,'' and pointed out that these variables are independent of position.
Following Bertin's definition, we use the term retinal variable to describe nonspatial graphical attributes.
Here we apply these attributes, however, to
\changed{the individual primitives that we locate on our lattice to create the specific pattern, as distinct}
from the variables we discussed in the \autoref{sec:spatial-relationship-variables} and \autoref{sec:pattern-with-internal-variation}. 

Bertin identified six initial retinal variables: \emph{shape}, \emph{size}, \emph{orientation}, \emph{value}, \emph{color}, and \emph{texture} (\emph{granularity}). \emph{Granularity}, notably, comprises two subdi\-men\-sions: \emph{size} and \emph{spacing}, as we discussed in \autoref{sec:texture-grain}. \emph{Granularity} is thus a composite visual variable rather than an atomic one. Strictly speaking, \emph{size} also comprises two independent com\-po\-nents---\emph{width} and \emph{length} (more in hi\-gher-di\-men\-sio\-nal space). Similarly, Bertin's \emph{color} combines \emph{hue}, \emph{saturation}, and \emph{value\discretionary{/}{}{/}light\-ness}, where the last component is essentially the already included \emph{value}. By removing these ambiguities as well as \emph{granularity}, we are left with \emph{shape}, \emph{size} (1D size), \emph{orientation} (pri\-mi\-tive-le\-vel orientation), and \emph{color} (\emph{hue}, \emph{saturation}, \emph{va\-lue\discretionary{/}{}{/}light\-ness}).
%
%
%
Unlike the spatial arrangement and group appearance relationship discussed before, these retinal variables are not new variables specific to \emph{patterns}. This list can thus be extended to include any visual variable that is applicable to individual marks. For instance, researchers have added variables such as \emph{resolution}, \emph{transparency}, and \emph{crispness}. For non-static charts, the list can also include \emph{motion} parameters \cite{MacEachren:2004:HowMapsWork,Ware:2004:motion}. 
%
Past work (\eg, \cite{Bertin:1998:SG, bertin:1983:semiology, MacEachren:2004:HowMapsWork,Munzner:2015:VisualizationAnalysisAndDesign,roth:2017:visual}) has investigated the use of these variables and proposed guidelines on their syntactics for mapping (such as which variable is suitable for which type of data). 
\changed{When applied repetitively to primitives, however, these variables can produce new effects, as we discuss next and in \autoref{sec:interactions-among-attributes}.}

\subsubsection{Regularity of retinal variables}
\label{sec:retinal-variables-regularity}

Similar to positional regularity (\autoref{sec:pattern-variation-positional-regularity}), any of the mentioned retinal variables can be applied with a varying degree of regularity.
These regularities are, in fact, a secondary characteristic for each of the retinal variables at the primitive level, for which we show examples in  \autoref{fig:pattern-procedure} (column ``varied mapping'').
For variables that can carry numerical values (\eg, size, orientation, value, lightness), similar to positional regularity, we can quantify the range using the maximum deviation and the dispersion level using the standard deviation. For variables that do not carry numerical values (\eg, hue, shape) the quantification of range and dispersion level depends on the variable. For example, we can use entropy to describe the degree of their regularity.\footnote{The (ir)regularity can also be intended to not be used as a visual variable but simply as an aesthetic criterion, such as in non-pho\-to\-rea\-li\-stic rendering (NPR). Here, the goal is the generation of (hu\-man-draw\-ing-like) non-re\-pe\-ti\-tive patterns (\eg, \cite{Barla:2006:SPA,Hurtut:2009:appearance, Martin:2017:SDS, salisbury:1994:interactive, Winkenbach:1994:CGP}), which have non-regularity both in their primitive shapes and in their placement.}

\subsection{\changed{Transform and fit pattern}}
\label{sec:transform-and-fit-pattern}
\changed{We can now apply transformations to the pattern, such as stretching or wrapping\footnote{\changed{Such transformations can theoretically be achieved by modifying the lattice primitives directly as we described in \autoref{sec:spatial-relationship-variables}. In practice, however, it is often more intuitive to apply an additional transformation to the entire pattern after its design, because this allows the designer to adjust the parameters with all decisions having been made already.}}
and fit it to the host symbol (\autoref{fig:pattern-procedure}--\fignumber{5}). 
For doing the latter, we may add further steps such as adjusting the relative position between pattern and host symbol, cropping, omitting incomplete primitives, or adding a halo next to the border of the host symbol.}

\section{\changed{Interactions among variables}}
\label{sec:interactions-among-attributes}
\changed{Beyond these three sets of variables that characterize the pattern we also need to discuss the interactions among the variables.}

\subsection{Dependency between variables}

Retinal variables are independent of variables in spatial and appearance relationships, but when they are used in the context of patterns they are constrained by the texture context. Size, \eg, is influenced by the spatial arrangement. Theoretically, a graphical element's \emph{size} can range from 0 to infinity, but in the context of a pattern a primitive cannot have 0 size and
\changed{increasing the size of primitives beyond a certain point (dependent on lattice configuration and primitive shape) will lead to overlap, which can mask the retinal properties of the primitives or make their characteristics difficult to determine, and ultimately may result in a solid fill (2D or 1D) or thick line (1D).}%
\footnote{Drawing on Gestalt principles, even as primitives merge and lose their individuality, our brain is often still capable of perceiving the shape of primitives to some extent through mental completion. Regardless of the specific primitive shape, however, as their size continues to increase the pattern ultimately becomes completely saturated and turns into a solid fill. The range of size variation that can be effectively used in visualization thus spans from just noticeable differences to a given threshold, at which the pattern is no longer identifiable. In some specific cases, such as perfectly aligned squares or dashed lines in a line pattern, primitives that touch each other immediately form a seamless tessellation and directly convert the pattern into a solid fill.}


Dependencies also exist between retinal variables.
A primitive's \emph{sha\-pe}, \eg, affects both the range and the possible step size of its \emph{orientation}. \changed{A round primitive is invariant} to \emph{orientation}. The more elongated the shape is, however, the better we can perceive its orientation \cite{Bertin:1998:SG,bertin:1983:semiology}. Orientation variation on line patterns thus works well.\footnote{\changed{For lines placed on 1D lattices, a change of orientation at the lattice level and at the primitive level lead to visually very similar results, with the latter also affecting the perceived density of the resulting pattern.}}
\changed{Reliable hue detection, however, is challenging for small primitives.}

\subsection{Using multiple visual variables to encode data}

Bertin \cite{Bertin:1998:SG, bertin:1983:semiology} introduced the concept of combination of variables but primarily focused on retinal variable combinations. The application of multiple variables in patterns beyond retinal variables, however, extends his explicit discussion; we can find examples in his own work. Ber\-tin's \emph{grain}, \eg, is a combination of \emph{primitive size} and \emph{spacing}\changed{, with the latter two varying consistently across all primitives. As we introduce the concept of multiple groups of primitives within a pattern (\autoref{sec:number-of-primitive-groups}), however, when there is more than one group, we can either covary multiple retinal variables using the same method of primitive grouping (\eg, \autoref{fig:combination-multiple-variables:b} in \autoref{sec:additional-figures-for-pattern-definition}), or we can decrease covariation by using different methods of primitive grouping (\eg, \autoref{fig:combination-multiple-variables:c} in \autoref{sec:additional-figures-for-pattern-definition}). With each additional visual variable, we gain an ``extent of co-variation'' channel that can convey information. This can be an emergent phenomenon (discussed in the next section) and can implicitly encode which visual variables are more closely correlated.}
In addition, when discussing the retinal variable \emph{shape}, Bertin presents patterns embedded with semantic meaning (\autoref{fig:BertinBook-PatternSymbol} in \autoref{sec:additional-figures-from-bertin}). 
\hty{These examples, however, vary not only primitive shapes but multiple visual variables.} 
\emph{Hat\-ching} patterns used in technical and architectural drawing similarly use multiple variables, such as some of those in \autoref{fig:BertinBook331-PreprintedPattern} in \autoref{sec:additional-figures-from-bertin}.

\subsection{Emergent phenomena}
Directly manipulating the pattern attributes we discussed may affect the appearance of the pattern beyond the attributes we control.
\changed{These \emph{emergent phenomena} are a result of the composite nature of pattern and can affect value, shade and shape and result in optical illusions.}\footnote{Repeated patterns can cause a sense of instability---the Moiré effect \cite{Bertin:1998:SG, bertin:1983:semiology} (\eg, \autoref{fig:BertinBook80-Moire} in \autoref{sec:additional-figures-from-bertin})---or even have neurological effects for some people \cite{Wilkins:2005:CPI}. There are also many Op artworks based on patterns (\eg, \href{https://bridget-riley.publications.britishart.yale.edu/catalogue/1/}{Movement in Squares} by Bridget Riley) as well as non-photorealistic (NPR) recreations of them (\eg, \cite{Elber:2001:RPS,Dodgson:2008:RRB,Inglis:2012:OAR}), which vary the size and spacing of pattern primitives to create a perception of regions (as we intend with patterns) or movement (which would be detrimental in our case).}



\textbf{\changed{Patterned area} value:}
%
Following Bertin's definition of value while reserving this concept for individual marks or primitives, we propose to use the term ``\changed{patterned area} value'' to specifically describe the ratio of black (or colored) to white---across an entire applied pattern.
%
Bertin controlled \emph{value} akin to traditional halftoning \cite{Ulichney:1987:DH}, which creates the illusion of various shades of gray \changed{by adjusting the lattice size and primitive shapes of numerous black dots on a white background}. Given its composite nature, a pattern can inherently produce a \emph{\changed{patterned area} value} that aligns with the logic of \emph{value} as perceived from halftoning (or from stippling \cite{Martin:2017:SDS,Deussen:2013:HS}).
Even though \emph{\changed{patterned area} value} is thus emergent and cannot be controlled directly,\footnote{Indirect control via machine leaning may work, akin to Da\-ta\-sau\-rus \cite{Cairo:2016:DDN,Matejka:2017:SSD}.} we need to pay attention to it when encoding data---specifically because \emph{value} variation is a dominant variation for conveying order \cite{Bertin:1998:SG,bertin:1983:semiology}. 
\hty{It is important to identify which independent pattern parameters affect the \emph{\changed{patterned area} value}.}

Among the variables we discussed, both \emph{orientation} (at the primitive and \changed{lattice} levels) and carefully designed \emph{shape} variation (\ie, maintaining a constant number of black pixels) can keep the area value constant. In addition, employing combinations of variables can also preserve \emph{\changed{patterned area} value}, such as simultaneously adjusting primitive size and spacing (\ie, Bertin's \emph{granularity} variation). 
\emph{Size} variation by itself usually affects \emph{\changed{patterned area} value}, only the special case of changing the subpa\-ra\-me\-ters \emph{width} and \emph{height} in opposite directions can maintain constant \emph{value}. Conversely, an isolated variation of \emph{spacing} or of the individual primitive's \emph{value} also directly affects \emph{\changed{patterned area} value}.

Understanding which visual variables can be used without causing \emph{\changed{patterned area} value} variation can help us to reason about the encoding of data.
One recommended use of patterns \cite{Ware:2019:InformationVisualization}, for instance, is to overlay a \emph{pattern} on a \emph{color} encoding to represent a bivariate scalar field, with one data dimension mapped to \emph{pattern} and the other to \emph{color}.
Using this concept, Retchless and Brewer \cite{Retchless:2015:Uncertainty} compared eight ways to show uncertainty (\autoref{fig:Retchless15Uncertaint} in \autoref{sec:more-examples-of-patterns}). Among them, most par\-ti\-ci\-pants preferred the design with a dot \emph{pattern} overlaid on \emph{color} \inlinevis{-1pt}{1em}{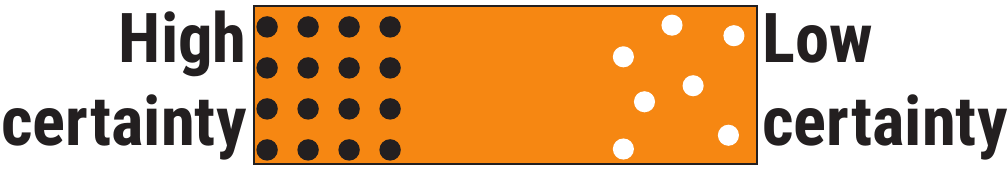} (\autoref{fig:Retchless15Uncertaint}(g)) using \emph{po\-si\-tio\-nal regularity} to encode uncertainty.
Yet Ware \cite{Ware:2019:InformationVisualization} pointed out that, when \emph{patterns} are overlaid on \emph{color}, the bandwidth of \emph{luminance} is shared between the two. 
When using a pattern to represent one of the bivariate variables; therefore, the (ordered) data dimension should be represented by a \emph{pattern} variation with a constant \emph{\changed{patterned area} value}, to minimize its impact on the perceived value of the \emph{color} layer, \eg, \inlinevis{-1pt}{1em}{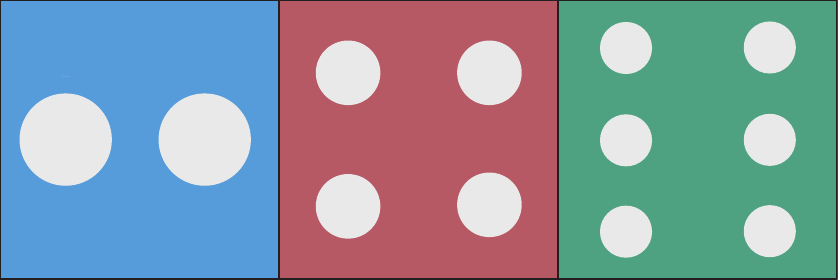} instead of \inlinevis{-1pt}{1em}{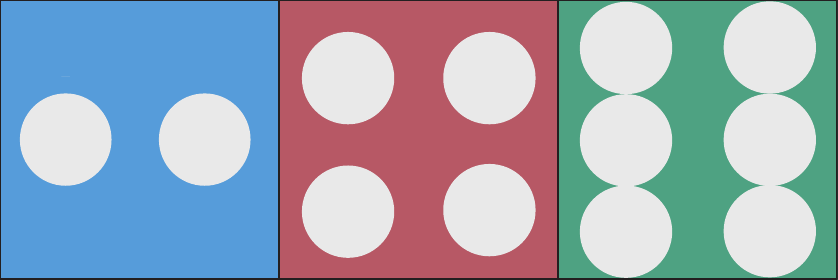} .

\textbf{Regional shade:} 
If all primitives in a pattern are the same color, the regional shade is simply the color of the primitives mixed with the (white) background. If there are multiple color primitives, however, we introduce a \emph{regional shade} to the pattern. 
Incorporating internal \emph{hue} variations among different primitives (\eg, some primitives are blue whereas others are yellow) can result in a different \emph{regional hue} (\eg, green) due to color mixing, similar to color halftoning \inlinevis{-1pt}{1em}{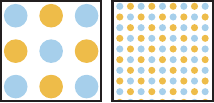}.


\section{\changed{More complex spatial arrangements}}
\label{sec:pattern-beyond-lattice}
\changed{Our exploration of the pattern design space also helps us see gaps for exploration and situate and relate existing methods and practices. In our discussion so far we explored patterns based on a regular arrangement of primitives. 
The regular lattice provides a convenient and consistent framework for expressing the spatial arrangement of primitives within a pattern. 
Bertin explored this approach and it is broadly accepted in the community for encoding data using patterns---yet many other possibilities exist.
We can use, for instance, irregular or nonuniform lattices: Many line dot patterns used to encode lines \cite{brath2005visualization} are defined by irregular spacing and, while lattice crystallography is constrained by nature, visualization design is not.\footnote{\changed{Visualization design may be constrained by convention, which is itself driven by technical capability, which changes. Our efforts here are to open up a broad theoretical design space that is not constrained by convention and is as such somewhat unoccupied and unknown. And thus somewhat exciting.}}
Fundamentally, we thus only need a set of rules for organizing primitives in space---a lattice or any other algorithm that can define the positions of primitives across a given area. In this section, we discuss two representative examples of more complex spatial arrangements that go beyond regular lattice-based configurations: data-driven patterns and nested patterns.}


\subsection{\changed{Data-driven patterns}}
\label{sec:internal-variation-from-geographical-information}


\changed{When we use a lattice to arrange primitives, their position has not been used to encode data---but our pattern system does not prevent us from doing so: we can use each primitive position to encode data directly.}

\setlength{\myfigbesidewidth}{.6\columnwidth}
\begin{figure}
\begin{minipage}[b]{.53\columnwidth}\includegraphics[width=\textwidth]{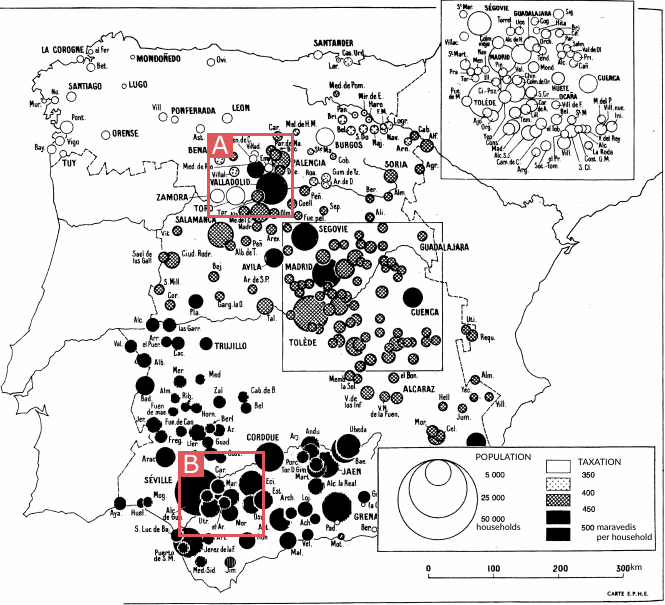}\vspace{8pt}\end{minipage}\hfill
\begin{minipage}[b]{.42\columnwidth}\caption{\textls[-14]{Symbol map example, edited version (regions A and B added and changes for space efficiency: legend moved moved to inside the figure) based on a map about Population and Taxation in Castille from Bertin's book \cite{Bertin:1998:SG,bertin:1983:semiology}. A and B can be considered as patterns, whose primitives encode geo\-gra\-phic information. Image from \cite{bertin:1983:semiology}, \textcopyright~EHESS (text translated), used with permission.}}\label{fig:BertinBook188-DotMap-edit}\end{minipage}\vspace{-2ex}
\end{figure}


One intuitive approach is to encode geographical position directly  (\autoref{fig:pattern-procedure}--last row) as the position of each primitive, yielding symbol maps.
Yet the created appearance is less intuitively a ``pattern'' in the canonical sense. Consider, \eg, the symbol map in \autoref{fig:BertinBook188-DotMap-edit}, in which each dot represents a city and its geographical location on the map. When we read an individual dot we can interpret a city's population (from the dot's \emph{size}) and its tax rate (from the dot's value). When we look at regions, such as Region A, highlighted with a red frame on the map in \autoref{fig:BertinBook188-DotMap-edit}, we see a pattern emerging consisting of a group of dots---the pattern's primitives. We can see that the pattern of Region A has internal variation and conveys comprehensive regional information. From the pattern, we can discern (1) \textbf{where} the cities are located in this region (from the dot \emph{positions}), (2) \textbf{how many} cities there are (from the dot \emph{density}), and (3) \textbf{what} their characteristics are (with dot \emph{size} representing population and \emph{value} representing tax rate). (1) and (2) are emergent variables that come from geographic information, which uses arrangement-level visual variables, whereas (3) is directly encoded on the primitive-level. This pattern also allows us to compare different regions. We can see, \eg, that the pattern in Region A is different from the one in Region B, and we can find that the taxation levels within Region A may have a greater diversity compared to those in Region B. 
We can also seek `patterns' across the map that have similar visual characteristics to that of Region A.
When we read the map, we can thus visually select different patterns at multiple scales and multiple places concurrently to understand geospatial data based on the emergent visual patterns. This process aligns with MacEachren's \cite{MacEachren:1990:pattern} goal of cartographic visualization: to ``assist an analyst in discovering patterns and relationships in the data'' at multiple locations and multiple scales.

\changed{We note that data-driven arrangements are not limited to accurate geographic locations (\autoref{fig:data-driven-geo-pattern:a}); they can also be inaccurate geographic (\ie, transformations applied to accurate geographic locations, \eg, to avoid overlap between dots, \autoref{fig:data-driven-geo-pattern:b}), or gridded geographic (\eg, \autoref{fig:grid-pattern-internal-variation:b} and \ref{fig:data-driven-geo-pattern:c} show an application of this type of pattern). In addition, such arrangements can even be nongeographic---derived from statistical data \hty{(\eg, \cite{Goertler:2019:S2S})}. Fundamentally, we encode the positional information of primitives relative to coordinate axes. In maps, we use geographical locations, but for a scatterplot, \eg, the point positions are the data's $x$- and $y$-dimensions. We can thus use a similar way to interpret emergent ``patterns'' on scatterplots. Our pattern system allows us to describe and differentiate patterns in such cases and facilitates further research on how people can accomplish data analysis tasks at these different pattern levels and with different pattern characteristics.}

\subsection{\changed{Nested patterns}}

\changed{As we discussed in \autoref{sec:pattern-design-space}, our pattern system characterizes a pattern by setting the rules for: (1) \emph{spatial relationships} of primitives (\autoref{sec:spatial-relationship-variables}), (2) \emph{appearance relationships} among primitives (\autoref{sec:pattern-with-internal-variation}), and (3) \emph{retinal visual variables} that define the appearance of each individual primitive (\autoref{sec:retinal}). For the last of these rules we can also treat patterns themselves as retinal variables and use a pattern as primitives of another pattern---a ``nested pattern.'' 
The primitives of the base pattern then serve as host symbols for the new pattern. We can characterize the new added pattern using the same three sets of rules from our overall system and, for the retinal variables of its primitives, we can again add another layer of pattern. We thus provide the possibility of adding multiple layers of additional patterns ad infinitum.\hty{\footnote{It is important to note, however, that, while our systematic description of the pattern design space allows us to describe this recursive application of pattern, its utility may be limited as, \eg, levels of visual complexity will likely be high and interpretation may well be extremely challenging.}}}

\changed{By varying how spatial relationships are defined, we can construct both lattice-based and data-driven nested patterns. Notably, either type can serve as the spatial rule for both the base pattern and the added pattern. A two-layer nested pattern thus yields four possible configurations: (1) lattice pattern on lattice pattern (\eg, \autoref{fig:NYCmap}); (2) lattice pattern on data-driven pattern (\eg, Pattern A in \autoref{fig:BertinBook188-DotMap-edit}); (3) data-driven pattern on lattice pattern (this configuration is less common, but we illustrate it in \autoref{fig:nested-pattern-geo-on-lattice} in \autoref{sec:additional-figures-for-pattern-definition}: a regular grid pattern hosts a dot pattern based on geographical data); and (4) data-driven pattern on data-driven pattern (\eg, \autoref{fig:nested_pattern_od_map_of_london} in \autoref{sec:more-examples-of-patterns}), where we can consider the entire map as a base pattern, and the small maps are the added pattern, and both of which use position to encode aspects of geography).}

\changed{Our new pattern system thus offers a theoretical lens through which we can not only discover new design possibilities but also compare, explain, and relate different designs, to
theoretically connect and perhaps understand idioms that we usually separate with different names.}



\section{Bringing it all together}


We deeply engaged with and discussed Bertin's use of texture and its various translations, interpretations, and applications and noted inconsistencies. This leads us to recommend avoiding the term ``texture'' in lists of visual variables, and to instead discuss the use of granularity, spacing, and shape of \emph{pattern} primitives. Patterns are ultimately a composite visual variable that consists of a series of expressive components whose design space we explored.  We are not the first to have attempted this exploration but our pattern system provides a new unifying angle. 
Specifically, we identified three sets of attributes of patterns. 
First, \emph{spatial arrangement} relationships of pattern primitives distinguish how pattern primitives are arranged, either through a static, dynamic, or irregular lattice or modified by data parameters. This conceptualization of arrangements encapsulates previously proposed pattern parameters \cite{brewer:2016:DesigningBetterMaps,slocum:2022:ThematicCartography,mackinlay:1986:automating,chung:2016:ordered,Morrison:1974:TFC} such as density or regularity.  
Second, \emph{group appearance relationships} introduce internal variation within patterns as a novel concept that includes the number of primitive groups, ratio between groups, and distribution style. 
Our group appearance \mbox{relationships} allow us to categorize and discuss pattern visualizations by Bertin and others. 
Examples in \autoref{fig:internal-variation-variables:b} and \ref{fig:internal-variation-distribution-style:b} in \autoref{sec:additional-figures-for-pattern-definition} also remind us of waffle charts, BallotMaps \cite{Wood:2011:BallotMaps}, or proportion visualizations more generally.
Finally, we discussed the use of \emph{retinal variables on pattern primitives}. Although retinal variables as such have been investigated before, we introduce new variables made possible by patterns, specifically the regularity of the re\-ti\-nal variables and the visual effects caused by using retinal variables repeatedly. 
Based on \hty{our} three attribute sets we further discussed the use of multiple variables and the creation of patterns through more complex spatial arrangements.

\changed{Our design space is deeply generative---it is easy to conceive of pattern variations using any of our dimensions. 
Existing examples (\eg, \autoref{fig:grid-pattern-internal-variation:a} and \ref{fig:NYCmap} as well as \autoref{fig:great-war-isotype} in \autoref{sec:more-examples-of-patterns}) demonstrate the potential of patterns in encoding data, making them worthy of further exploration.
Creative design experiments are needed to establish possibilities and rigorous empirical work required to understand how these variations are perceived.
One aspect that we have not at all touched upon, however, is the use of patterns decidedly \emph{\textbf{not}} for encoding data.
Think, for example, about the use of non-lattice arrangements or intentional irregularity in a lattice-based pattern being used simply for aesthetic purposes---to make the visualization more interesting. \autoref{fig:pat_example_brinton_4} and~\ref{fig:land_in_farms}, for instance, vary the dot placement in an even yet irregular fashion, which shows the overall density well yet in a visually more interesting fashion than a regular lattice-based placement---linking our system to approaches used in non-photorealistic rendering on stippling \cite{Martin:2017:SDS,Deussen:2013:HS} and non-re\-pe\-ti\-tive patterns (\eg, \cite{Barla:2006:SPA,Hurtut:2009:appearance, Martin:2017:SDS, salisbury:1994:interactive, Winkenbach:1994:CGP}). Also parameters other than position can be manipulated purely for aesthetic purposes, such as line width to simulate hand-drawn lines as in traditional data graphics \cite{Bach:2013:IDG,Wood:2012:SRI}.}



\changed{The degree to which such uses of pattern for aesthetics \cite{He:2023:BVS} are effective should be further investigated,} \changed{as well as their effect on readability \cite{Cabouat:2025:PPR}. 
In addition, we need to study the encoding size of types of pattern variations---that is, how many different variations of a texture are even noticeable by a viewer.  Any of these metrics can be applied to a number of specific questions. For example, it would be interesting to explore whether regular patterns create more readable visualizations than random placements or if and how semantically resonant patterns can be created with abstract pattern designs \cite{Lu:2025:DSP}. Specifically intriguing is the use of combinations of patterns with other retinal variables as well as nested patterns. We see few examples of their application in practice. Is it because they are a bad idea? Because common visualization libraries do not allow or facilitate pattern encoding? Or because people have lacked the conceptual framing to explore and advocate for or study their use? Ultimately, our design space is thus also evaluative. It aids in the design of studies comparing different pattern parameters to ultimately build up a more comprehensive understanding of what makes patterns interesting, effective and appreciated. In addition, given the limited pattern support in current visualization tools (\autoref{sec:pattern-design-tools}), one could build a more flexible pattern library for based on our pattern system to support pattern design and empirical studies.}


Finally, having described a wide design space, we do not know the various limits of the use of patterns. As we use more parameters and visual variables for encoding in composites, the resulting intricate patterns are unlikely to be easily interpretable or reliably decoded. Like hand-drawn stipple images, or efforts to convey surface material characteristics, the properties of \emph{patterns} may then become more akin to those of natural \emph{textures} \cite{Maciejewski:2007:AHD,Maciejewski:2008:MSA}, suggesting that our system may have potential for bridging conceptual, linguistic, and perhaps practical gaps in data visualization design.

\acknowledgments{We thank all members of the Aviz team and the University of Utah's SCI Institute for their insightful input throughout this theory-building process, especially A.-F.\ Cabouat, F.\ Cabric, C.\ Han, and Y.\ Lu.}

\section*{Supplemental Material Pointers}

We share our additional material (author version of the paper including the appendix with footnotes, self-created figures) at \osfrepo. 

\nocite{Goertler:2019:S2S}

\section*{Images/figures license/copyright}
With the exception of images whose li\-cen\-ses\discretionary{/}{}{/}co\-py\-rights we have specified in the respective captions, we as authors state that all our own figures in this article (\ie, \autoref{fig:BertinBook-TranslatorNote} and~\ref{fig:pattern-procedure} and all word-scale visualizations we embedded in the text) are and remain under our own personal copyright, with the permission to be used here. We also make them available under the \href{https://creativecommons.org/licenses/by/4.0/}{Creative Commons At\-tri\-bu\-tion 4.0 International (\ccLogo\,\ccAttribution\ \mbox{CC BY 4.0})} license and share them at \osfrepo.

\nocite{Hunter:2007:Matplotlib,plotly,ggpattern,mdn:svg-patterns,He:2024:DCB,Zhong:2020:BWT,caivano:1994:towards,caivano:1990:visual,bertin:1983:semiology,Bertin:1998:SG,NYCmap,brinton:1919:graphic,Brinton:1939:GP,neurath:2010:hieroglyphics,Retchless:2015:Uncertainty,brewer:2016:DesigningBetterMaps,Zhang:2024:OCD,deBrouwer:2017:FTL,Jo:2019:DRM,Kumpf:2019:VAT,Kumpf:2018:VCC,Chan:2019:ViBr}

\bibliographystyle{abbrv-doi-hyperref-narrow}
\bibliography{abbreviations,template}
\flushcolsend
\raggedend

\appendix 

\clearpage

\begin{strip} 
\noindent\begin{minipage}{\textwidth}
\makeatletter
\centering%
\sffamily\bfseries\fontsize{15}{16.5}\selectfont
\papertitle \\[.5em]
\large Appendix\\[.75em]
\makeatother
\normalfont\rmfamily\normalsize\noindent\raggedright In this appendix we provide additional figures and some detailed discussion that we could include in the main paper due to space limitations or because it was not essential for explaining our approach. 
\end{minipage}
\end{strip}

\appendix

\newpage

\section{Caivano's system for texture description}
\label{sec:more-on-caivano}
\begin{figure}[!t]
\vspace{0.5ex}
    \centering
        (a)\,\includegraphics[width=0.18\columnwidth]{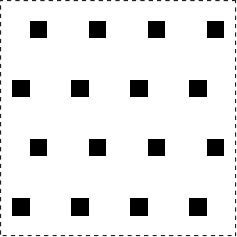}\hfill%
        (b)\,\includegraphics[width=0.18\columnwidth]{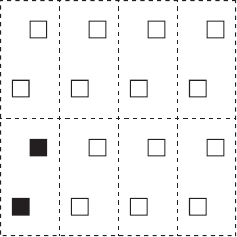}\hfill%
        (c)\,\includegraphics[width=0.18\columnwidth]{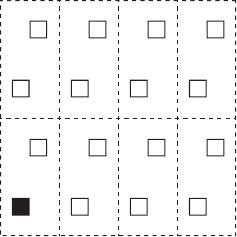}\hfill%
        (d)\,\includegraphics[width=0.18\columnwidth]{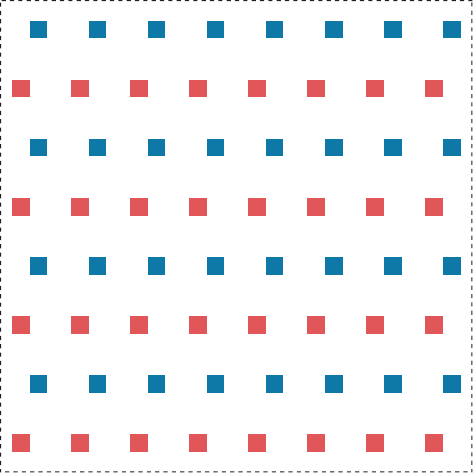}%
    \caption{Recreated schematic drawing based on Caivano's diagram of composition of a simple texture \cite{caivano:1994:towards, caivano:1990:visual}. (a) A texture, (b) a texture unit, (c) a texture element, (d) two subsets of textures identified from this simple texture composition, colored blue and red, respectively.}
    \label{fig:Caivano-SimpleTextureComposition}
\end{figure}

From the field of architecture, Caivano \cite{caivano:1994:towards, caivano:1990:visual} adopts a design approach to describe patterns (although under the term ``texture''). He classifies \emph{simple textures} and \emph{complex textures}, defining the former as ``the uniform repetition of a certain element'' and the latter as combinations of multiple sets of simple textures \cite{caivano:1990:visual}. His simple tex\-tures are essentially two elements within a tiling unit. Caivano constructs his simple texture through the tiling of a \emph{texture unit} (the minimal entity for repetition; see \autoref{fig:Caivano-SimpleTextureComposition} (b)). In Caivano's model, a texture unit comprises a pair of texturing elements (see \autoref{fig:Caivano-SimpleTextureComposition} (c)). He then treats texture as a tripartite variable, including size of the texture elements, directionality (the unit's width-height ratio), and density (the overall black-to-white ratio). Later \cite{caivano:1994:towards}, he refines his theory to describe pattern variation through the shape of texture elements, organization (the relative positions of the two texture elements within the tiling unit), proportionality (the tiling unit's width-height ratio), and density (the overall black-to-white ratio). 

Cavaino, however, did not intend to use texture as a visual variable for data encoding. As a result, not all the dimensions he identifies are directly manipulable, and his composition of simple textures is unsuitable for our purpose.
In addition, we interpret Caivano's classification such that a simple texture should be the most basic form of texture---without any subsets of textures (``uniform repetition of a certain element''). If a texture is a combination of multiple sets of textures, we should categorize it as a complex texture. Upon analysis of Caivano's simple texture composition, however, we identify two subsets of texture within it, which appears to contradict our interpretation of his definition of a simple texture. \autoref{fig:Caivano-SimpleTextureComposition}(d) illustrates the two subtextures identified in a simple texture according to Caivano's composition, with blue and red highlighting, respectively. We thus develop our own pattern configuration and offer an alternative that covers a wider design space of \emph{pattern}, specifically aimed at encoding data, which we present in \autoref{sec:pattern-design-space}.

\section{\changed{Terms translated as ``pattern'' in Bertin's book}}
\label{sec:more-on-pattern}
\begin{figure}
    \centering
    \includegraphics[width=\columnwidth]{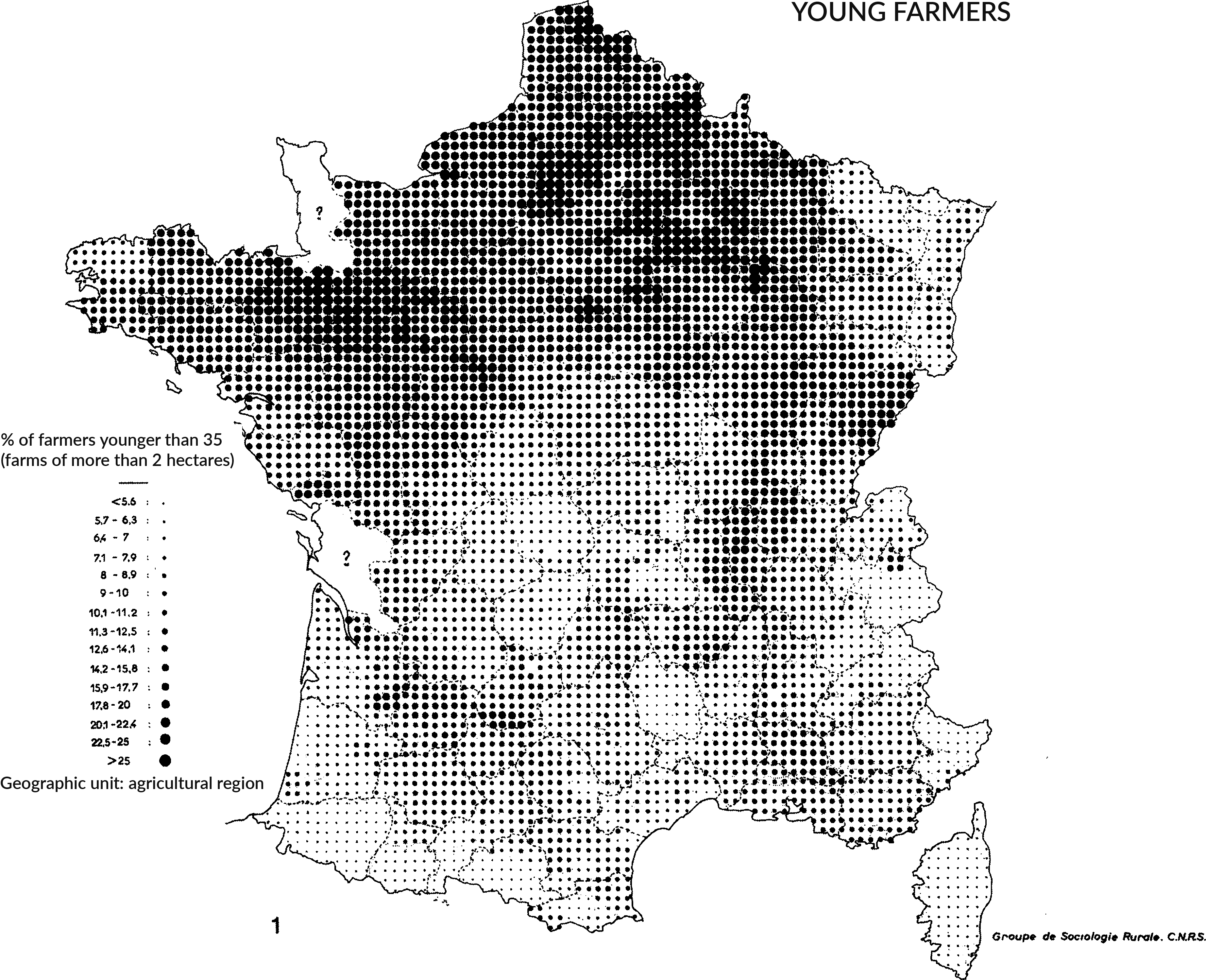}
    \caption{Examples of ``semis'' from Bertin's book. \cite{bertin:1983:semiology,Bertin:1998:SG}; \textcopyright\ EHESS (text translated), \todo{used with permission.}}
    \label{fig:BertinBook-Semis}
\end{figure}

\changed{In the translator's note (\autoref{fig:BertinBook-TranslatorNote}), the translator mentions that ``pattern'' is the English equivalent for the French word ``texture.'' However, in the book, the term ``pattern'' is not actually translated from ``texture'' in French. The translator translated the following French terms from Bertin's book \cite{Bertin:1998:SG} into English as ``pattern'':}
\changed{\begin{description}
    \item[``Semis,''] which means ``seedbed.'' Bertin extends this term to refer to a ``dot pattern.'' He consistently uses it in phrases like ``semis régulier (a regular pattern)'' or ``semis de point (a dot pattern),'' but it always refers to something similar, as shown in \autoref{fig:BertinBook-Semis}. In the Lexicon of the book \cite{Bertin:1998:SG}, ``semis'' is explained as ``Type d'imposition qui disperse, sur le plan, les éléments d'une variable,'' which translates to a type of graphic that spreads the elements of a variable across a plane. ``Imposition'' in Bertin's book refers to a graphic type (\eg, a map). This reflects the characteristic regularity of patterns, which are governed by rules of arrangement. Here, ``semis'' serves as the rule for generating a pattern in terms of the spatial arrangement of graphical primitives across a plane.
    \item[``Baguettage/ligne''] refers to a single line but is translated as ``line pattern.'' Similarly, ``pointillé'' has been translated as ``dot pattern,'' although it actually means ``dotted'' and, by extension, can denote a dotted line.''
    \item[``Trame''] is a term used in drawing or technical drawing, with a meaning similar to ``hatching,'' a specific type of pattern used in technical drawing for indicating materials. The word ``trame'' also conveys a sense of a grid. In Bertin's book, it appears in phrases such as ``des trames mécaniques'' (mechanical hatching) and ``des trames préfabriquées'' (prefabricated hatching), referring to preprinted hatchings. \autoref{fig:BertinBook331-PreprintedPattern} shows examples of these preprinted hatchings.
\end{description}}

\section{\changed{Additional figures for our new pattern definition}}
\label{sec:additional-figures-for-pattern-definition}

\changed{\autoref{fig:TextureComposition}--\ref{fig:PatternVariation-2Dp2Da-P-regularity} illustrate some aspects of our new pattern definition in more detail, beyond our summary in \autoref{fig:pattern-procedure}.}

\setlength{\examplewidth}{0.15\columnwidth}
\begin{figure}[t]
    \centering
        \setlength{\subfigcapskip}{-3ex}
        \subfigure[\hspace{\columnwidth}]{\label{fig:TextureComposition:a}~~~~~\includegraphics[width=\examplewidth]{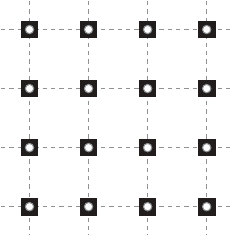}}\hfill%
        \subfigure[\hspace{\columnwidth}]{\label{fig:TextureComposition:b}~~~~~\includegraphics[width=\examplewidth]{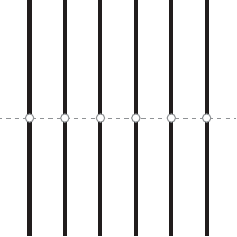}}\hfill%
        \subfigure[\hspace{\columnwidth}]{\label{fig:TextureComposition:c}~~~~~\includegraphics[width=\examplewidth]{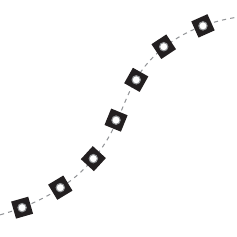}}\hfill%
        \subfigure[\hspace{\columnwidth}]{\label{fig:TextureComposition:d}~~~~~\includegraphics[width=\examplewidth]{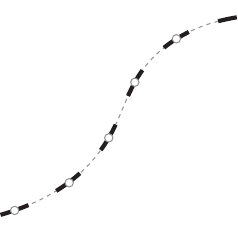}}
    \caption{Configuration of \emph{pattern} with (a) 2D primitives on a 2D lattice, tiling across a 2D area\changed{---\texttt{2\texttimes2\texttimes2}};
    (b) 1D primitive on a 1D lattice, tiling across a 2D area\changed{---\texttt{1\texttimes1\texttimes2}};
    (c) 2D primitive on a 1D lattice, tiling along a 1D line \changed{(that fills a 2D area)---\texttt{2\texttimes1\texttimes1}}; and (d) 1D primitive on a 1D lattice, tiling along a 1D line \changed{(that fills a 2D area)---\texttt{1\texttimes1\texttimes1}}. \changed{We can apply (a) and (b) to area symbols, or to point symbols with area (\eg, circles) and line symbols with width. We can apply (c) and (d) only to line symbols, as their lattices follow the line symbol's direction.}
    \changed{Here, the primitives are black. The gray dashed lines and white dots represent the lattice and the lattice points, which we use only for descriptive purposes and they are not part of the \emph{pattern} itself.}}
    \label{fig:TextureComposition}
\end{figure}

\begin{figure}
    \centering
        \setlength{\subfigcapskip}{-3ex}
    ref: \includegraphics[width=\examplewidth]{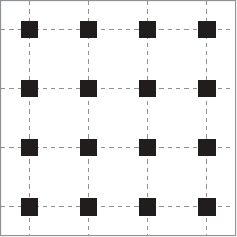}\hspace{5mm}%
        \subfigure[\hspace{\columnwidth}]{\label{fig:lattice-parameters:a}~~~~~\includegraphics[width=\examplewidth]{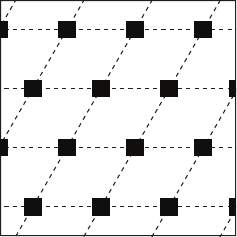}}\hspace{5mm}%
        \subfigure[\hspace{\columnwidth}]{\label{fig:lattice-parameters:b}~~~~~\includegraphics[width=\examplewidth]{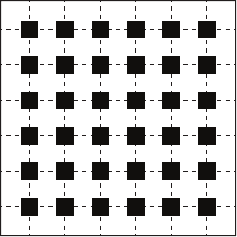}}
    \caption{Variation on (a) shape and (b) size of unit cells.}
    \label{fig:lattice-parameters}
\end{figure}

\begin{figure}[t]
    \centering
        \setlength{\subfigcapskip}{-3ex}
        ref: \includegraphics[width=\examplewidth]{figures/system_2Dp2Da-default.pdf}\hfill%
        \subfigure[\hspace{\columnwidth}]{\label{fig:PatternVariation-2Dp2Da-orientation:a}~~~~~\includegraphics[width=\examplewidth]{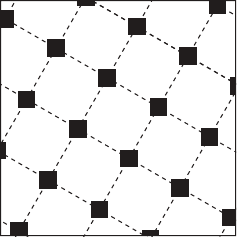}}\hfill%
        \subfigure[\hspace{\columnwidth}]{\label{fig:PatternVariation-2Dp2Da-orientation:b}~~~~~\includegraphics[width=\examplewidth]{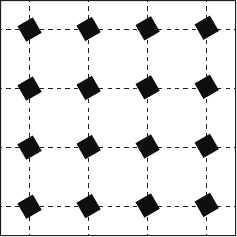}}\hfill%
        \subfigure[\hspace{\columnwidth}]{\label{fig:PatternVariation-2Dp2Da-orientation:c}~~~~~\includegraphics[width=\examplewidth]{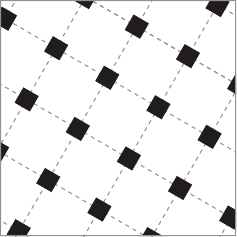}}
    \caption{Orientation at different levels, compared to the left: (a) at arrangement-level, (b) at primitive level, and (c) at both levels (we can call it orientation of the whole pattern), all with same degrees. }
    \label{fig:PatternVariation-2Dp2Da-orientation}
\end{figure}

\begin{figure}[t]
    \centering
        \setlength{\subfigcapskip}{-3ex}
        ref: \includegraphics[width=\examplewidth]{figures/system_2Dp2Da-default.pdf}\hfill%
        \subfigure[\hspace{\columnwidth}]{\label{fig:PatternVariation-2Dp2Da-A-regularity:a}~~~~~\includegraphics[width=\examplewidth]{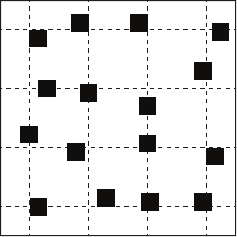}}\hfill%
        \subfigure[\hspace{\columnwidth}]{\label{fig:PatternVariation-2Dp2Da-A-regularity:b}~~~~~\includegraphics[width=\examplewidth]{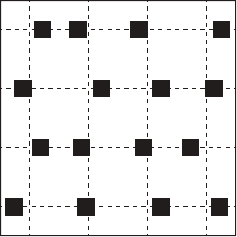}}\hfill%
        \subfigure[\hspace{\columnwidth}]{\label{fig:PatternVariation-2Dp2Da-A-regularity:c}~~~~~\includegraphics[width=\examplewidth]{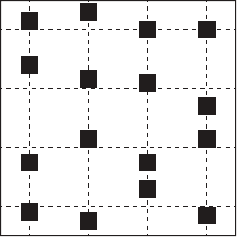}}
    \caption{Positional regularity variation, compared to the reference on the left: (a) in both directions or (b, c) only in one direction.}
    \label{fig:PatternVariation-2Dp2Da-A-regularity}
\end{figure}

\begin{figure}
    \centering%
        \setlength{\subfigcapskip}{-3ex}%
    ref: \includegraphics[width=\examplewidth]{figures/system_2Dp2Da-default.pdf}\hspace{5mm}%
        \subfigure[\hspace{\columnwidth}]{\label{fig:combination-and-internal-variation:a}~~~~~\includegraphics[width=\examplewidth]{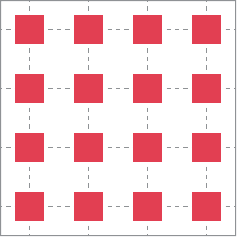}}\hspace{5mm}%
        \subfigure[\hspace{\columnwidth}]{\label{fig:combination-and-internal-variation:b}~~~~~\includegraphics[width=\examplewidth]{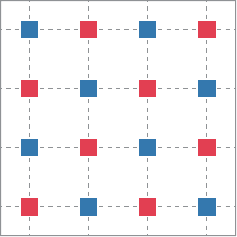}}
    \caption{The reference (left) has one primitive group: (a) global encoding with hue \& size, with one primitive group; (b) pattern with internal variation for hue (one subset blue, another subset red), with two primitive groups.}%
    \label{fig:combination-and-internal-variation}%
\end{figure}

\begin{figure}
    \centering%
        \setlength{\subfigcapskip}{-3ex}%
    ref: \includegraphics[width=\examplewidth]{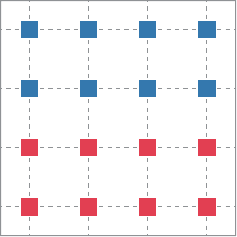}\hspace{5mm}%
        \subfigure[\hspace{\columnwidth}]{\label{fig:internal-variation-variables:a}~~~~~\includegraphics[width=\examplewidth]{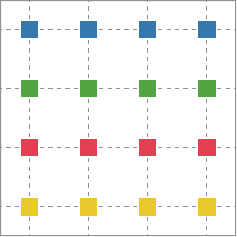}}\hspace{5mm}%
        \subfigure[\hspace{\columnwidth}]{\label{fig:internal-variation-variables:b}~~~~~\includegraphics[width=\examplewidth]{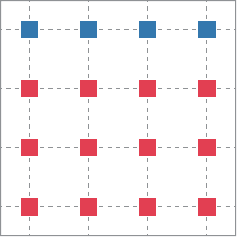}}
    \caption{Compared to the reference (left), (a) increase of primitive groups from 2 to 4, with even primitive count per group; (b) variation of the ratio between each group (from 1:1 to 1:3), with a constant number of primitive groups (2). Here, the different primitive groups are differentiated by hue, but any primitive-level variable can be used, \ie, size, shape, etc.}%
    \label{fig:internal-variation-variables}
\end{figure}

\begin{figure}
    \centering%
        \setlength{\subfigcapskip}{-3ex}%
        \subfigure[\hspace{\columnwidth}]{\label{fig:internal-variation-distribution-style:b}~~~~~\includegraphics[width=\examplewidth]{figures/system_2Dp2Da-internalvariation-color-PosStacked.pdf}}\hspace{5mm}%
        \subfigure[\hspace{\columnwidth}]{\label{fig:internal-variation-distribution-style:c}~~~~~\includegraphics[width=\examplewidth]{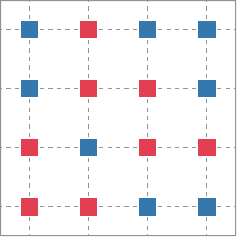}}\hspace{5mm}
        \subfigure[\hspace{\columnwidth}]{\label{fig:internal-variation-distribution-style:a}~~~~~\includegraphics[width=\examplewidth]{figures/system_2Dp2Da-internalvariation-color-PosUniform.pdf}}
    \caption{Pattern with internal variation with different primitive group arrangement: \subref{fig:internal-variation-distribution-style:b} grouped, \subref{fig:internal-variation-distribution-style:c} interspersed, and \subref{fig:internal-variation-distribution-style:a} dispersed.}%
    \label{fig:internal-variation-distribution-style}%
\end{figure}

\begin{figure}
    \centering%
        \setlength{\subfigcapskip}{-3ex}%
    ref: \includegraphics[width=\examplewidth]{figures/system_2Dp2Da-default.pdf}\hspace{5mm}%
        \subfigure[\hspace{\columnwidth}]{\label{fig:combination-multiple-variables:a}~~~~~\includegraphics[width=\examplewidth]{figures/system_2Dp2Da-combination-size-color.pdf}}\hspace{5mm}%
        \subfigure[\hspace{\columnwidth}]{\label{fig:combination-multiple-variables:b}~~~~~\includegraphics[width=\examplewidth]{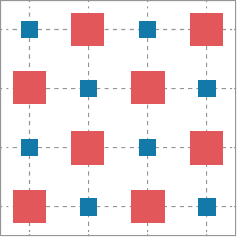}}\hspace{5mm}%
        \subfigure[\hspace{\columnwidth}]{\label{fig:combination-multiple-variables:c}~~~~~\includegraphics[width=\examplewidth]{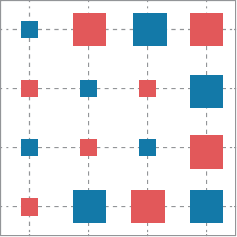}}
    \caption{The reference (left) has one primitive group: (a) global encoding with hue and size, with one primitive group (``combination of variables'' in Bertin's book); (b) Primitives' hue and size covary in the pattern, with same two primitive groups for the two variables; (c) Primitives' hue and size do not covary in the pattern, two primitive groups for hue (\ie, two hues) and size (\ie, two sizes), but they are not the same two groups.}%
    \label{fig:combination-multiple-variables}%
\end{figure}

\begin{figure}
    \centering
        \setlength{\subfigcapskip}{-2.5ex}
        ref: \includegraphics[width=\examplewidth]{figures/system_2Dp2Da-default.pdf}\hfill%
        \subfigure[\hspace{\columnwidth}]{\label{fig:PatternVariation-2Dp2Da-P-regularity:a}~~~~~\includegraphics[width=\examplewidth]{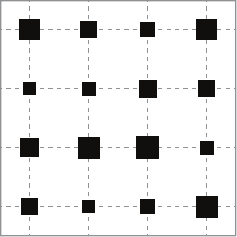}}\hfill%
        \subfigure[\hspace{\columnwidth}]{\label{fig:PatternVariation-2Dp2Da-P-regularity:b}~~~~~\includegraphics[width=\examplewidth]{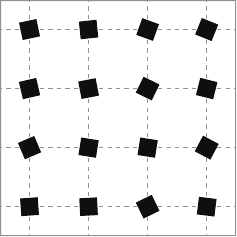}}\hfill%
        \subfigure[\hspace{\columnwidth}]{\label{fig:PatternVariation-2Dp2Da-P-regularity:c}~~~~~\includegraphics[width=\examplewidth]{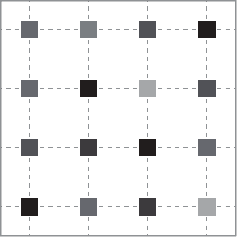}}
    \caption{Primitive regularity variation, compared to the left: (a) for size, (b) for orientation, and (c) for value.}
    \label{fig:PatternVariation-2Dp2Da-P-regularity}
\end{figure}


\section{Additional figures from Bertin's book that we used in our discussion}
\label{sec:additional-figures-from-bertin}
\begin{figure}[!t]
    \centering
        \includegraphics[width=\columnwidth]{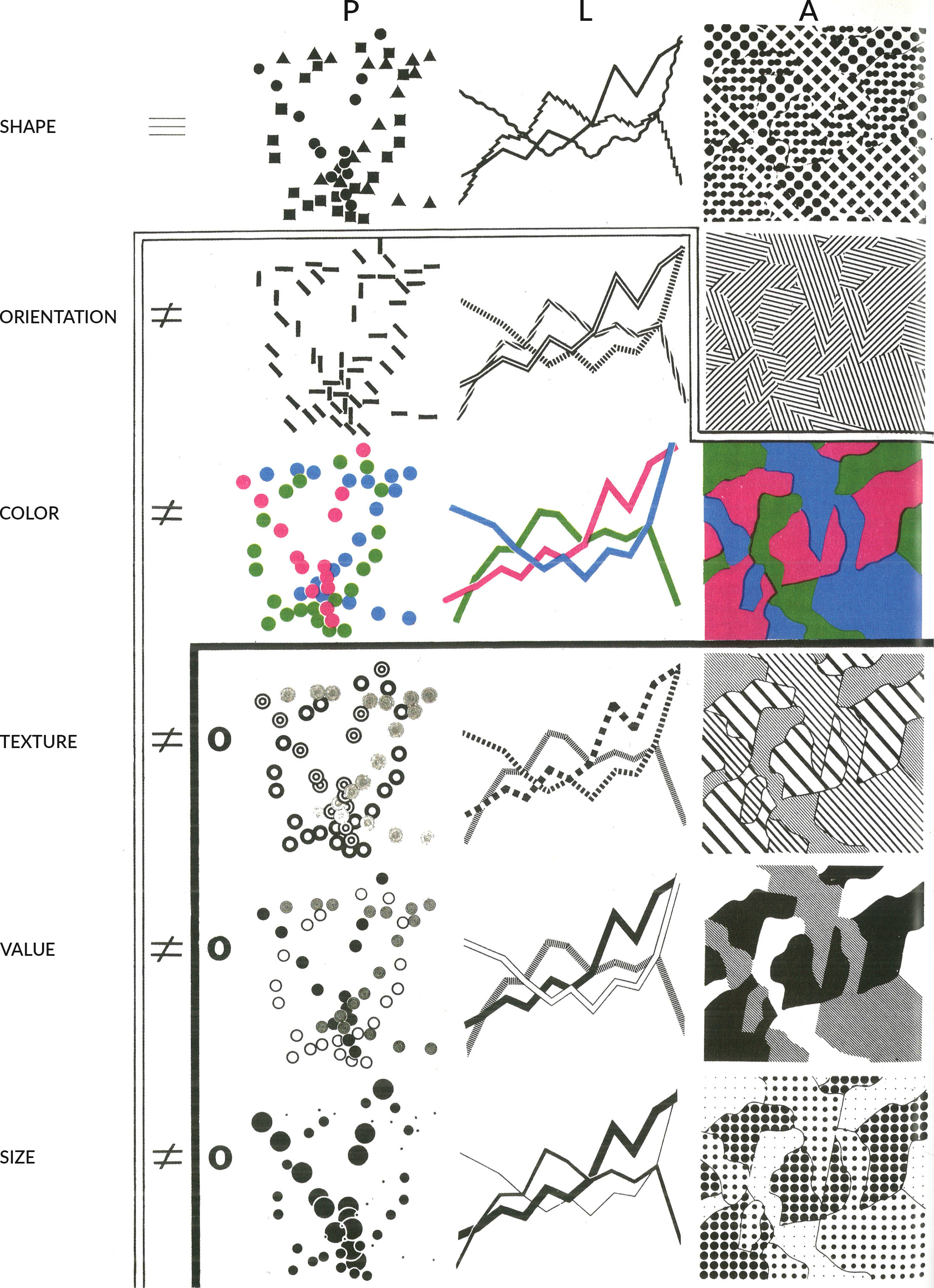}
    \caption{Bertin's diagram for visual variables across three mark types \cite{Bertin:1998:SG,bertin:1983:semiology}. From left to right, the columns represent point mark, line mark, and area mark; image \textcopyright~EHESS (text translated), used with permission.}
    \label{fig:bertin-variables-on-three-mark-types}
\end{figure}

\begin{figure}[!t]
    \centering
    \includegraphics[width=\columnwidth]{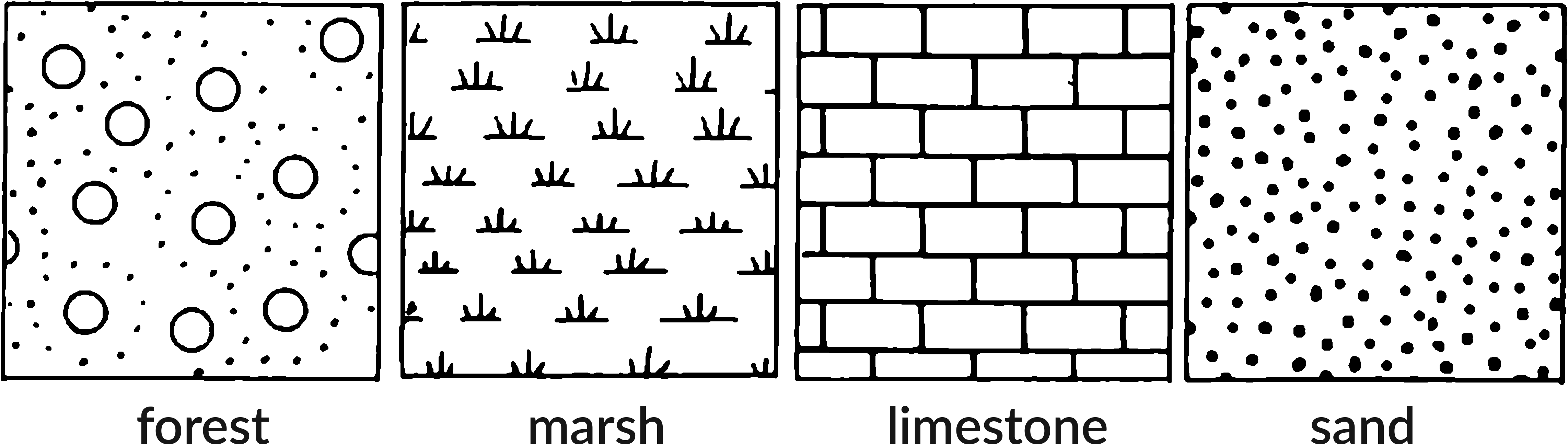}
    \caption{Bertin's examples of ``patterns have achieved the status of symbols'' from \cite{Bertin:1998:SG,bertin:1983:semiology}; image \textcopyright\ EHESS (text translated), used with permission.}
    \label{fig:BertinBook-PatternSymbol}
\end{figure}

\begin{figure}[!t]
    \centering
    \includegraphics[width=\columnwidth]{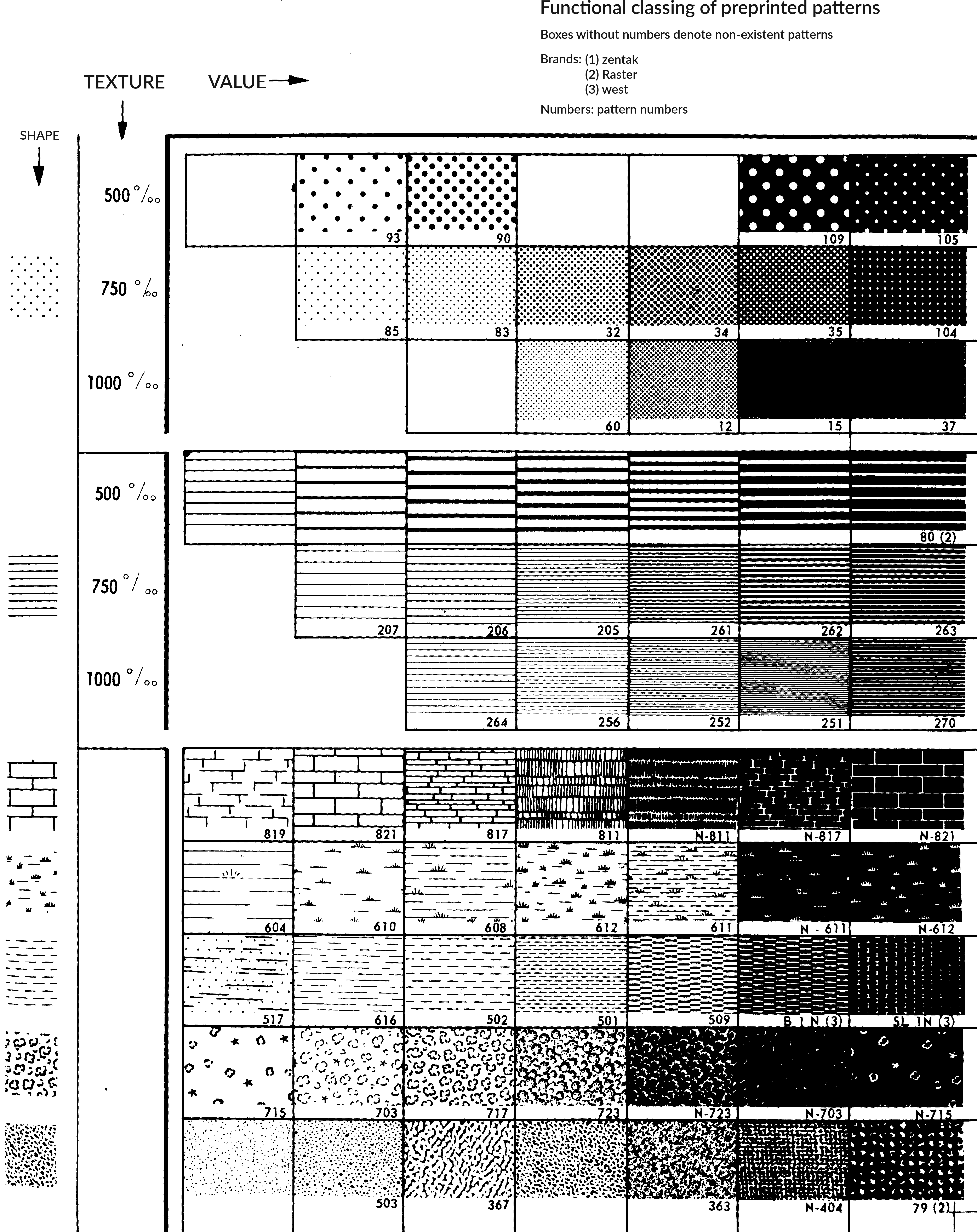}
    \caption{Examples of pre-printed hatchings from Bertin's book \cite{Bertin:1998:SG,bertin:1983:semiology}; image \textcopyright~EHESS (text translated), used with permission.}
    \label{fig:BertinBook331-PreprintedPattern}
\end{figure}

\begin{figure}[!t]
    \centering
    \includegraphics[width=0.8\columnwidth]{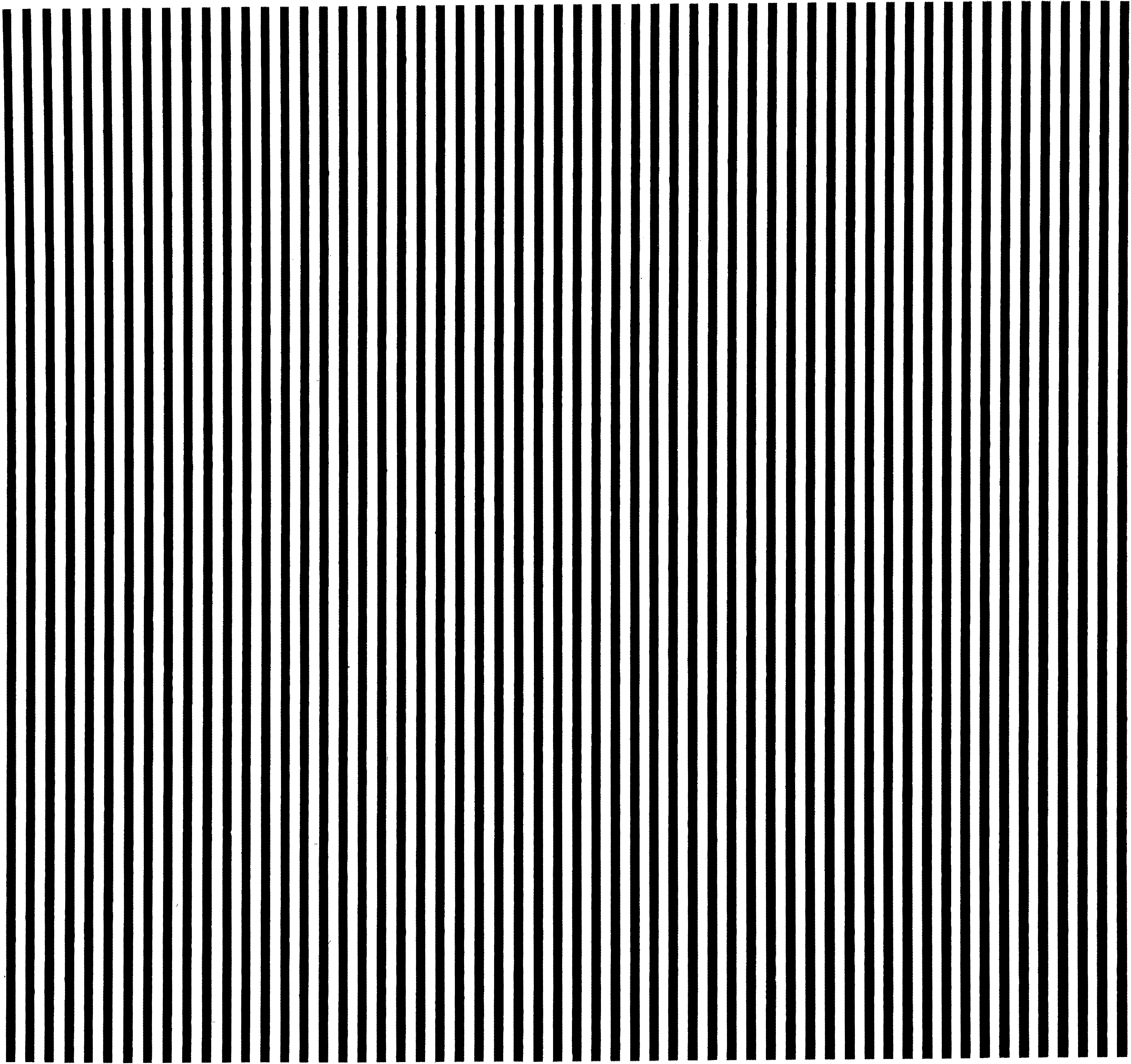}
    \caption{An example of Moiré effect from Bertin's book \cite{Bertin:1998:SG,bertin:1983:semiology}; image \textcopyright~EHESS, used with permission.}
    \label{fig:BertinBook80-Moire}
\end{figure}

This section includes supplemental examples (\autoref{fig:bertin-variables-on-three-mark-types}--\ref{fig:BertinBook80-Moire}) from Bertin's book \cite{bertin:1983:semiology,Bertin:1998:SG}, to which we referred in our discussion in the main paper. 
Specifically, \autoref{fig:bertin-variables-on-three-mark-types} shows Bertin's six retinal visual variables as they are applied onto line and area marks (as well as point marks). Next, \autoref{fig:BertinBook-PatternSymbol} shows an example by Bertin for patterns with an embedded semantic meaning---here from the field of cartography---that have achieved the status of symbols in this domain. More generally, specific \emph{hat\-ching} patterns are often used in technical and architectural drawing as we show in \autoref{fig:BertinBook331-PreprintedPattern} from Bertin's book. Finally, \autoref{fig:BertinBook80-Moire} shows an example from Bertin for repeated, regular patterns that can cause a visual sense of instability, or Moiré effect.

\section{More examples of patterns}
\label{sec:more-examples-of-patterns}
\begin{figure}[!t]
     \centering
         \includegraphics[width=\columnwidth]{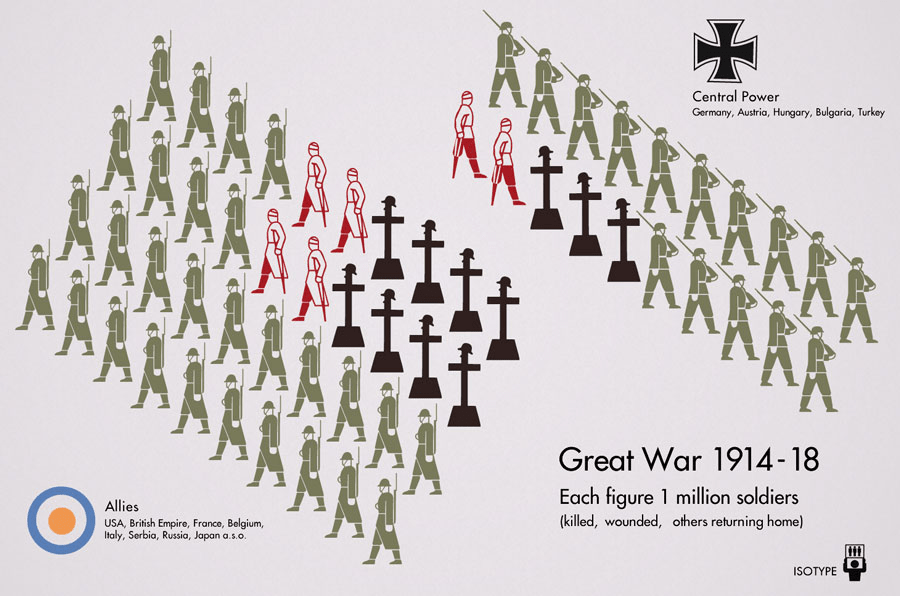}
     \caption{Unit visualization (IsoType, \cite{neurath:2010:hieroglyphics}) that can be considered to be using internal variation. Image `The Great War' by Otto Neurath; \ccPublicDomain\ the image is in the public domain.}
     \label{fig:great-war-isotype}
 \end{figure}

\begin{figure}[!t]
    \centering
    \includegraphics[width=\columnwidth]{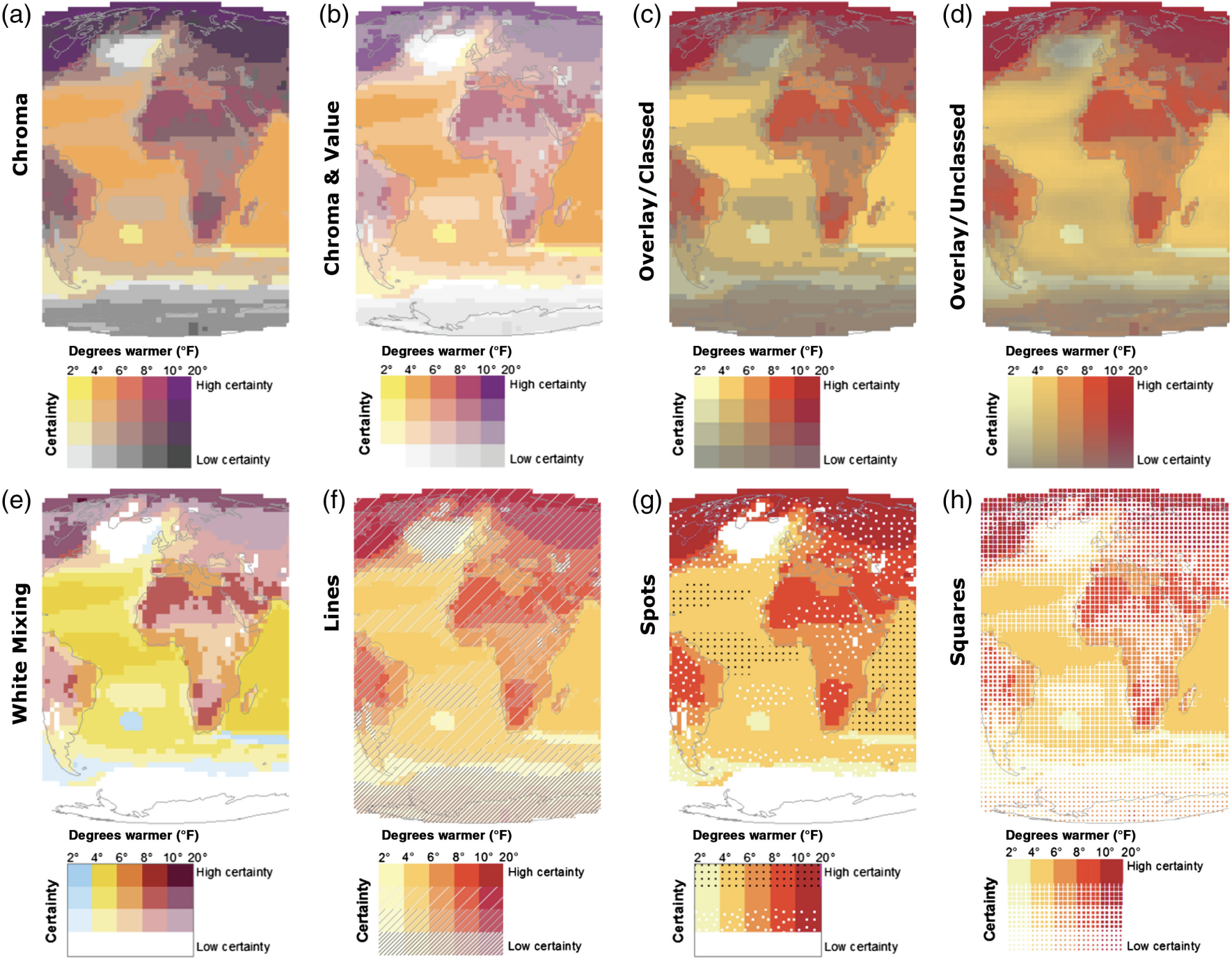}
    \caption{Comparison of eight uncertainty representations by Retchless and Brewer. Most participants prefer (g). Image reproduced from \cite[Fig.~11]{Retchless:2015:Uncertainty}, used under the \href{https://creativecommons.org/licenses/by-nc/4.0/}{Creative Commons At\-tri\-bu\-tion-Non\-Com\-mer\-cial 4.0 International (\ccLogo\,\ccAttribution\,\ccNonCommercialEU\ \mbox{CC BY-NC}) license}.}
    \label{fig:Retchless15Uncertaint}
\end{figure}

\begin{figure}[!t]
	\centering
	\includegraphics[width=\columnwidth]{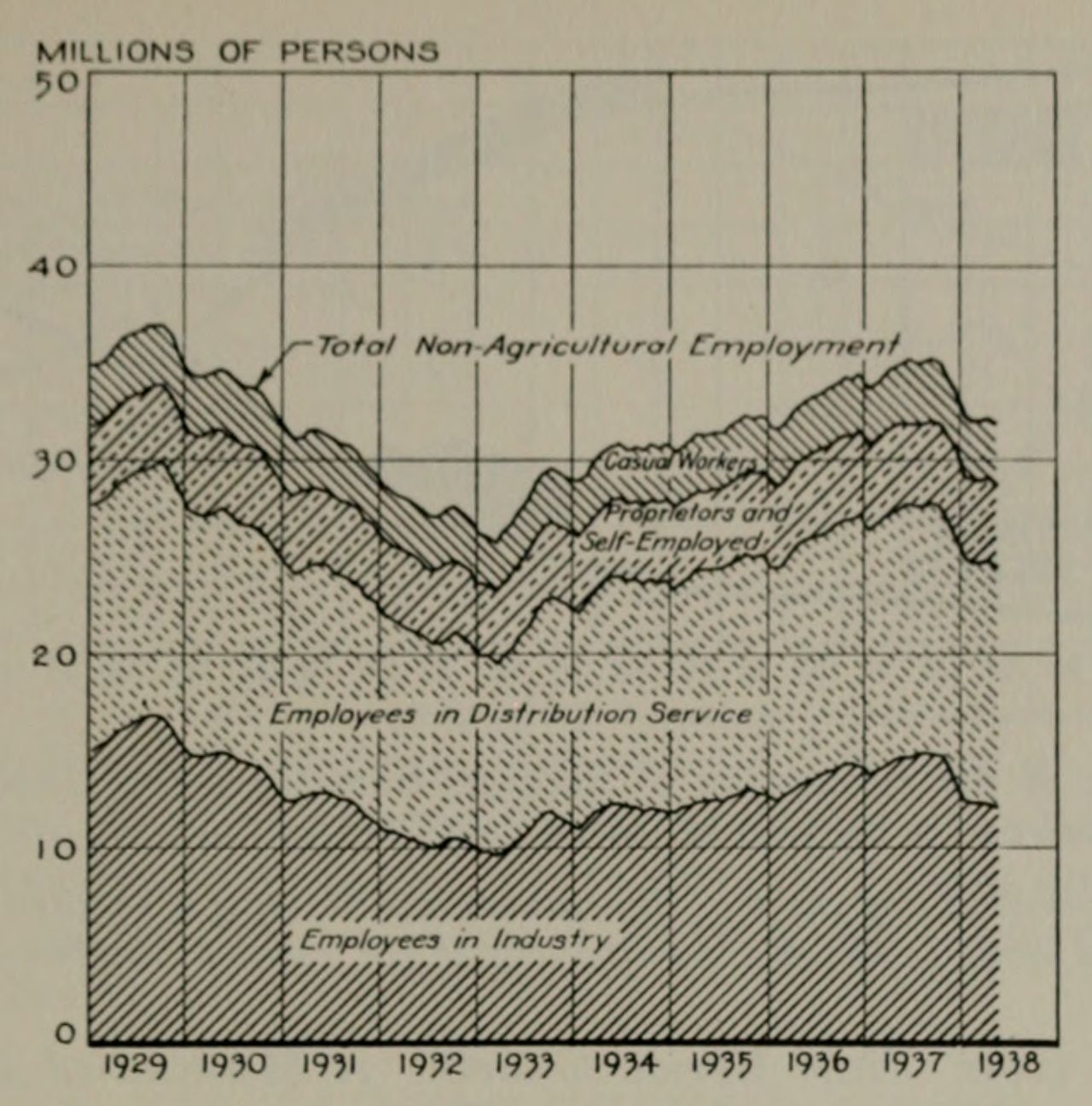}
	\caption{\changed{Historic example of multiple 1D lattice-based patterns used for qualitatively marking regions; by Willard C.\ Brinton \cite[page 297]{Brinton:1939:GP}; \ccPublicDomain\ the image is in the public domain.}}
	\label{fig:pat_example_brinton_7}
\end{figure}

\begin{figure}[!t]
	\centering
	\includegraphics[width=\columnwidth]{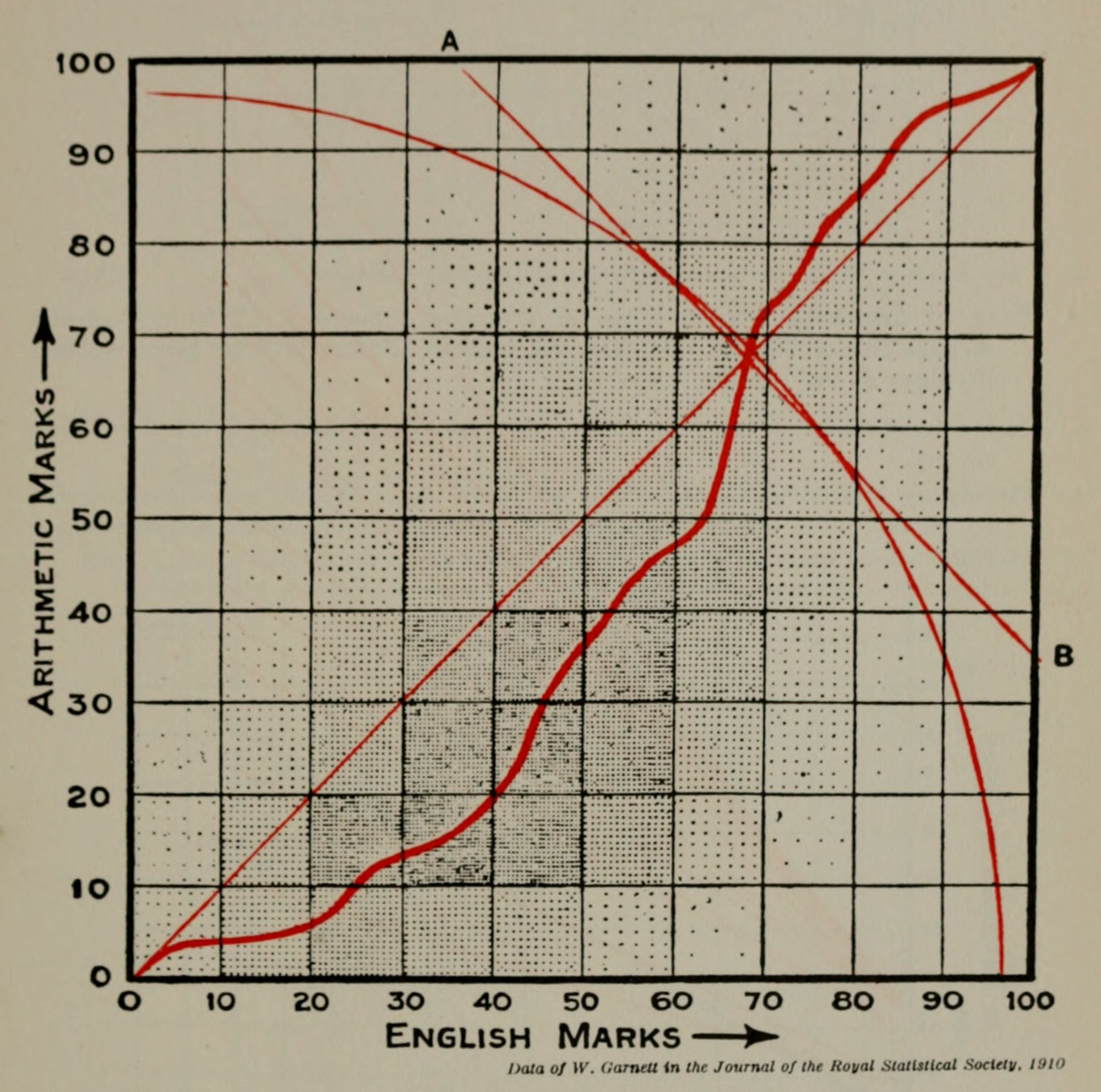}
	\caption{\changed{Historic example of 2D lattice-based pattern for encoding data values (notice that the reproduction of the pattern is partially flawed, likely due to problems with the used plate); by Willard C.\ Brinton \cite[page 327]{Brinton:1939:GP}; \ccPublicDomain\ the image is in the public domain.}}
	\label{fig:pat_example_brinton_8}
\end{figure}

\begin{figure*}
	\centering
	\includegraphics[width=\linewidth]{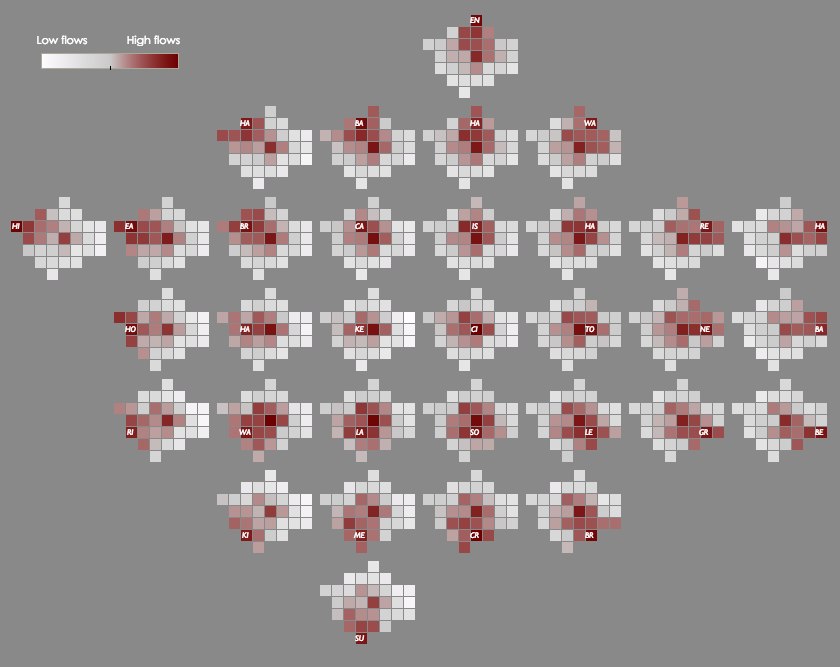}
	\caption{\changed{\href{https://mappinglondon.co.uk/2016/od-map-of-london-commuting/}{Origin Destination Maps (OD Maps) \cite{Wood:2010:Visualisation} of London Commuting}: an example of a nested pattern in which both the top-level pattern and the embedded patterns are data-driven. Image \textcopyright{} by Robert Radburn, used with permission.}}
	\label{fig:nested_pattern_od_map_of_london}
\end{figure*}

\begin{figure}[!t]
	\centering
	\includegraphics[width=\columnwidth]{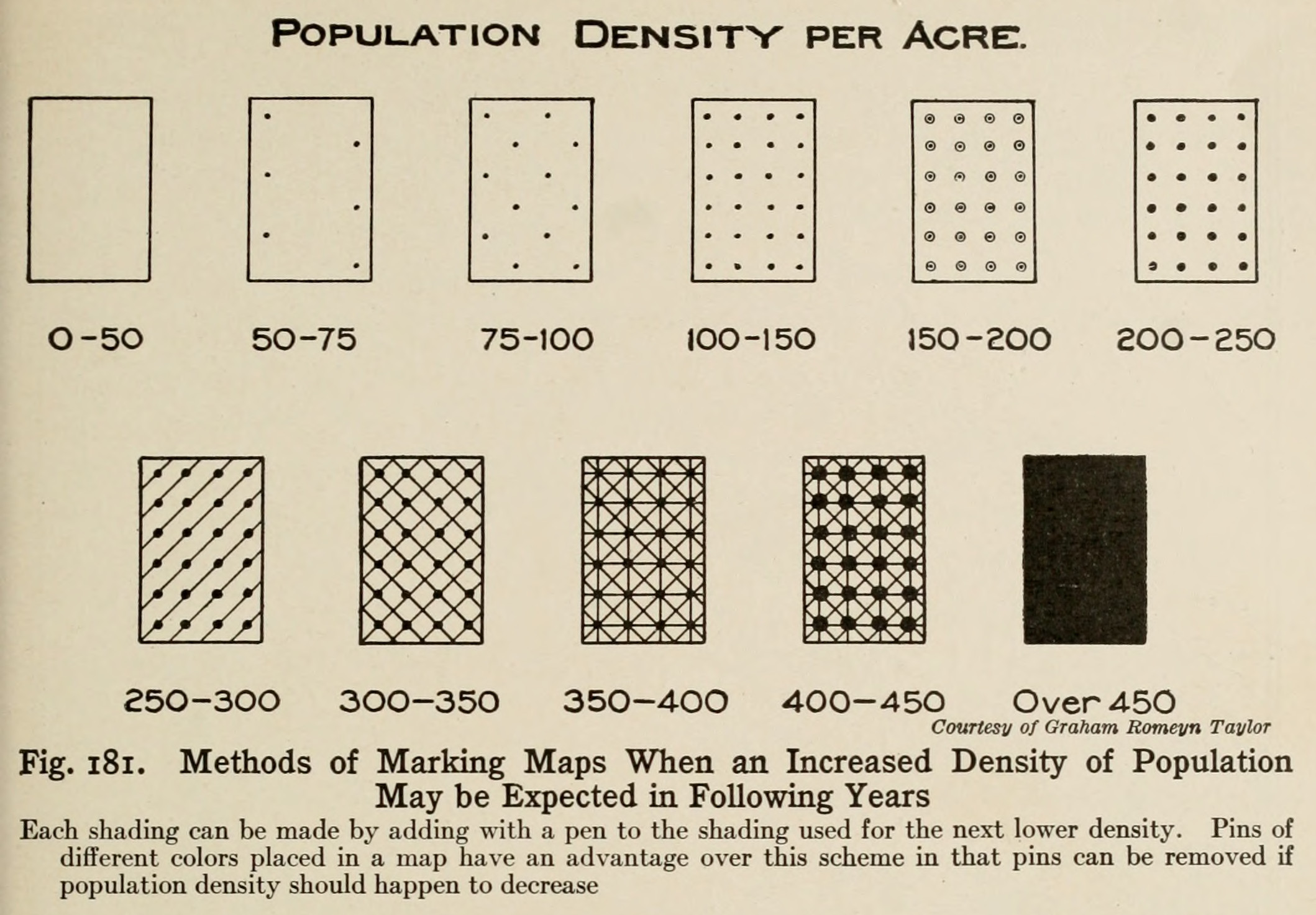}
	\caption{Historic example pattern with \emph{lattice size} (density) sequence by Willard C.\ Brinton \cite[page 221]{brinton:1919:graphic}; \ccPublicDomain\ the image is in the public domain.}
	\label{fig:pat_example_brinton_1}
\end{figure}

\begin{figure}[!t]
	\centering
	\includegraphics[width=\columnwidth]{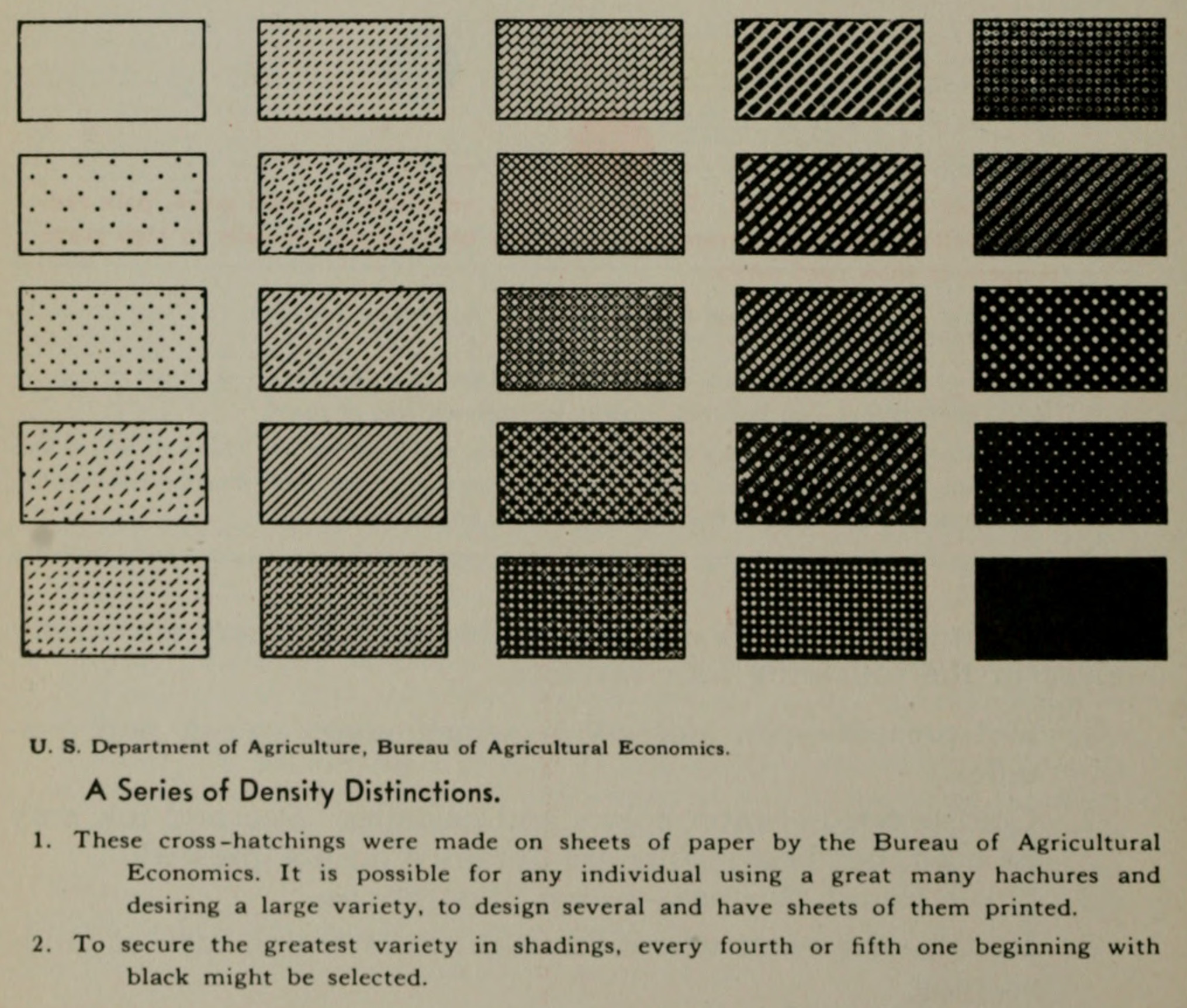}
	\caption{Historic example pattern with \emph{lattice size} (density) sequence by Willard C.\ Brinton \cite[page 422]{Brinton:1939:GP}; \ccPublicDomain\ the image is in the public domain.}
	\label{fig:pat_example_brinton_2}
\end{figure}

\begin{figure}[!t]
	\centering
	\includegraphics[width=\columnwidth]{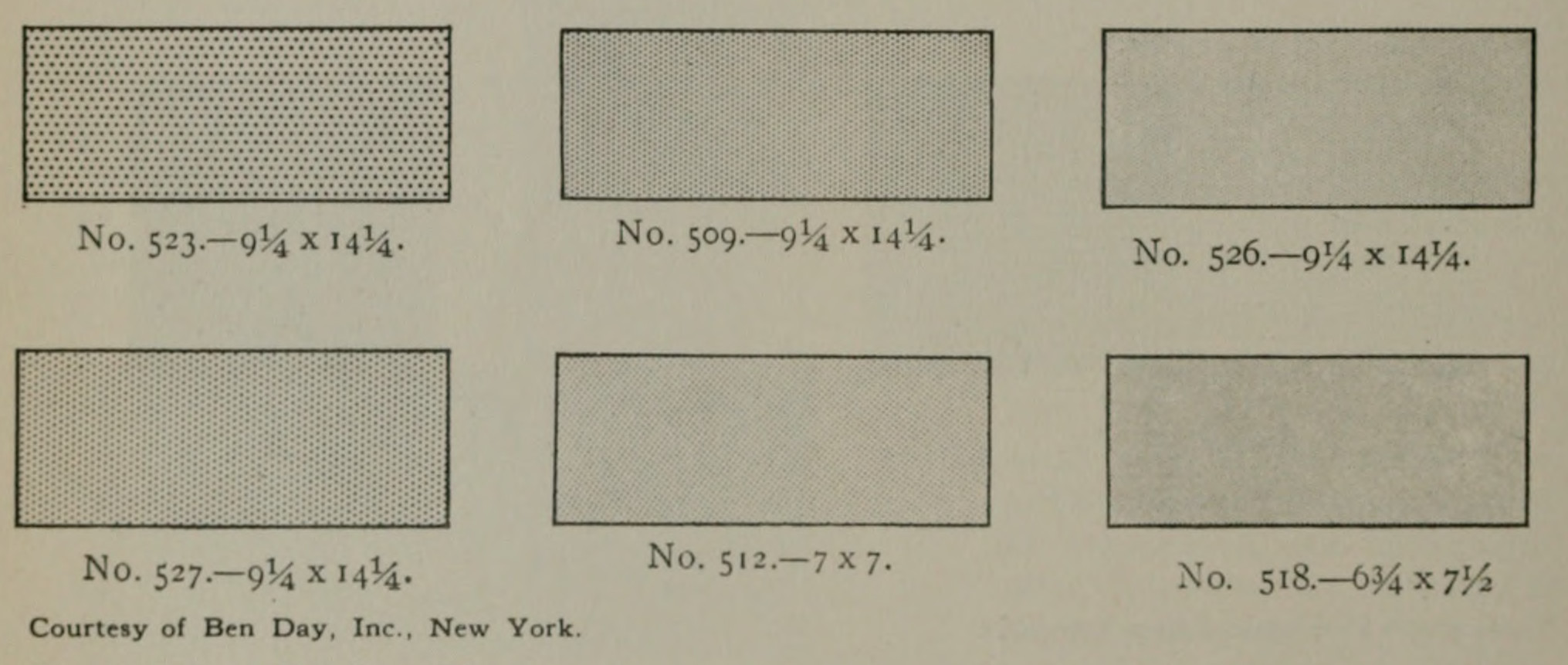}
	\caption{Historic example pattern with \emph{granularity} sequence by Willard C.\ Brinton \cite[page 420]{Brinton:1939:GP}; \ccPublicDomain\ the image is in the public domain.}
	\label{fig:pat_example_brinton_3}
\end{figure}

\begin{figure}[!t]
	\centering
	\includegraphics[width=\columnwidth]{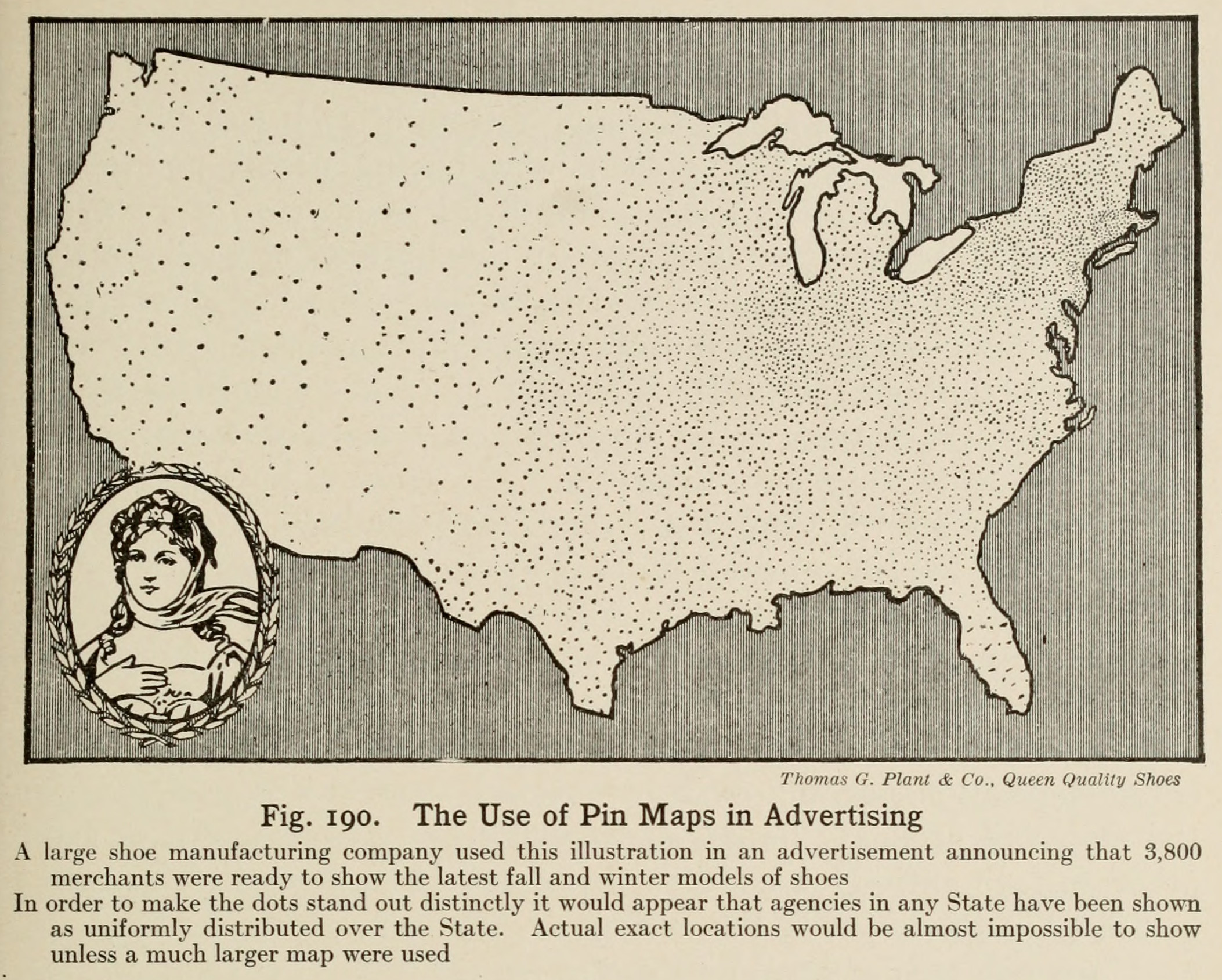}
	\caption{Historic example of using \emph{lattice size} (density) to illustrate the, in fact, density of merchants of a given product in the US; ``pin map'' by Willard C.\ Brinton \cite[page 233]{brinton:1919:graphic}; notice that the use of the pattern can transition from a density pattern encoding in regions with high merchant density to a location-based encoding with discrete, actual locations in regions with lower merchant density (but we are not actually sure whether this was done in this illustration); \ccPublicDomain\ the image is in the public domain.}
	\label{fig:pat_example_brinton_4}
\end{figure}

\begin{figure}[!t]
	\centering
	\includegraphics[width=\columnwidth]{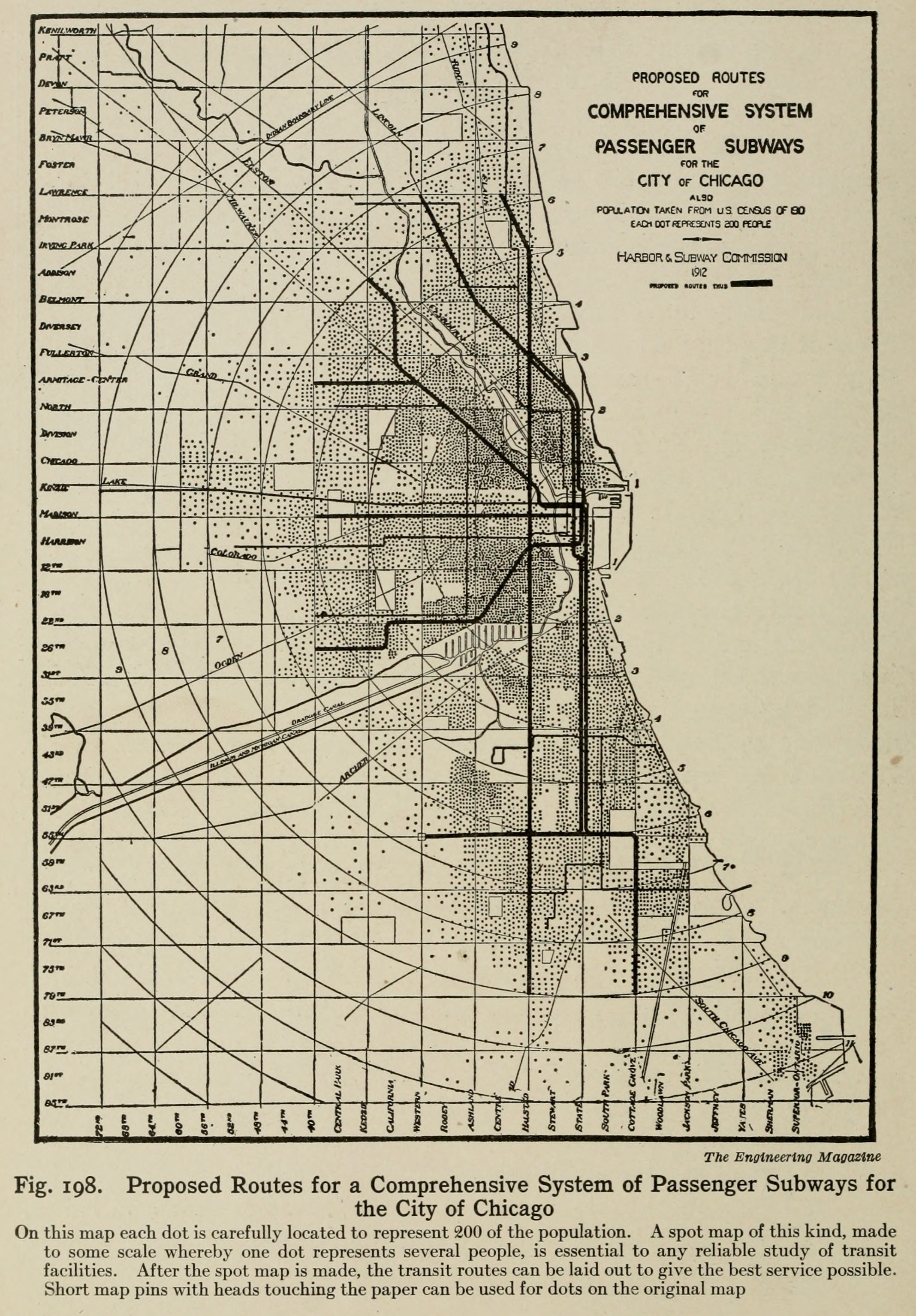}
	\caption{Another historic example of using \emph{lattice size} (density) to illustrate a density of people in a city (Chicago in the US) by Willard C.\ Brinton \cite[page 246]{brinton:1919:graphic}; \ccPublicDomain\ the image is in the public domain.}
	\label{fig:pat_example_brinton_5}
\end{figure}

\begin{figure*}[!t]
	\centering
	\includegraphics[width=\linewidth]{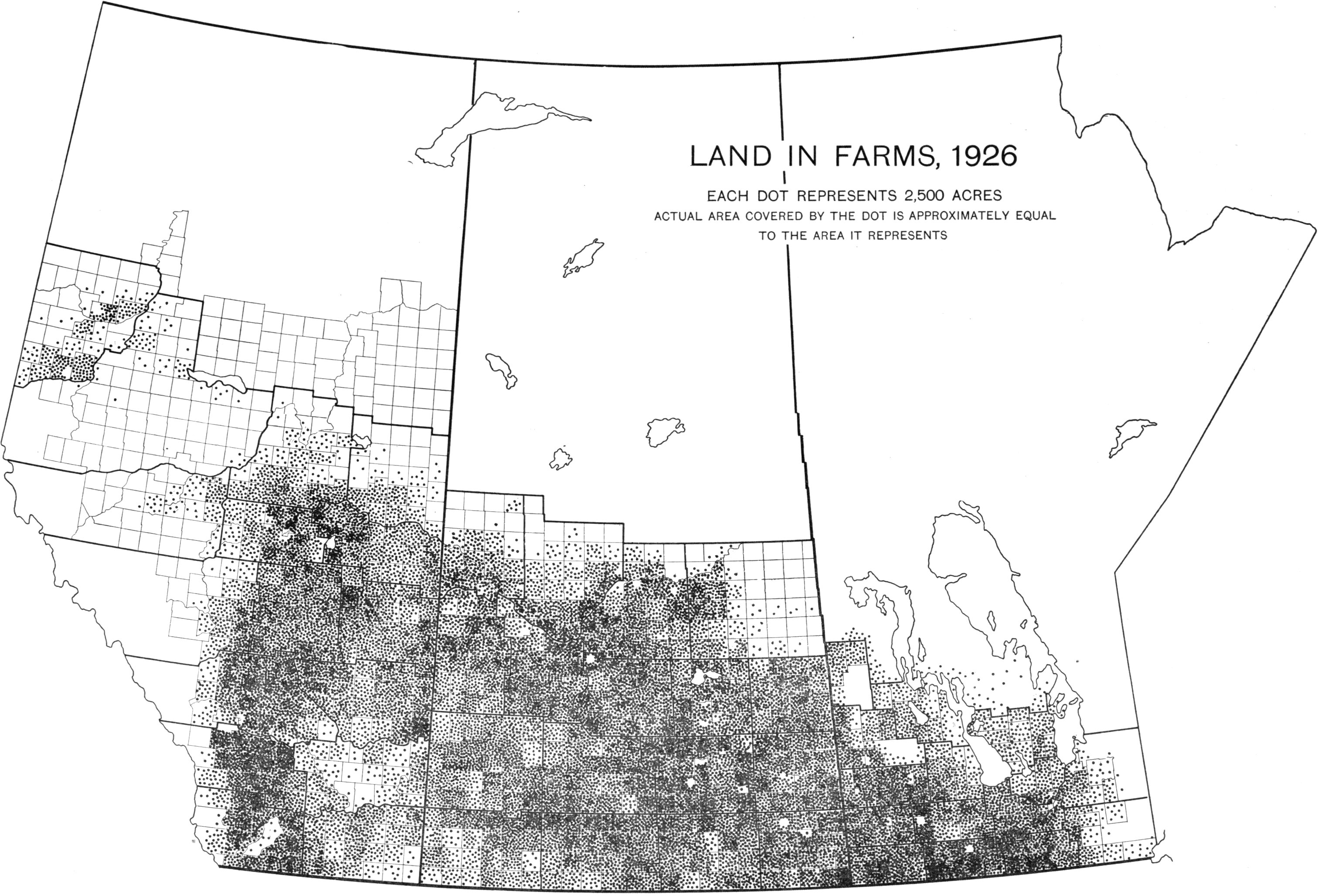}
	\caption{Another historic example of dot patterns based on \emph{lattice size} (density); notice the remark that the size of each dot represents approximately the area of the actual space to which it refers; \href{https://archive.org/details/1926981926m1931eng/page/28/mode/2up}{image published in 1931 by the Canada Dominion Bureau of Statistics, Ottawa} (Fig.~29 on page~28 of their ``\href{https://archive.org/details/1926981926m1931eng/}{Agriculture, Climate and Population of the Prairie Provinces of Canada}'' statistical atlas), \ccPublicDomain\ the image is in the public domain.}
	\label{fig:land_in_farms}
\end{figure*}

\begin{figure*}[!t]
	\centering
	\includegraphics[width=\linewidth,trim={10mm 195mm 10mm 10mm},clip]{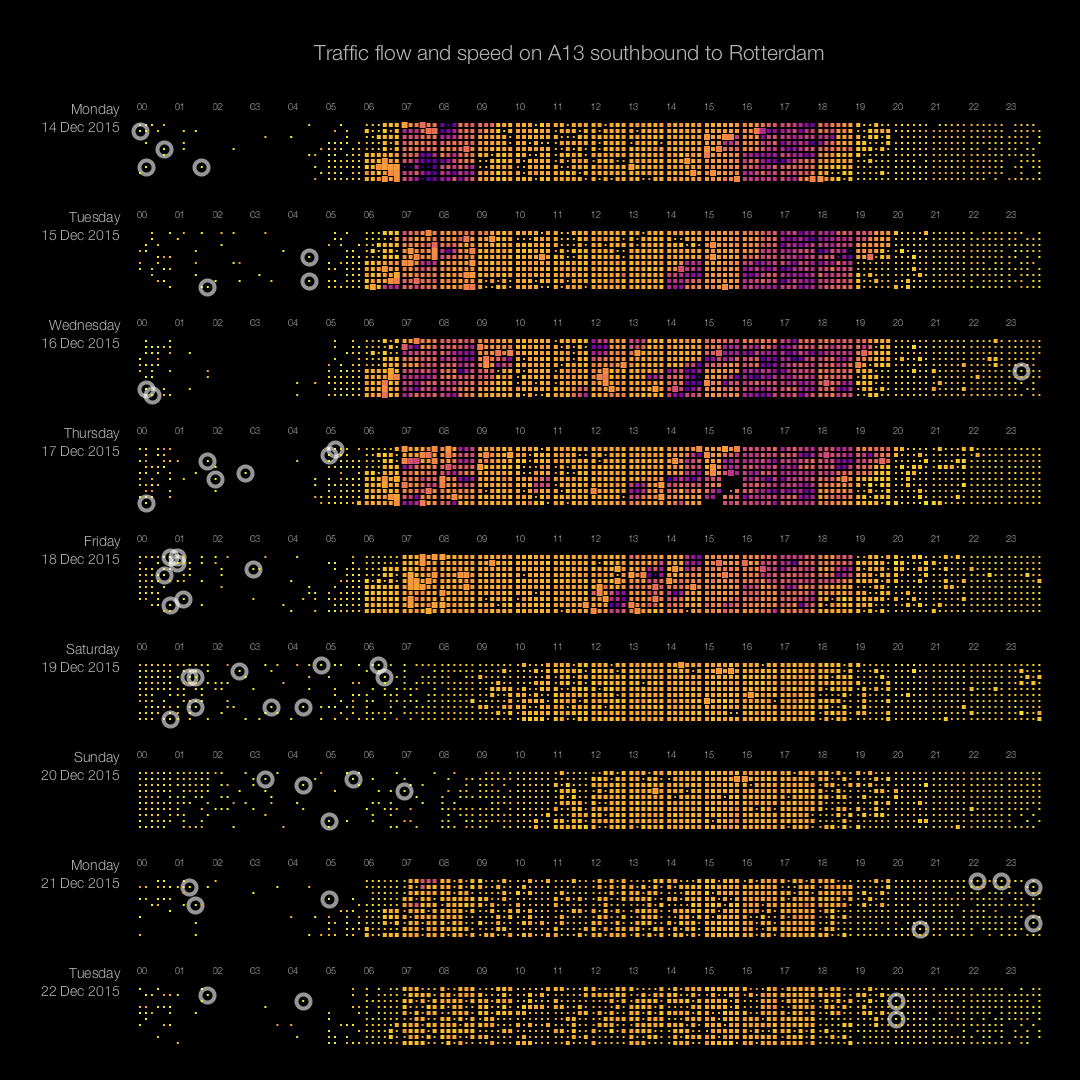}
	\caption{Contemporary pixel-based visualization of traffic data, technically not a pattern (because the data marks are each individual minutes in the day, to which a size and a color hue is mapped based on that minute's actually captured data), this visualization still has characteristics of a pattern if viewed, \eg, per hour. \href{https://social-glass.tudelft.nl/visualising-traffic-data/}{Image \textcopyright{} Erik Boertjes}, used with permission.}
	\label{fig:pixel_patterns}
\end{figure*}

\begin{figure}[!t]
	\centering
	\includegraphics[width=\columnwidth]{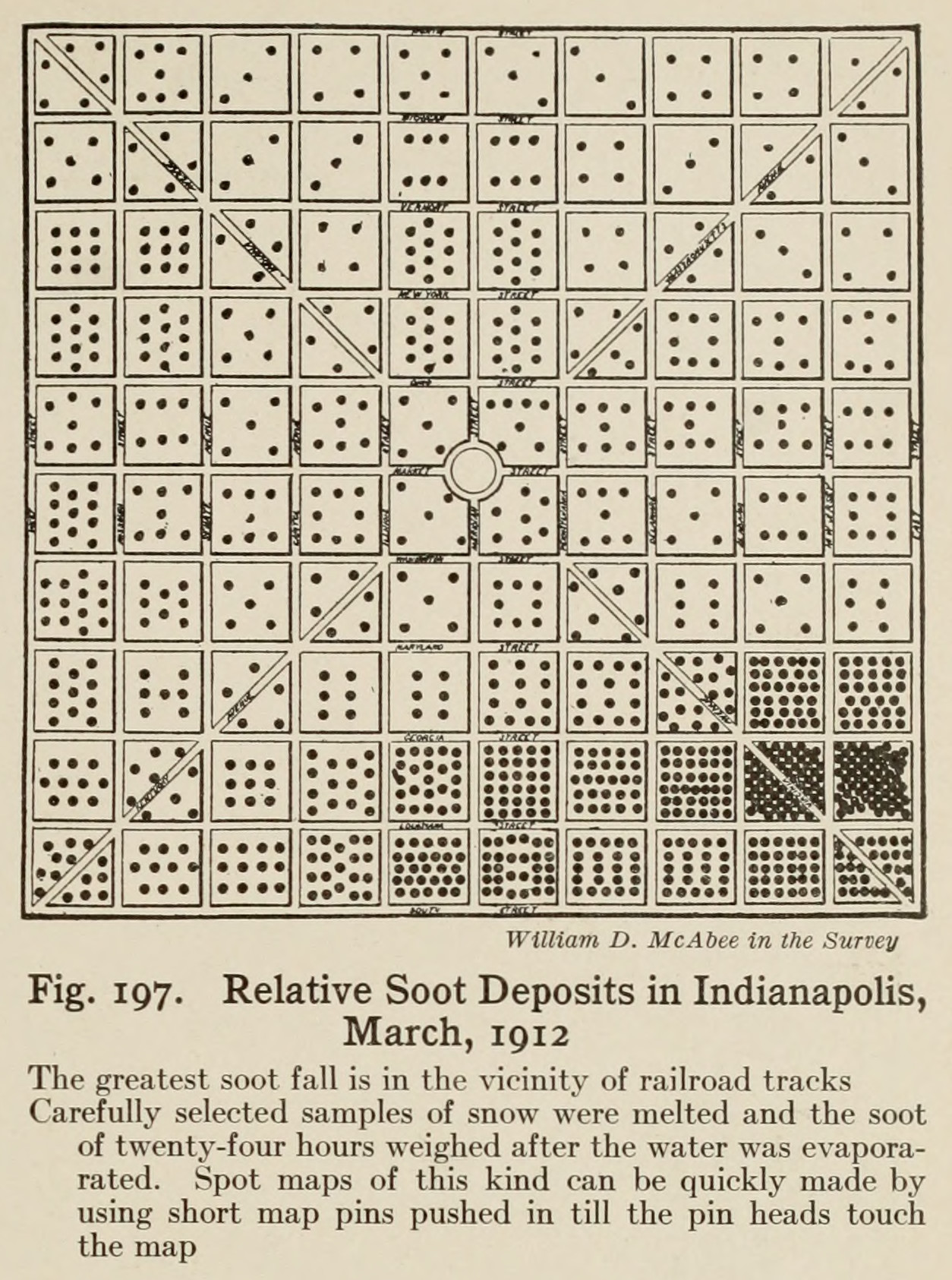}
	\caption{Historic example of grid-based (pollution) data for city blocks in Indianapolis in 1912 being visualized by \emph{lattice size} (density); by Willard C.\ Brinton \cite[page 245]{brinton:1919:graphic}; \ccPublicDomain\ the image is in the public domain.}
	\label{fig:pat_example_brinton_6}
\end{figure}

\begin{figure}[!t]
	\centering
	\includegraphics[width=\columnwidth]{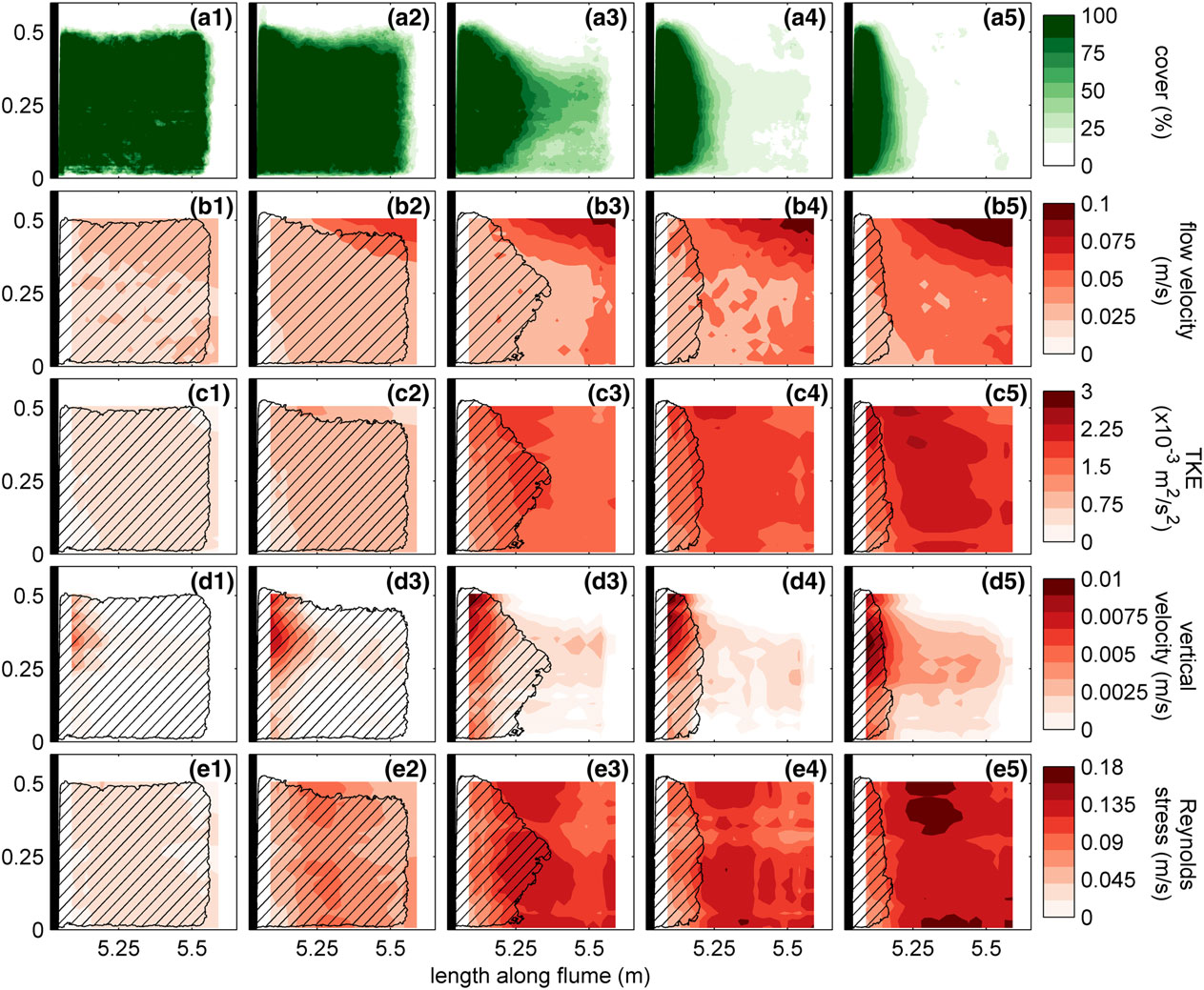}
	\caption{Contemporary example of a pattern being overlaid on a color-based representation to encode a second quantity (in this example the pattern only has two levels---with pattern and without it). Image reproduced from \cite[Fig.\ 5]{deBrouwer:2017:FTL}, used under the \href{https://creativecommons.org/licenses/by/4.0/}{Creative Commons At\-tri\-bu\-tion 4.0 International (\ccLogo\,\ccAttribution\ \mbox{CC BY-NC}) license}.}
	\label{fig:example_color_plus_pattern}
\end{figure}

\begin{figure}[!t]
	\centering
	\includegraphics[width=\columnwidth]{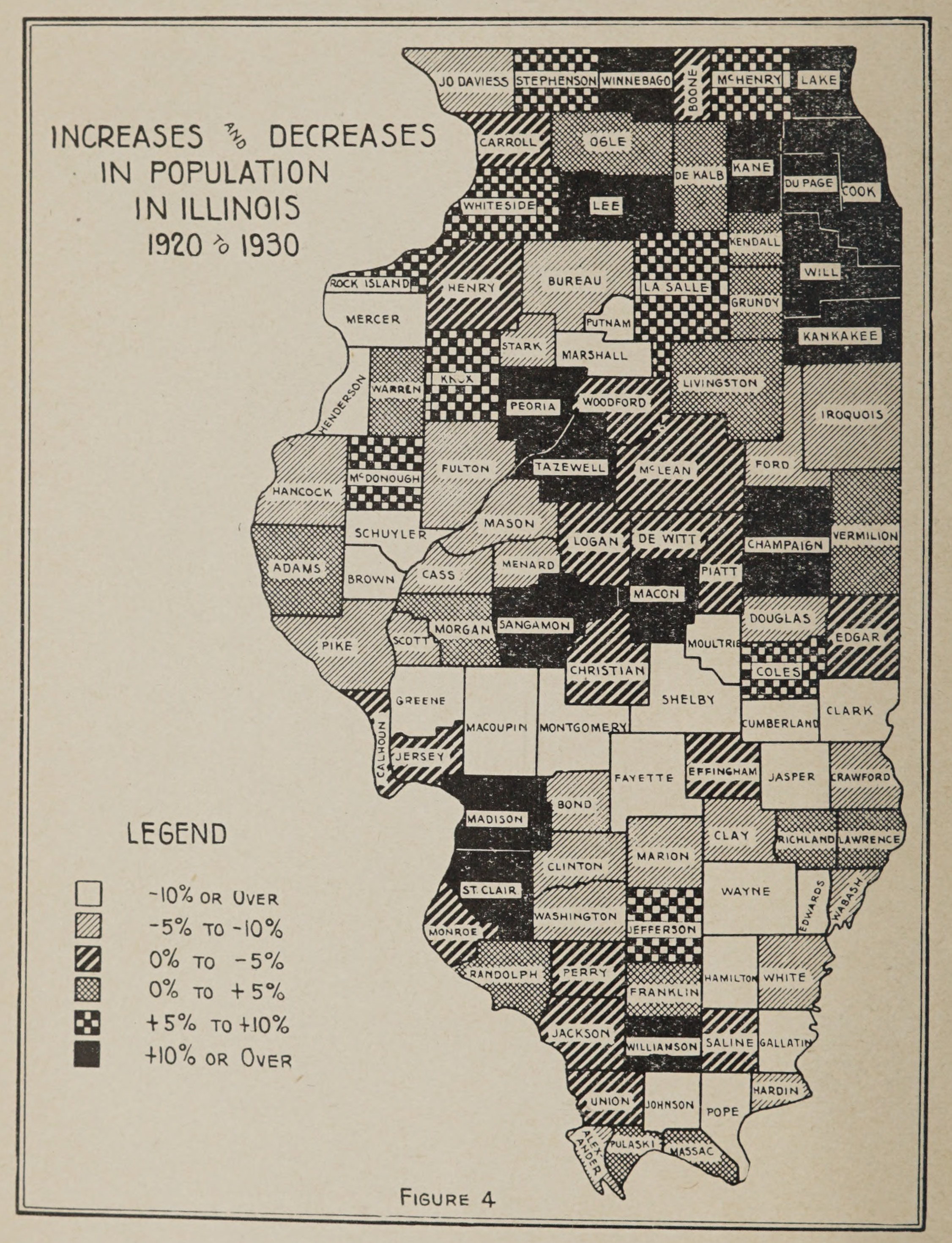}
	\caption{Historic example of a pattern being used to encode a diverging scale: line patterns for negative values and crossing lines resp.\ dot patterns for positive values; \href{https://archive.org/details/CAT10507566/page/24/mode/2up}{image published in 1933 by the State of Illinois, Department of Public Works and Buildings, Division of Highways} (Fig.~4 on page~24 of their ``\href{https://archive.org/details/CAT10507566/}{Economic Survey of Illinois}''), \ccPublicDomain\ the image is in the public domain.}
	\label{fig:pat_example_diverging_scale}
\end{figure}

\begin{figure*}[!t]
	\centering
	\includegraphics[width=\linewidth]{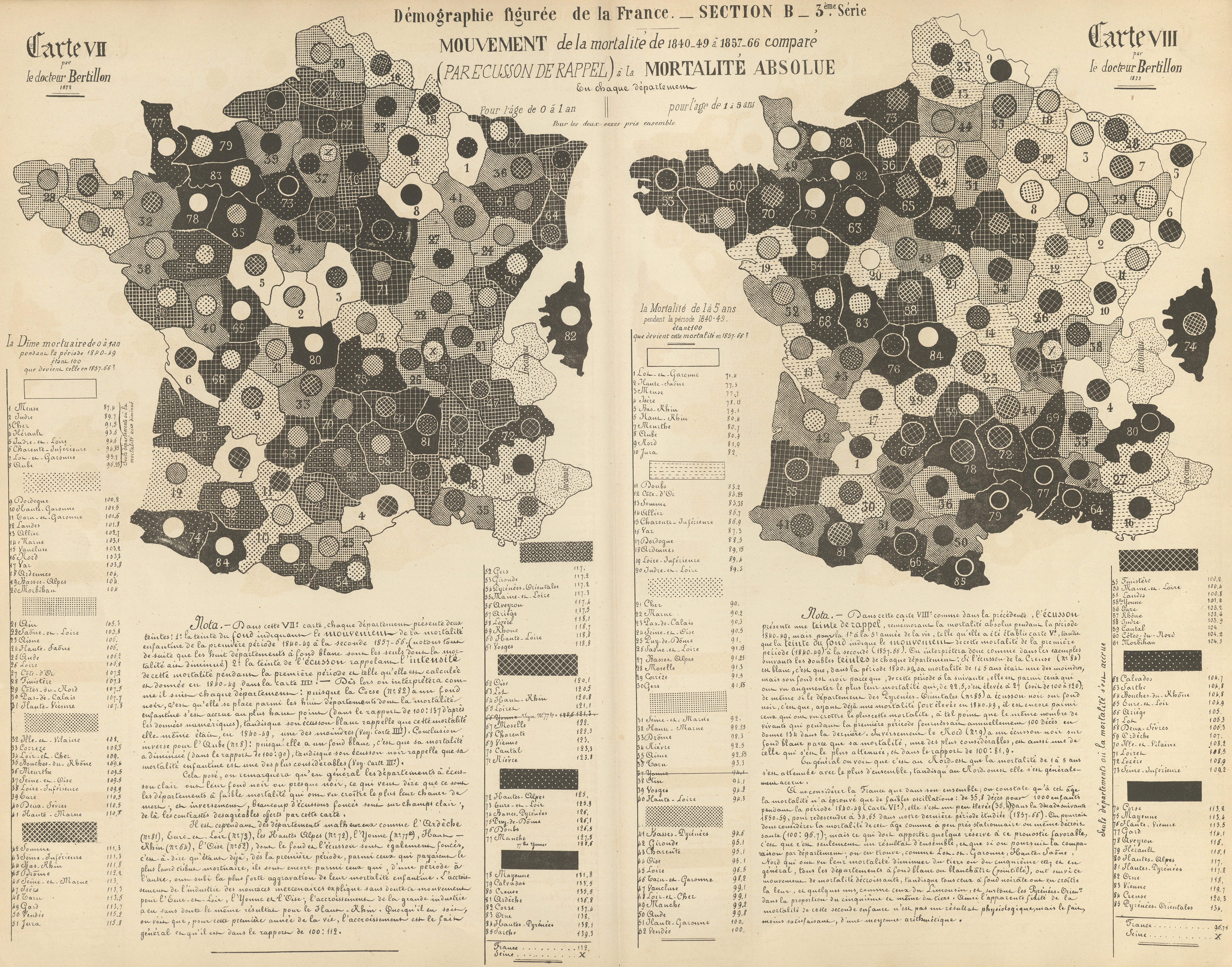}
	\caption{Historic example for a dual encoding with patterns, where one pattern fills the area mark, while another fills an inset circle (which, of course, requires large-enough area marks to begin with); image \href{https://www.davidrumsey.com/luna/servlet/detail/RUMSEY~8~1~338039~90106023}{\emph{``Mouvement de la mortalité de 1840-49 à 1857-66 comparé (parecusson de Rappel) à la mortalité absolue en chaque départemens les écussons rappelant la mortalité absolue des deux sexes de 0 à 1 an : Carte VII. Par le docteur Bertillon. 1872. -- Carte VIII. Par le docteur Bertillon. 1872.''}}; \ccPublicDomain\ the image is in the public domain.}
	\label{fig:example_pattern_on_pattern}
\end{figure*}

\begin{figure*}[!t]
	\centering
	\includegraphics[width=\linewidth]{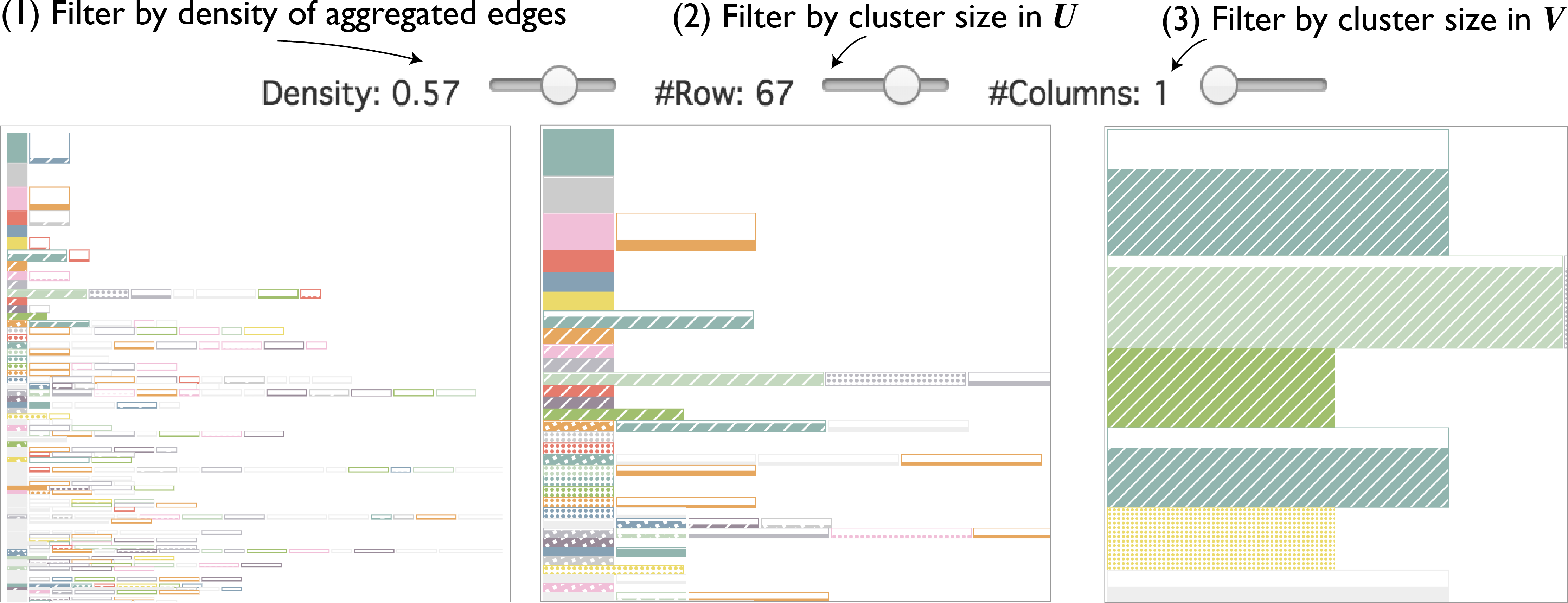}
	\caption{Contemporary example of combining color scales with patterns to improve the discriminability of a categorical encoding. Image reproduced from \cite[Fig.\ 6]{Chan:2019:ViBr}; \textcopyright\ IEEE, used with permission.}
	\label{fig:example_pattern_plus_color}
\end{figure*}

\begin{figure*}[!t]
	\centering
	\includegraphics[height=.285\linewidth]{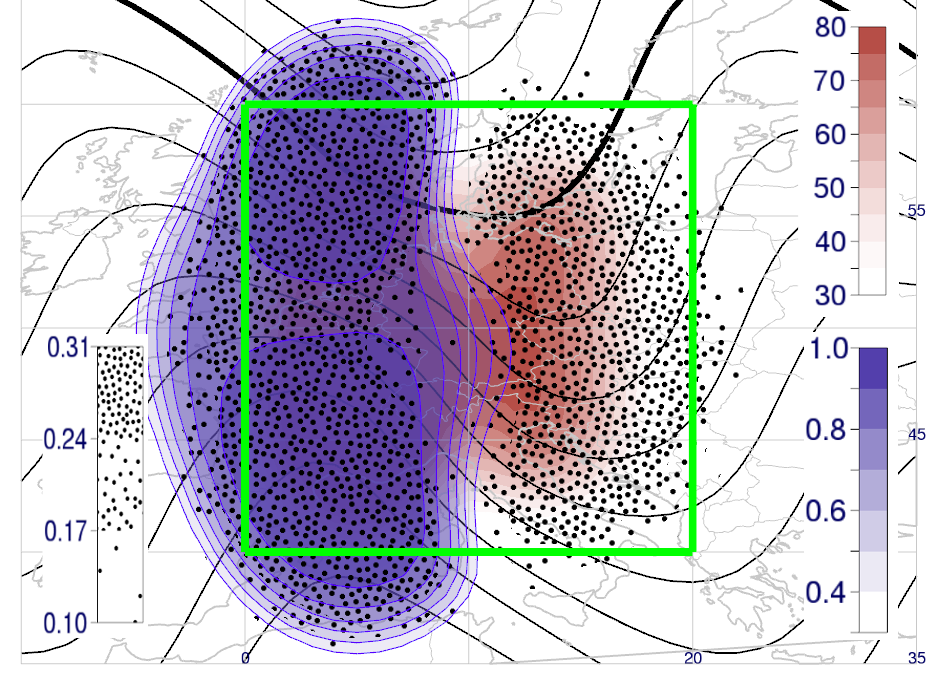}\hfill%
	\includegraphics[height=.285\linewidth]{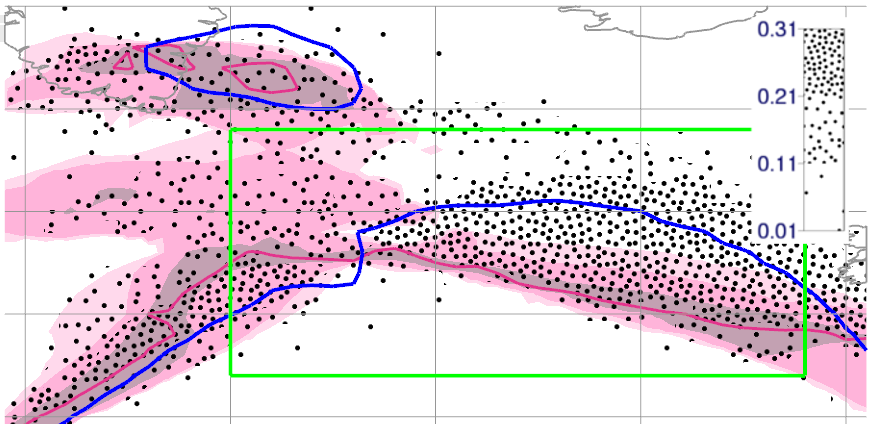}%
	\caption{Contemporary examples of combining color scales with patterns (here point densities of a stippling pattern) to encode an additional spatially changing attribute (here: robustness of the shown cluster). Images reproduced from \cite[Fig.\ 11(a) and 16(a)]{Kumpf:2018:VCC}; \textcopyright\ IEEE, used with permission.}
	\label{fig:example_pattern_plus_color2}
\end{figure*}

\begin{figure*}[!t]
	\centering
	\includegraphics[width=\linewidth]{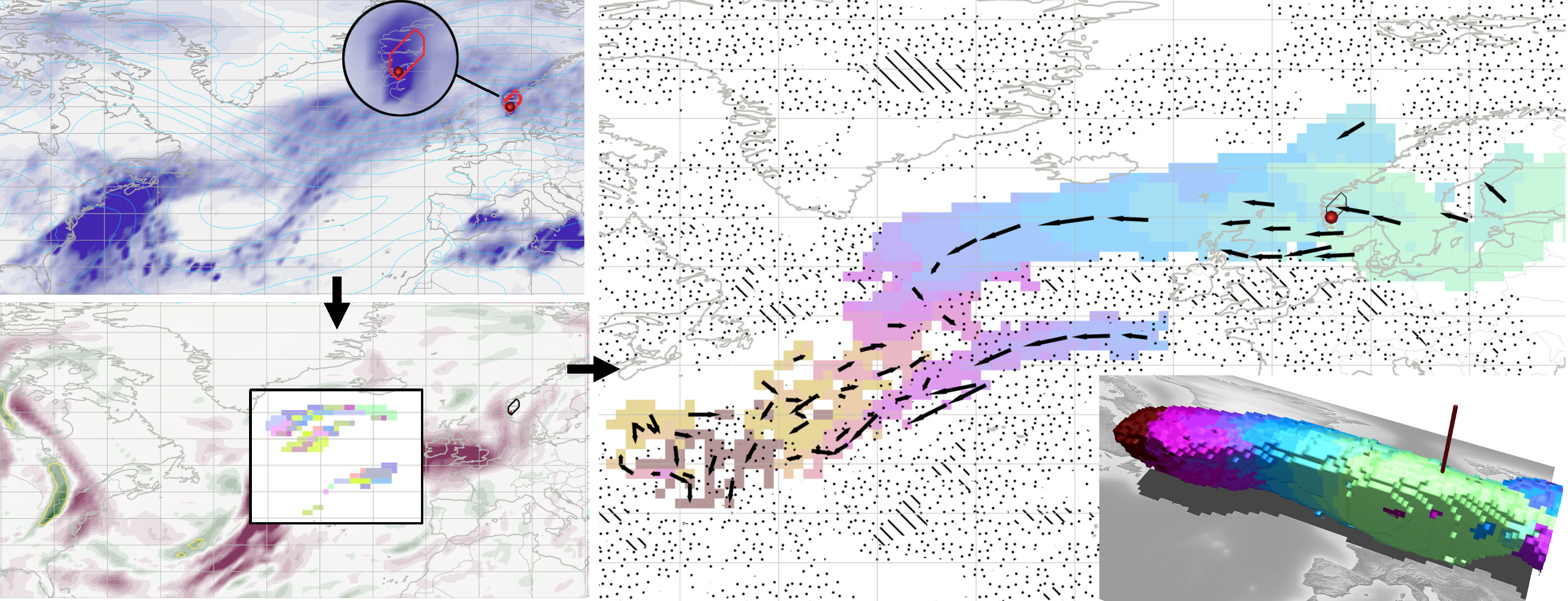}
	\caption{Contemporary example of combining color scales with patterns to visualize two levels of information in weather forecast time series. Image reproduced from \cite[Fig.\ 1]{Kumpf:2019:VAT}; \textcopyright\ IEEE, used with permission.}
	\label{fig:example_pattern_plus_color3}
\end{figure*}

\begin{figure*}[!t]
	\centering
	\includegraphics[height=.205\linewidth]{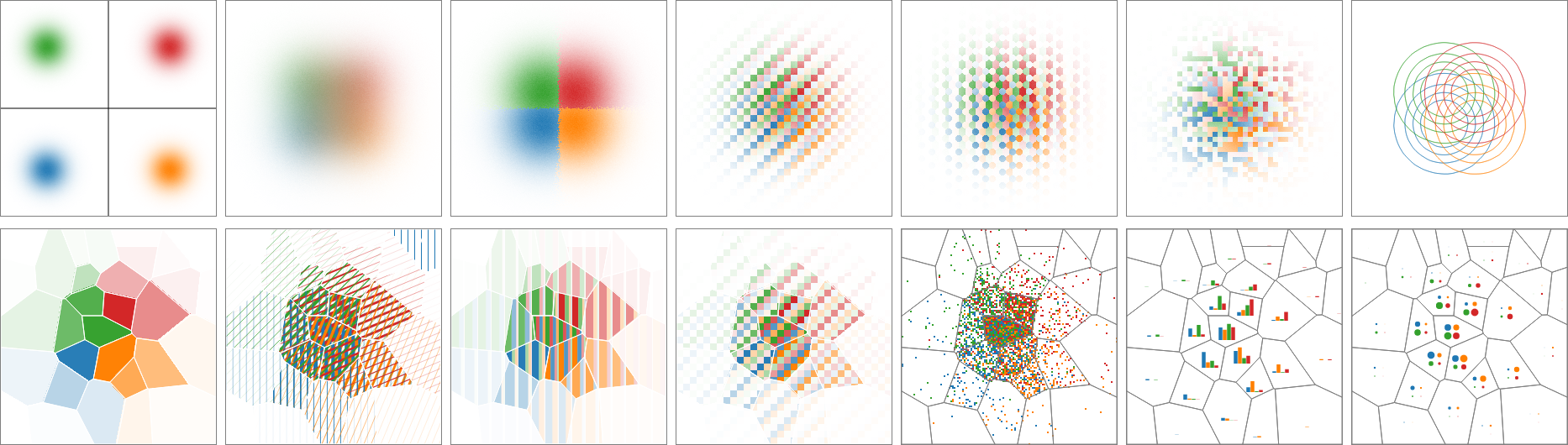}\hfill%
	\includegraphics[height=.205\linewidth]{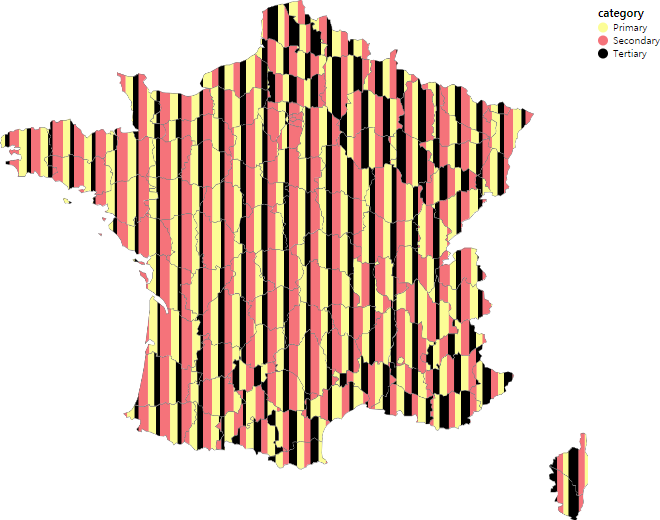}%
	\caption{Contemporary examples of Jo et al.'s \cite{Jo:2019:DRM} declarative rendering model for multi-class density maps, some of which rely on patterns (see \autoref{sec:pattern-design-tools}). The larger example on the right demonstrates the ability of the approach to reproduce Bertin's example we showed in \autoref{fig:grid-pattern-internal-variation:a}. Images reproduced from \cite[Fig.\ 1 and 6(c)]{Jo:2019:DRM}; \textcopyright\ IEEE, used with permission.}
	\label{fig:example_multiclass}
\end{figure*}

\begin{figure*}[!t]
	\centering
	\setlength{\subfigcapskip}{-3ex}%
	\subfigure[\hspace{\columnwidth}]{\label{fig:example_stippling_modern:a}~~~~~\includegraphics[height=.22\linewidth]{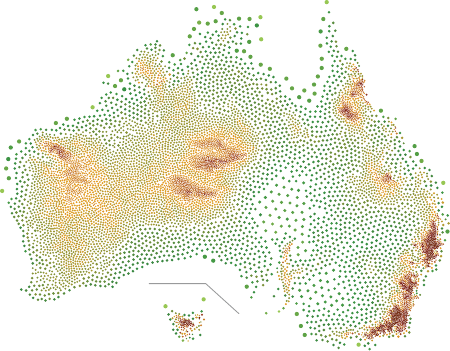}}\hspace{10mm}
	\subfigure[\hspace{\columnwidth}]{\label{fig:example_stippling_modern:b}~~~~~~~~~~~~~~~\includegraphics[height=.22\linewidth]{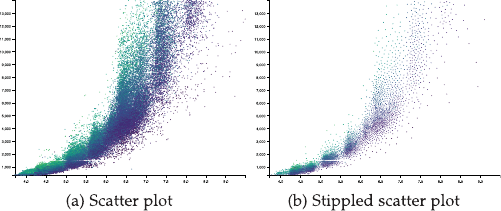}}%
	\caption{\hty{Contemporary examples by Görtler et al.\ \cite{Goertler:2019:S2S} of using stippling \cite{Martin:2017:SDS,Deussen:2013:HS} to show data, both for \subref{fig:example_stippling_modern:a} spatial and \subref{fig:example_stippling_modern:b} nonspatial datasets---similar to the historic examples in \autoref{fig:pat_example_brinton_4}--\ref{fig:land_in_farms}. Images reproduced from \cite[Fig.\ 8 and 11]{Goertler:2019:S2S}; \textcopyright\ IEEE, used with permission.}}
	\label{fig:example_stippling_modern}
	\label{fig:last_figure}
\end{figure*}

This section includes additional examples (\autoref{fig:great-war-isotype}--\ref{fig:last_figure}) from authors other than Bertin that were partially already discussed in the main paper---some of which we found via \href{https://oldvis.github.io/gallery/}{OldVisOnline} \cite{Zhang:2024:OCD}. 
Specifically, the historical example in \autoref{fig:great-war-isotype}---created before the the availability of computers---illustrates the use of pattern groups and of the relationship of their respective appearances to encode data (by shape and color in \autoref{fig:great-war-isotype}; another example that uses line orientation and secondary line width is included in the main paper as \autoref{fig:NYCmap}). 
\autoref{fig:Retchless15Uncertaint}, in contrast, is a recent example that illustrates the use of retinal visual variables on each of the pattern's primitives.

\changed{\autoref{fig:pat_example_brinton_7} and \ref{fig:pat_example_brinton_8} demonstrate what we refer to as patterns based on 1D lattices (\autoref{fig:pat_example_brinton_7}) and 2D lattices (\autoref{fig:pat_example_brinton_8}). \autoref{fig:nested_pattern_od_map_of_london}--\ref{fig:data-driven-geo-pattern} then showcase various examples of non-lattice-based patterns.}

\autoref{fig:pat_example_brinton_1}--\ref{fig:pat_example_brinton_3} show historical examples of pattern legends to be used in visual representations, for example \autoref{fig:pat_example_brinton_1}--\ref{fig:pat_example_brinton_2} based on \emph{density} for encoding increasing value ranges.
\autoref{fig:pat_example_brinton_4}--\ref{fig:land_in_farms} and \ref{fig:pat_example_brinton_6} use the \emph{density} of simple dot patterns to encode values, with interesting aspects such as a (possible) transition from a \emph{density} to a \emph{position} encoding (\autoref{fig:pat_example_brinton_4}) or the use of dot sizes that accurately represent the dot's referents in a map (\autoref{fig:land_in_farms}).
\autoref{fig:pixel_patterns} then technically does not show a pattern but each (minute) mark gets only one (square) primitive, but due to the composition of the marks we see patterns at a coarser (hour) data scale.



\autoref{fig:example_color_plus_pattern} shows a contemporary example, where a pattern is used as a secondary data layer to a color-based visualization. We found examples such as this one, \eg, to show uncertainty for the color-encoded data, or as in this case a meaningful subarea of the plot was marked.

\autoref{fig:pat_example_diverging_scale} shows a pattern design for a diverging scale, but we would argue that it could be improved for a better support of preattentive value reading. We argue, \eg, that the neutral point should be white for both the positive and the negative part of the scale, and then the negative part could use an increasing line width or increasing line frequency of diagonal lines \inlinevis{-1pt}{1em}{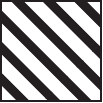} (but with a negative slope orientation to suggest negativity), and the positive part could either use crossing diagonal lines \inlinevis{-1pt}{1em}{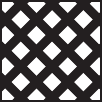} or, potentially even better, crossing horizontal and vertical lines \inlinevis{-1pt}{1em}{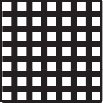} to suggest the shape of a plus---again with an increasing line width or increasing line frequency for more positive values. Both ``sides'' of the scale should be designed in such a way that their perceived \emph{value} at any given absolute level is the same, regardless of whether it is the positive or the negative side of the scale. Ultimately, further design and empirical research would be needed to identify suitable scales (and not only for diverging scales), to result in something along the lines of what ColorBrewer \cite{brewer:2016:DesigningBetterMaps} provides us with for color scales in visualization.

\autoref{fig:example_overlapping_marks} comes from a tutorial (see below) on how to fill plots with patterns in Matplotlib, and demonstrates cleverly designed patterns that clearly indicate where two data marks overlap---in a similar way as transparent colors would do for overlapping marks.
\autoref{fig:example_pattern_on_pattern} shows again a historic example where two layers of patterns have been used to encode data. The first layer completely fills the mark, but the second layer relies on additional circular area marks being overlaid (or inset) on each main mark (which, of course, requires the main marks to be of sufficient size).
Next, \autoref{fig:example_pattern_plus_color}--\ref{fig:example_pattern_plus_color3} show contemporary examples of combining color maps with patterns to either make categories more distinguishable (\autoref{fig:example_pattern_plus_color}) or to visualize two layers of information at the same time (\autoref{fig:example_pattern_plus_color2}, \ref{fig:example_pattern_plus_color3}). \autoref{fig:example_multiclass} demonstrates Jo et al.'s \cite{Jo:2019:DRM} declarative rendering model for multiclass density maps, some of whose approaches rely on patterns to make the different data classes dinstinguisheable from each other. The larger example on the right side of the figure demonstrates the ability of this approach to reproduce Bertin's technique of color-based categorical encoding for geographic data we showed in \autoref{fig:grid-pattern-internal-variation:a} and discussed in \autoref{sec:ratio-between-each-group}. \hty{Finally, \autoref{fig:example_stippling_modern} show contemporary examples of using stippling as a non-lattice arrangement.}

\section{Discussion of current solutions for implementing patterns in visualizations}
\label{sec:pattern-design-tools}
Current visualization tools and graphical drawing libraries offer limited support for patterns. Most tools provide only a few options to vary patterns, so users cannot fully use all pattern attributes. Achieving complex patterns often requires programming skills, making their implementation challenging, especially for designers with limited technical backgrounds. For example, one of the most popular visualization tools, Tableau, does not officially support pattern fills at this point in time. There have been requests for this feature in the forums since 10 years ago. For example, such as the posts ``Pattern fill'' (\href{https://community.tableau.com/s/question/0D54T00000C5nGlSAJ/pattern-fill}{\texttt{community\discretionary{}{.}{.}tableau.com\discretionary{/}{}{/}s\discretionary{/}{}{/}question\discretionary{/}{}{/}0D54T00000C5nGlSAJ\discretionary{/}{}{/}pattern\discretionary{}{-}{-}fill}}) and ``Fill Patterns (Dots and Stripes)'' (\href{https://community.tableau.com/s/question/0D54T00000C5s3GSAR/fill-patterns-dots-and-stripes}{\texttt{community\discretionary{}{.}{.}tableau.com\discretionary{/}{}{/}s\discretionary{/}{}{/}question\discretionary{/}{}{/}0D54T00000C5s3GSAR\discretionary{/}{}{/}fill\discretionary{}{-}{-}patterns\discretionary{}{-}{-}dots\discretionary{}{-}{-}and\discretionary{}{-}{-}stripes}}).
This functionality, however, has not yet been integrated. Users can use workarounds. For example, A.\ McCann shared two tutorials in 2018: ``Multiple pattern fill bar charts'' (\href{https://duelingdata.blogspot.com/2018/06/multiple-pattern-fill-bar-charts.html}{\texttt{duelingdata\discretionary{}{.}{.}blogspot\discretionary{}{.}{.}com\discretionary{/}{}{/}2018\discretionary{/}{}{/}06\discretionary{/}{}{/}multiple\discretionary{}{-}{-}pattern\discretionary{}{-}{-}fill\discretionary{}{-}{-}bar\discretionary{}{-}{-}charts\discretionary{}{.}{.}html}}) and ``Pattern fill bar chart in Tableau'' (\href{https://duelingdata.blogspot.com/2018/06/pattern-fill-bar-chart-in-tableau.html}{\texttt{duelingdata\discretionary{}{.}{.}blogspot\discretionary{}{.}{.}com\discretionary{/}{}{/}2018\discretionary{/}{}{/}06\discretionary{/}{}{/}pattern\discretionary{}{-}{-}fill\discretionary{}{-}{-}bar\discretionary{}{-}{-}chart\discretionary{}{-}{-}in\discretionary{}{-}{-}tableau\discretionary{}{.}{.}html}}).
These operations, however, are far more complicated than using other visual variables such as color, size, or shape.
Some graphical drawing libraries offer pattern fills, but users can only select from default patterns or create repetitive tilings of shapes on a grid. Examples include Matplotlib \cite{Hunter:2007:Matplotlib} and Plotly \cite{plotly} for Python, and ggpattern \cite{ggpattern} for R. Below we list additional tools or source code resources we found during our explorations for creating patterns that offer more flexibility:

\begin{itemize}
\item SVG's \texttt{<pattern>} Element: If we want to create patterns in charts, one of the most flexible options is using the SVG \texttt{<pattern>} element, which offers a high degree of customization. It is described, however, as ``arguably one of the more confusing fill types to use in SVG'' \cite{mdn:svg-patterns}. It is based on, and thus confined to, the repetitive tiling of shapes in vertical and horizontal directions. Therefore, even the creation of a frequently used diagonal line pattern such as \inlinevis{-1pt}{1em}{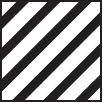} does not follow the native logic of SVG \texttt{<pattern>} and can confuse people. For example, a question on Stack Overflow highlights this issue: ``Simple fill pattern in SVG: diagonal hatching'' (\href{https://stackoverflow.com/questions/13069446/simple-fill-pattern-in-svg-diagonal-hatching}{\texttt{stackoverflow\discretionary{}{.}{.}com\discretionary{/}{}{/}questions\discretionary{/}{}{/}13069446\discretionary{/}{}{/}simple\discretionary{-}{}{-}fill\discretionary{-}{}{-}pattern\discretionary{-}{}{-}in\discretionary{-}{}{-}svg\discretionary{-}{}{-}diagonal\discretionary{-}{}{-}hatching}}).

\item \texttt{Textures.js}---SVG patterns for Data Visualization: \href{https://riccardoscalco.it/textures/}{\texttt{riccardoscalco\discretionary{}{.}{.}it\discretionary{/}{}{/}textures}} and \href{https://github.com/riccardoscalco/textures}{\texttt{github\discretionary{}{.}{.}com\discretionary{/}{}{/}riccardoscalco\discretionary{/}{}{/}textures}}; this software package aims to make SVG's \texttt{<pattern>} easier to use, but it still requires extensive manual coding and does not fully break the constraint of repetitive tiling shapes. In addition, its options are limited. For example, although it supports line patterns, the lines cannot fully rotate 180 degrees and have only several predefined rotation options, such as 3/8.

\item \emph{Design Characterization for Black-and-White [sic!] Textures}: our own tool for the generation of simple iconic and abstract patterns from our previous work on patterns \cite{He:2024:DCB,Zhong:2020:BWT}; \href{https://github.com/tingying-he/design-characterization-for-black-and-white-textures-in-visualization}{\texttt{github\discretionary{}{.}{.}com\discretionary{/}{}{/}tingying\discretionary{}{-}{-}he\discretionary{/}{}{/}design\discretionary{}{-}{-}characterization\discretionary{}{-}{-}for\discretionary{}{-}{-}black\discretionary{}{-}{-}and\discretionary{}{-}{-}white\discretionary{}{-}{-}textures\discretionary{}{-}{-}in\discretionary{}{-}{-}visualization}}; it offers greater freedom and ease of use compared to existing libraries and tools. Since we developed this tool earlier than we proposed our pattern design space in this paper, however, the tool does not cover all possible pattern variations. In addition, the tool is currently built around a web design interface, with which users can only design patterns in the given charts on the web, rather than integrate patterns into their own visualizations (\autoref{fig:patterns-in-vis:c}).

\item Jo et al.'s \cite{Jo:2019:DRM} \emph{Declarative Rendering Model for Multiclass Density
Maps}: \href{https://www.jaeminjo.com/Multiclass-Density-Maps/}{\texttt{jaeminjo\discretionary{}{.}{.}com\discretionary{/}{}{/}Multiclass\discretionary{}{-}{-}Density\discretionary{}{-}{-}Maps}} and \href{https://github.com/e-/Multiclass-Density-Maps}{\texttt{github\discretionary{}{.}{.}com/e\discretionary{}{-}{-}\discretionary{/}{}{/}Multiclass\discretionary{}{-}{-}Density\discretionary{}{-}{-}Maps}}; this tool focuses on maps and some of its techniques rely on patterns (\autoref{fig:example_multiclass})

\item \emph{Line Textures} by Richard Brath: source code for creating line patterns using D3 and dasharray; \href{https://observablehq.com/@richardbrath/line-textures}{\texttt{observablehq\discretionary{}{.}{.}com\discretionary{/}{}{/}@richardbrath\discretionary{/}{}{/}line\discretionary{}{-}{-}textures}}

\item \emph{Remaking Figures from Bertin's \textit{Semiology of Graphics}} by Nicolas Kruchten: code in Python, using Plotly Express, to recreate some of Bertin's patterns; \href{https://nicolas.kruchten.com/semiology_of_graphics/}{\texttt{nicolas\discretionary{}{.}{.}kruchten\discretionary{}{.}{.}com\discretionary{/}{}{/}semiology\discretionary{\_}{}{\_}of\discretionary{\_}{}{\_}graphics}}

\item \emph{Generating different spatial patterns in R and their visualization using ggplot2} by Muhammad Mohsin Raza: \href{https://www.datawim.com/post/generating-different-patterns-in-r/}{\texttt{datawim\discretionary{}{.}{.}com\discretionary{/}{}{/}post\discretionary{/}{}{/}generating\discretionary{}{-}{-}different\discretionary{}{-}{-}patterns\discretionary{}{-}{-}in\discretionary{}{-}{-}r}}

\item \emph{How To Fill Plots With Patterns In Matplotlib} by Elena Kosourova: \href{https://towardsdatascience.com/how-to-fill-plots-with-patterns-in-matplotlib-58ad41ea8cf8/}{\texttt{towardsdatascience\discretionary{}{.}{.}com\discretionary{/}{}{/}how\discretionary{}{-}{-}to\discretionary{}{-}{-}fill\discretionary{}{-}{-}plots\discretionary{}{-}{-}with\discretionary{}{-}{-}patterns\discretionary{}{-}{-}in\discretionary{}{-}{-}matplotlib\discretionary{}{-}{-}58ad41ea8cf8}} (\autoref{fig:example_overlapping_marks})
\end{itemize}

%

\section{Footnotes}

\theendnotes

\section*{Images/figures license/copyright}
\begin{figure}
    \centering%
        \setlength{\subfigcapskip}{-2.5ex}%
    \subfigure[\hspace{\columnwidth}]{\label{fig:nested-pattern-geo-on-lattice:a}~~~~~\includegraphics[width=0.25\columnwidth]{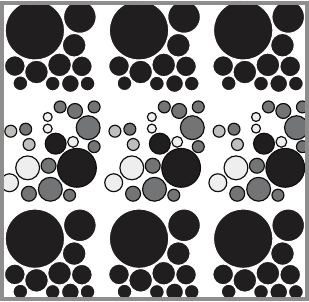}}\hfill
    \subfigure[\hspace{\columnwidth}]{\label{fig:nested-pattern-geo-on-lattice:b}~~~~~\includegraphics[width=0.6\columnwidth]{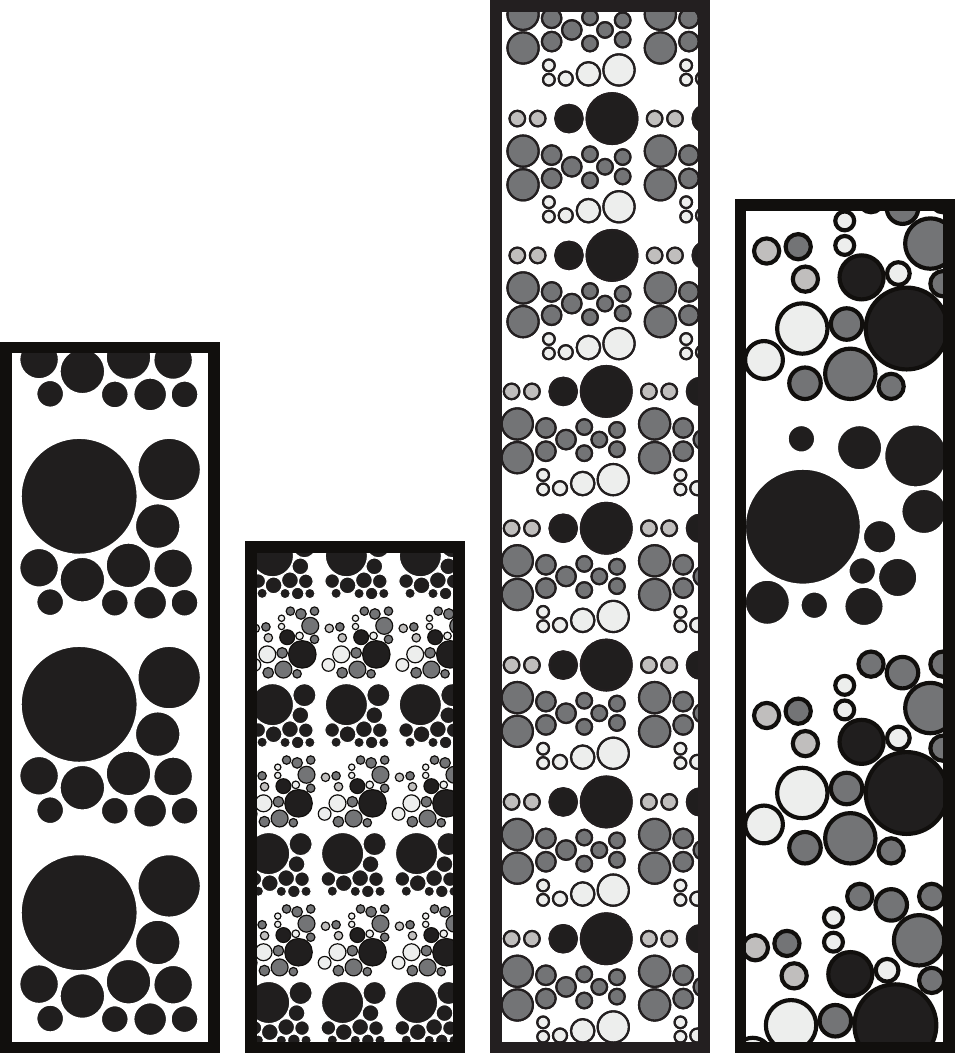}}
    \caption{\changed{(a) An example of a type of nested pattern: adding data-driven patterns on top of a lattice-based pattern.(b) A potential application of the nested pattern shown in (a) within a bar chart. This design shows a possibility opened by our pattern system; future work can explore which types of data are best suited for this encoding format.}}%
    \label{fig:nested-pattern-geo-on-lattice}
\end{figure}

\begin{figure}
    \centering%
        \setlength{\subfigcapskip}{-3ex}%
        \subfigure[\hspace{\columnwidth}]{\label{fig:data-driven-geo-pattern:a}~~~~~\includegraphics[width=\examplewidth]{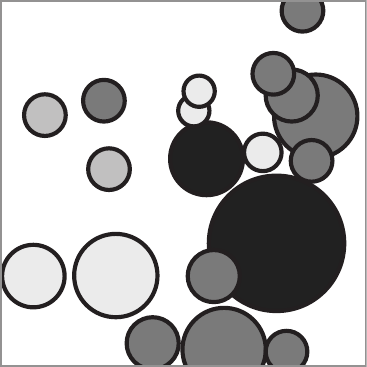}}\hspace{5mm}%
        \subfigure[\hspace{\columnwidth}]{\label{fig:data-driven-geo-pattern:b}~~~~~\includegraphics[width=\examplewidth]{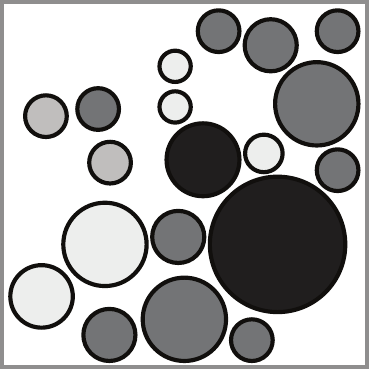}}\hspace{5mm}
        \subfigure[\hspace{\columnwidth}]{\label{fig:data-driven-geo-pattern:c}~~~~~\includegraphics[width=\examplewidth]{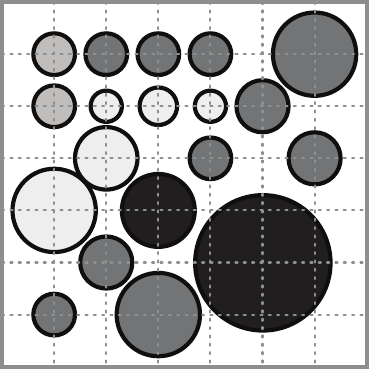}}
    \caption{\changed{Patterns with primitives based on geographically data-driven spatial arrangements: (a) accurate geographic data; (b) inaccurate geographic data, with dot overlap avoided; (c) gridded geographic data.}}%
    \label{fig:data-driven-geo-pattern}%
\end{figure}

\begin{figure}[!t]
	\centering
	\includegraphics[width=\columnwidth]{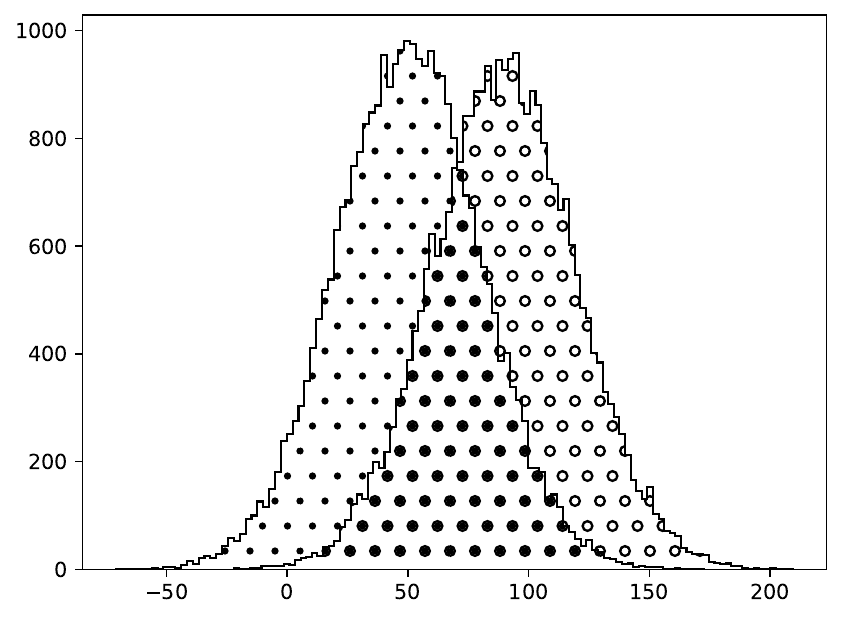}
	\caption{Contemporary example of two merging area marks, with cleverly designed patterns that highlight the overlap between both marks; image created with Python and Matplotlib \cite{Hunter:2007:Matplotlib}, following the \href{https://towardsdatascience.com/how-to-fill-plots-with-patterns-in-matplotlib-58ad41ea8cf8/}{tutorial by Elena Kosourova}, see \autoref{sec:pattern-design-tools}.}
	\label{fig:example_overlapping_marks}
\end{figure}

With the exception of those images from external authors whose li\-cen\-ses\discretionary{/}{}{/}co\-py\-rights we have specified in the respective figure captions, we as authors state that all of our own figures in this appendix (\ie, those not marked: \autoref{fig:Caivano-SimpleTextureComposition}, \autoref{fig:TextureComposition}--\ref{fig:PatternVariation-2Dp2Da-P-regularity}, \autoref{fig:nested-pattern-geo-on-lattice}--\ref{fig:data-driven-geo-pattern}, and \autoref{fig:example_overlapping_marks} as well as the inline word-scale graphics) are and remain under our own personal copyright, with the permission to be used here. We also make them available under the \href{https://creativecommons.org/licenses/by/4.0/}{Creative Commons At\-tri\-bu\-tion 4.0 International (\ccLogo\,\ccAttribution\ \mbox{CC BY 4.0})} license and share them at \osfrepo.

%

\end{document}